\documentstyle[seceq,supplement]{ptptex}

\title{
Post-Newtonian Approximation}
\subtitle{Its Foundation and Applications 
}    

\author{
Hideki Asada$^{*}$
and Toshifumi Futamase 
}

\inst{
$^*$
Yukawa Institute for Theoretical Physics, 
Kyoto University, Kyoto 606-01 \\
Astronomical Institute, Graduate School of 
Science, Tohoku University \\
Sendai 980-77, Japan
}

\abst{
We discuss various aspects of the post-Newtonian approximation in
general relativity. After presenting the foundation based on the
Newtonian limit, we use the (3+1) formalism to formulate the
post-Newtonian approximation for the perfect fluid. As an application
we show the method for constructing the equilibrium configuration of
nonaxisymmetric uniformly rotating fluid. We also discuss the
gravitational waves including tail from post-Newtonian systems. 
}

\begin{document}

\maketitle

\def\pa{\partial}
\def\bI{\hbox{$\,I\!\!\!\!-$}}
\def\two{\hbox{$_{(2)}$}}
\def\three{\hbox{$_{(3)}$}}
\def\four{\hbox{$_{(4)}$}}
\def\five{\hbox{$_{(5)}$}}
\def\six{\hbox{$_{(6)}$}}
\def\seven{\hbox{$_{(7)}$}}
\def\eight{\hbox{$_{(8)}$}}
\newcommand{\lsim}{\raisebox{0.3mm}{\em $\, <$} \hspace{-3.3mm}
\raisebox{-1.8mm}{\em $\sim \,$}}

\section{Introduction}
The motion and associated emission of gravitational waves (GW) of 
self gravitating systems have been a main research interest 
in general relativity.
The problem is complicated conceptually as well as mathematically 
because of the nonlinearity of Einstein's equations. 
There is no hope in any foreseeable future to have exact solutions 
describing motions of arbitrary shaped, massive bodies so that 
we have to adopt some sort of approximation scheme for solving 
Einstein's equations to study such problems.
In the past years many types of approximation schemes have been developed 
depending on the nature of the system under consideration. 
Here we shall focus on a particular scheme called the post-Newtonian (PN) 
approximation. There are many systems in astrophysics 
where Newtonian gravity is dominant, but general relativistic gravity plays 
also important roles in their evolution. 
For such systems it would be nice to have an approximation scheme which 
gives Newtonian description in the lowest order and  
general relativistic effects as higher order perturbations. 
The post-Newtonian approximation is perfectly suited for this purpose. 
Historically Einstein computed first the post-Newtonian effects, 
the precession of the perihelion\cite{einstein}, but a systematic 
study of the post-Newtonian approximation was not made until 
the series of papers by Chandrasekhar and associates 
\cite{chandra65,chandra67,chandra69a,cn,ce}.
Now it is widely known that the post-Newtonian approximation is important 
in analyzing a number of relativistic problems, such as the 
equation of motion of binary pulsar\cite{bt,epstein,ht,dt}, 
solar-system tests of general relativity\cite{will87,will94}, 
and gravitational radiation reaction\cite{ce,burke}.
 
Any approximation scheme nesseciates a  small 
parameter(s) characterizing the nature of the system under consideration.
Typical parameter which most of schemes adopt is the magnitude of 
the metric deviation away from a certain background metric. 
In particular if the background is Minkowski spacetime and 
there is no other parameter, the scheme is sometimes called as 
the post-Minkowskian approximation in the sense that 
the constructed spacetime reduces to Minkowski spacetime 
in the limit as the parameter tends to zero. 
This limit is called as the weak field limit.  
In the case of the post-Newtonian approximation 
the background spacetime is also 
Minkowski spacetime, but there is another small parameter, that is, 
the typical velocity of the system divided by the speed of light 
which we call $\epsilon$ henceforth. 
These two parameters 
(the deviation away from the flat metric and the velocity) 
have to have a certain relation in the following sense. 
As the gravitational field gets weaker, all velocities and forces 
characteristic of the material systems become smaller, 
in order to permit the weakening gravity to remain 
an important effect on the system's dynamics. 
For example in case of a binary system, the typical velocity 
would be the orbital velocity $v/c \sim \epsilon$ and 
the deviation from the flat metric would be the Newtonian potential, 
say $\Phi$. 
Then these are related by $ \Phi /c^2 \sim v^2/c^2 \sim \epsilon^2$ 
which guarantees that the system is bounded by its own gravity.

In the post-Newtonian approximation, the equations in general relativity 
take the form of Newton's equations in an appropriate limit 
as $\epsilon \to 0$. Such a limit is called as the Newtonian limit 
and it will be the bases of constructing the post-Newtonian approximation. 
However the limit is not in any sense trivial 
since the limit may be thought of two limits tied together as just described.
It is also worth noting that the Newtonian limit cannot be uniform 
everywhere for all time. For example any compact binary systems, 
no matter how weak the gravity between components and slow its orbital motion, 
will eventually spiral together due to backreaction of the emission 
of gravitational waves. As the result the effects of its Newtonian gravity  
will be swamped by those of its gravitational waves. 
This will mean that higher order effect of the post-Newtonian 
approximation eventually dominates the lowest order Newtonian dynamics 
and thus if the post-Newtonian approximation is not carefully constructed, 
this effect can lead to many formal problems, such as divergent
integrals \cite{ergh}. 
It has been shown that such divergences may be 
avoided by carefully defining Newtonian limit\cite{futamase83}. 
Moreover, the use of such limit provides us a strong indication that the 
post-Newtonian hierarchy is an asymptotic approximation to 
general relativity\cite{fs83}.
Therefore we shall first discuss in this paper the Newtonian limit and 
how to construct the post-Newtonian hierarchy before attacking
practical problems in later sections. 

Before going into the details, we mention the reason for the 
growth of interest in the post-Newtonian approximation in recent years.  
Certainly the discovery of the binary neutron star system 
PSR 1913+16 was a strong reason to have renewed interest 
in the post-Newtonian approximation since it is 
the first system in which general relativitistic gravity plays 
fundamental roles in its evolution\cite{ht}. Particularly 
indirect discovery of gravitational wave by the observation of 
the period shortening led to many fruitful studies of 
the equation of motion with gravitational radiation emission 
in 1980's\cite{damour82,schutz84,walker}.
The effect of radiation reaction appears as the form of the potential force 
at the order of $\epsilon^5$ higher than Newtonian force in the equation 
of motion. There have been various attempts to show the validity of 
the so-called quadrupole formula for the radiation reaction
\cite{ad,futamase83,iww,kates80a,kates80b,kerlick80a,kerlick80b,schutz80,ww80a,ww80b}.

At that time, however, no serious attempts had been made for 
the study of higher order effects in the equation of motion. 
The situation changed gradually in 1990's because of the increasing 
expectation of direct detection of gravitational waves by
Kilometer-size interferometric gravitational wave detectors, 
such as LIGO\cite{abramovici}, VIRGO\cite{bradaschia} and 
TAMA\cite{kuroda} now under construction. 
Coalescing binary neutron stars are the most promising candidate of 
sources of gravitational waves for such detectors.  
The reasons are that (1) we expect to detect the signal of coalescence of
binary neutron stars about several times per year \cite{phinney}, and
(2) the waveform from coalescing binaries can be predicted with a high
accuracy compared with other sources \cite{abramovici,thorne94,will94}. 
Informations carried by gravitational waves tell us not only 
various physical parameters of neutron
star\cite{thorne87,thorne94}, but also the cosmological 
parameters\cite{schutz86,mark,finn,wst} 
if and only if we can make a detailed comparison 
between the observed signal with theoretical prediction during 
the epoch of the so-called inspiraling phase where 
the orbital separation is much larger than the radius of component stars 
\cite{cab}. 
This is the place where the post-Newtonian approximation may be applied to 
make a theoretical template for gravitational waves. 
The problem is that in order to make any meaningful comparison between 
theory and observation  we need to know the detailed waveforms generated by  
the motion up to 4PN order which is of order $\epsilon^8$ 
higher than the Newtonian order \cite{tn}. 
This request from gravitational wave
astronomy forces us to construct higher order post-Newtonian
approximation. 
 
It should be mentioned that Blanchet and Damour have developed a systematic 
scheme to calculate the waveform at higher order where 
the post-Minkowskian approximation is used to construct the external field and 
the post-Newtonian approximation is used to construct the field 
near the material source\cite{bd84a,bd84b,bd86,bd88,bd92}. 
Blanchet has obtained the waveform up to the 2.5 PN order which is of 
order  $\epsilon^5$ higher than the lowest qudraupole wave 
\cite{bdiww,blan95,blan96} by using the equation of motion up to 
that order \cite{dd81a,dd81b,dd81c,gk,kopejkin}. 
We shall also study the waveform from different point of view because
of its importance. 
We obtain essentially the same results including
the tail term with them.  

There is another phase in the evolution of the binary neutron star which we 
can withdraw useful informations from the observed signals 
of gravitational waves. 
That is the intermediate phase between the inspiralling phase and the 
event of merging. In the inspiralling phase the components are usually 
treated as point particles, while the extendedness becomes important 
in the intermediate phase and the merging. Full general relativistic 
hydrodynamic simulation is needed for the understanding of the merging.
The material initial data for the simulation will be the density and the 
velocity distribution of the fluid. The post-Newtonian approximation 
will be used to construct the data. Furthermore the post-Newtonian 
approximation is able to take into account of the tidal effect 
which is important for the orbital instability. 
Recently,  Lai, Rasio and Shapiro \cite{lrs93,lrs94} have pointed out 
that such a tidal coupling of binary neutron stars is very important 
for their evolution in the final merging phase because the tidal
effect causes the instability of the circular motion 
even in the Newtonian gravity. 
Also important is the general relativistic gravity 
because in the intermediate phase, the orbital separation is 
about ten times as small as the Schwarzschild radius of the system. 
Thus, we need not only a hydrodynamic treatment, but also 
general relativistic one in order to study the final stage of 
the evolution of binary neutron stars.
Since the usual formulation of the post-Newtonian approximation is based 
on the particular gauge condition which is not so convenient 
for numerical purposes, we shall reformulate it by the (3+1) 
formalism  often used in numerical approach of general relativity.
     
This paper is organized as follows. 
In section 2 we introduce the Newtonian limit in general relativity 
and present how to construct the post-Newtonian hierarchy. There we 
mention how to avoid divergent integrals which appears at higher order 
in the previous treatments, and we also discuss how to incorporate 
strong internal gravity in the post-Newtonian approximation.  
Next we reformulate the post-Newtonian approximation appropriate for 
numerical treatment in section 3. 
There we shall adopt the (3+1) formalism which is frequently used 
in numerical relativity. 
Based on the formalism developed in section 3, 
we present a formulation for constructing numerically equilibrium
solutions of uniformly rotating fluid in the 2PN approximation in section 4. 
We shall also discuss the propagation of gravitational waves from slow 
motion systems in section 5.

\section{Foundation of the post-Newtonian approximation}

Since the Newtonian limit is the basis of the post-Newtonian approximation, 
we shall first formulate the Newtonian limit. 
We shall follow the  formulation by Futamase and Schutz\cite{fs83}.  
We will not mention other formulation of Newtonian limit by Ehlers 
\cite{ehlers81,ehlers97},
because it has not yet used to construct the post-Newtonian 
approximation.

\subsection{Newtonian limit along a regular asymptotic Newtonian sequence}

The formulation is based on the observation that any asymptotic approximation 
of any theory need a sequence of solutions of the basic equations of 
the theory\cite{sw,wald}. Namely, if we write the equations 
in abstract form as
\begin{equation}
E(g) = 0 ,
\end{equation}
for an unknown function $g$, one would like to have a one-parameter 
(or possibly multiparameter) family of solutions,
\begin{equation}
E(g(\lambda)) = 0 , 
\end{equation}
where $\lambda$ is some parameter. Asymptotic approximation then says that 
function $f(\lambda)$ approximates $g(\lambda)$ to order $\lambda^p$ 
if $|f(\lambda) - g(\lambda)|/\lambda^p \to 0 $ as $ \lambda \to 0$. 
We choose the sequence of solutions with appropriate properties 
in such a way that the properties reflect the character of the system 
under consideration. 

We shall formulate the post-Newtonian approximation according to 
the general idea just described. 
As stated in the introduction, we would like to have an approximation 
which applies the systems whose motion is described almost by Newtonian theory. 
Thus we need a sequence of solutions of Einstein's equations
parameterized by $\epsilon$ (the typical velocity of the system divided 
by the speed of light) which has the Newtonian character as $ \epsilon \to 0$.

The Newtonian character is most conveniently described by the following 
scaling law.  
The Newtonian equations involve  six variables, density $(\rho)$, 
pressure $(P)$, gravitational potential $(\Phi)$, 
and velocity ($v^i, i =1,2,3$):
\begin{equation}
\nabla^2 \Phi - 4 \pi \rho = 0 , 
\label{NewtonA}
\end{equation}
\begin{equation}
\partial_t \rho + \nabla_i ( \rho v^i)  = 0 , 
\label{NewtonB}
\end{equation}
\begin{equation}
\rho \partial_t v^i + \rho v^j \nabla_j v^i 
+ \nabla^i P + \rho \nabla^i \Phi = 0 , 
\label{NewtonC}
\end{equation}
supplemented by an equation of state. For simplicity we 
have considered perfect fluid. Here
we have set $G = 1$. 

It can be seen that the variables $\{ \rho(x^i, t), P(x^i, t), 
\Phi(x^i, t), v^i(x^j, t) \}$ obeying the above equations 
satisfy the following scaling law.
\begin{eqnarray}
\rho(x^i, t) \rightarrow \epsilon^2 \rho(x^i, \epsilon t) , \nonumber \\
P(x^i, t ) \rightarrow \epsilon^4  P(x^i, \epsilon t)  , \nonumber \\
v^i(x^k, t) \rightarrow \epsilon v^i(x^k, \epsilon t) , \nonumber \\
\Phi(x^i, t ) \rightarrow \epsilon^2  \Phi(x^i, \epsilon t) . 
\label{Scaling}
\end{eqnarray}
One can easily understand the meaning of this scaling by noticing 
that $\epsilon$ is the magnitude of typical velocity (divided by 
the speed of light). Then the magnitude of the gravitational potential 
will be of order $\epsilon^2$ because of the balance between gravity and 
centrifugal force. 
The scaling of the time variable expresses the fact that 
the weaker the gravity $( \epsilon \rightarrow 0)$ the longer the time scale.

Thus we wish to have a sequence of solutions of Einstein's equations 
which has the above scaling as $\epsilon \to 0$. 
We shall also take the point of view that the sequence of 
solutions is determined by the appropriate sequence of initial data. 
This has a practical advantage because there will be no solution 
of Einstein's equations which satisfy the above scaling  $(\ref{Scaling})$ 
even as $\epsilon \to 0$. 
This is because Einstein's equations are nonlinear in the field variables 
so it will not be possible to enforce the scaling everywhere 
in spacetime. We shall therefore impose it only on the initial data 
for the solution of the sequence. 
 
Here we shall first make a general discussion on the formulation of
the post-Newtonian approximation independent of any initial value
formalism  and then present the concrete treatment in 
the harmonic coordinate. The condition is used because of 
its popularity and some advantages in the generalization to the
systems with strong internal gravity. 

As the initial data for the matter we shall take the same data set 
in Newtonian case, namely, the density $\rho$, the pressure $P$, 
and the coordinate velocity $v^i$. In most of the application, 
we usually assume a simple equation of state which relates the pressure 
to the density. 
The initial data for the gravitational field are  
$g_{\mu\nu}, \pa{g}_{\mu\nu} /\pa t$. 
Since general relativity is
overdetermined system, there will be constraint equations among 
the initial data for the field. 
We shall write the free data for the field as $(Q_{ij}, P_{ij})$ whose 
explicit forms depend on the coordinate condition one assumes. 
In any coordinate we shall assume these data for the field vanish 
since we are interested in the evolution of an isolated system by its 
own gravitational interaction. It is expected that these 
choice corresponds to the absence of radiation far away from the source.
Thus we choose the following initial data which is indicated 
by Newtonian scaling:
\begin{eqnarray}
\rho(t=0, x^i, \epsilon) = \epsilon^2 a(x^i), \nonumber \\
P(t=0, x^i, \epsilon) = \epsilon^4 b(x^i), \nonumber \\  
v^i(t=0, x^k, \epsilon) = \epsilon c^i(x^k), \nonumber \\
Q_{ij} (t=0, x^i, \epsilon) = 0, \nonumber \\  
P_{ij} (t=0, x^i, \epsilon) = 0,
\label{Initial}
\end{eqnarray}  
where the functions $ a, b$ , and $c^i$ are $C^\infty$ functions 
that have compact support contained entirely within a sphere of a
finite radius.

Corresponding to the above data, we have a one-parameter set of 
spacetime parameterized by $\epsilon$. 
It may be helpful to visualize the set as a fiber bundle,  
with base space $R$ the real line (coordinate $\epsilon$) and 
fiber $R^4$ the spacetime (coordinates $t, x^i$).
The fiber $\epsilon=0$ is Minkowski spacetime since it is defined 
by zero data. In the following we shall assume that the solutions 
are sufficiently smooth functions of $\epsilon$ 
for small $\epsilon \neq 0$. 
We wish to take the limit $\epsilon \to 0$ along the sequence. 
The limit is, however, not unique and is defined by giving a smooth 
nowhere vanishing vector field on the fiber bundle which is nowhere 
tangent to each fiber\cite{geroch,sw}. 
The integral curves of the vector field give a correspondence 
between points in different fibers, namely events in different 
spacetime with different values of $\epsilon$. 
Remembering the Newtonian scaling of the time variable in the limit, 
we introduce the Newtonian dynamical time:
\begin{equation}
\tau = \epsilon t,
\label{Ntime}
\end{equation}
and define the integral curve as the curve on which $\tau$ and $x^i$
stay constant. 
In fact if we take the limit $\epsilon \to 0$ along this curve, 
the orbital period of the binary system with $\epsilon = 0.01$ is 
10 times of that of the system with $\epsilon = 0.1$ as expected from
the Newtonian scaling.  This is what we define as the Newtonian limit. 
Notice that this map never reaches the fiber $\epsilon = 0$ 
(Minkowski spacetime). 
There is no pure vacuum Newtonian limit as expected.

In the following we assume the existence of such a sequence of solutions 
constructed by the initial data satisfying the above scaling 
with respect to $\epsilon$. We shall call such a sequence is a regular  
asymptotically Newtonian sequence. 
We have to make further mathematical assumptions about the sequence 
to make explicit calculations. 
We will not go into details of them partly because to prove the assumptions 
we need a deep understanding of the existence and uniqueness
properties of Cauchy problem of Einstein's equations 
with perfect fluids of compact support which are not available at present.

\subsection{Post-Newtonian Hierarchy}

We shall now define the Newtonian, post-Newtonian and higher approximations 
of various quantities as the appropriate higher tangents of the corresponding 
quantities to the above integral curve at $\epsilon = 0$. 
For example the hierarchy of approximations for 
the spacetime  metrics can be expressed as follows: 
\begin{eqnarray}
g_{\mu\nu}(\epsilon, \tau, x^i) 
&=& g_{\mu\nu}(0, \tau, x^i) 
+ \epsilon ({\cal L}_V g_{\mu\nu})(0, \tau, x^i) 
\hspace*{5.5cm} \nonumber\\ 
&&+{1\over 2}\epsilon^2  ({\cal L}^2_V g_{\mu\nu})(0, \tau, x^i) 
+ ...
.. + {\epsilon^n\over {n!}}({\cal L}^n_V g_{\mu\nu})(0, \tau, x^i) 
+ R_{n+1} , 
\label{Taylor}
\end{eqnarray}
where ${\cal L}_V$ is the Lie derivative with respect to the tangent vector 
of the curves defined above, and the remainder term $R^{\mu\nu}_{n+1}$ is   
\begin{equation}
R^{\mu\nu}_{n+1} = {\epsilon^{n+1}\over {(n+1)!}} 
\int^1_0 d\ell ( 1 - \ell)^{n+1} 
({\cal L}^{n+1}_V g_{\mu\nu}) (\ell \epsilon, \tau, x^i) . 
\end{equation}
Taylor's theorem guarantees that the series is asymptotic expansion about 
$\epsilon = 0$ under certain assumptions mentioned above. 
It might be useful to point out that the above definition of 
the approximation scheme may be formulated purely geometrically 
in terms of jet bundle. 

The above definition of the post-Newtonian hierarchy 
gives us asymptotic series in which each term in the series 
is manifestly finite.  
This is based on the $\epsilon$ dependence of the domain 
of dependence of the field point $(\tau, x^k)$. 
The region is finite with finite values of $\epsilon$, 
and the diameter of the region increases as $\epsilon^{-1}$ 
as $\epsilon \to 0$. 
Without this linkage of the region with the expansion parameter 
$\epsilon$, the post-Newtonian approximation leads to 
divergences in the higher orders. 
This is closely related with the retarded expansion. 
Namely, it is assumed that the slow motion assumption enabled one 
to Taylor expand the retarded integrals in retarded time such as
\begin{equation}
\int dr f(\tau - \epsilon r, ...) 
= \int dr f(\tau, ...) - \epsilon \int dr r f(\tau, ...)_{,\tau} + ..
\label{Retarded}
\end{equation}
and assign the second term to a higher order because of its explicit 
$\epsilon$ in front. This is incorrect because $ r \to \epsilon^{-1}$ 
as $\epsilon \to 0$ and thus $\epsilon r$ is not uniformly small in the 
Newtonian limit. Only if the integrand falls off sufficiently fast, 
can retardation be ignored. This happens in the lower-order PN terms. 
But at some higher order there appear many terms which do not fall off 
sufficiently fast because of the nonlinearity of Einstein's equations. 
This is the reason that the formal PN approximation produces the divergent 
integrals. It turns out that such divergence appears at 3PN order 
indicating the breakdown of the PN approximation in the harmonic coordinate. 
This sort of divergence may be eliminated if we remember that 
the upper bound of integration does depend on $\epsilon$ as $\epsilon^{-1}$.  
Thus we would get something like $\epsilon^n \ln \epsilon $ 
instead of $\epsilon^n \ln \infty$ in the usual approach. 
This shows that the asymptotic Newtonian sequence is not differentiable 
in $\epsilon$ at $\epsilon = 0$, but there are no divergence in the expansion 
and it has still an asymptotic approximation in $\epsilon$ that 
involves logarithms.

\subsection{Explicit calculation in Harmonic coordinate}

Here we shall use the above formalism to make explicit calculation in 
the harmonic coordinate. 
The reduced Einstein's equation in the harmonic condition is written as 
\begin{equation}
{\tilde g}^{\alpha\beta} {\tilde g}^{\mu\nu}_{\;\;\; ,\alpha\beta} 
= 16\pi \Theta^{\mu\nu} 
- {\tilde g}^{\mu\alpha}_{\;\;\; ,\beta} 
{\tilde g}^{\nu\beta}_{\;\;\; ,\alpha} , 
\label{Reduce}
\end{equation}
\begin{equation}
\partial_\mu [{\tilde g}^{\mu\nu} \partial_\nu x^\alpha ] = 0 , 
\label{Guage}
\end{equation}
where 
\begin{equation}
{\tilde g}^{\mu\nu} = (-g)^{1/2} g^{\mu\nu} , 
\end{equation}
\begin{equation}
\Theta^{\alpha\beta}= (-g)( T^{\alpha\beta}+t^{\alpha\beta}_{LL} ) , 
\end{equation}
where $t^{\mu\nu}_{LL}$ is the Landau-Lifshitz pseudotensor \cite{ll}. 
In the following we shall choose an isentropic perfect fluid for 
$T^{\alpha\beta}$ which is enough for most of applications.
\begin{equation}
T^{\alpha\beta} = ( \rho + \rho \Pi + P ) u^\alpha u^\beta + 
P g^{\alpha\beta} , 
\label{stressenergy}
\end{equation}
where $\rho$ is the rest mass density, $\Pi$ the internal energy, $P$ the 
pressure, $u^\mu$ the four velocity of the fluid with normalization;
\begin{equation}
g_{\alpha\beta}u^\alpha u^\beta = -1 . 
\label{Norm}
\end{equation}
The conservation of the energy and momentum is expressed as 
\begin{equation}
\nabla_\beta T^{\alpha\beta} =0 . 
\label{emconsv}
\end{equation}

Defining the gravitational field variable as
\begin{equation}
{\bar h}^{\mu\nu} = \eta^{\mu\nu} - ( - g)^{1/2} g^{\mu\nu} , 
\label{Variable}
\end{equation}
where $\eta^{\mu\nu}$ is the Minkowski metric, 
the reduced Einstein's equation $(\ref{Reduce})$ 
and the gauge condition $(\ref{Guage})$ take the following form; 
\begin{equation}
 ( \eta^{\alpha\beta} - {\bar h}^{\alpha\beta} )
{\bar h}^{\mu\nu}_{\;\;\; ,\alpha\beta} 
= - 16 \pi \Theta^{\mu\nu} 
+ {\bar h}^{\mu\alpha}_{\;\;\; ,\beta}
{\bar h}^{\nu\beta}_{\;\;\; ,\alpha} , 
\label{Reduce1}
\end{equation}
\begin{equation}
{\bar h}^{\mu\nu}_{\;\;\; ,\nu} = 0 . 
\label{Harmonic}
\end{equation}
Thus the characteristics is determined by the operator 
$ ( \eta^{\alpha\beta} - {\bar h}^{\alpha\beta} ) 
\partial_\alpha \partial_\beta $, and thus the light cone deviates 
from that in the flat spacetime. We shall use this form 
of the reduced Einstein's equations in the calculation of wave form far away 
from the source because the deviation plays a fundamental role there. 
However in the study of the gravitational field near the 
source it is not neccesary to consider the deviation of the light cone 
away from the flat one and thus it is convenient to use the following form of the 
reduced Einstein's equations\cite{ad}.
\begin{equation}
\eta^{\mu\nu}{\bar h}^{\alpha\beta}_{\;\;\; ,\mu\nu} 
= - 16 \pi \Lambda^{\alpha\beta} , 
\label{Einstein}
\end{equation}
where 
\begin{eqnarray}
\Lambda^{\alpha\beta} 
&=& \Theta^{\alpha\beta} + \chi^{\alpha\beta\mu\nu}
_{\;\;\;\;\;\;\; ,\mu\nu} , \\
\chi^{\alpha\beta\mu\nu} 
&=& (16\pi)^{-1} 
( {\bar h}^{\alpha\nu} {\bar h}^{\beta\mu} 
- {\bar h}^{\alpha\beta} {\bar h}^{\mu\nu} ) . 
\end{eqnarray}
Equations $(\ref{Harmonic})$ and $(\ref{Einstein})$ together imply 
the conservation law
\begin{equation}
\Lambda^{\alpha\beta}_{\, \, \, ,\beta} = 0 . 
\label{Consv}
\end{equation}
We shall take as our variables 
the set $\{\rho, P, v^i, {\bar h}^{\alpha\beta} \}$, with the definition
\begin{equation}
v^i = u^i/u^0 . 
\end{equation}
The time component of 4-velocity $u^0$ is determined from Eq.$(\ref{Norm})$. 
To make a well-defined system of equations we must add the conservation law 
for number density $n$ which is some function of the density $\rho$ and 
pressure $P$:

\begin{equation}
\nabla_\alpha(nu^\alpha) = 0 . 
\label{Nconsv}
\end{equation}
Equations $(\ref{Consv})$ and $(\ref{Nconsv})$ imply that 
the flow is adiabatic. 
The role of the equation of state is played by the arbitrary function 
$n(\rho, p)$. 

Initial data for the above set of equations  are $ {\bar h}^{\alpha\beta}, 
{\bar h}^{\alpha\beta}_{\;\;\; ,0}, \rho, P $, and $v^i$, but not all these 
data are independent because of the existence of constraint equations.
Equations.$(\ref{Harmonic})$ and $(\ref{Einstein})$
imply the four constraint equations among the initial data for the field.
\begin{equation}
\Delta{\bar h}^{\alpha 0}
+ 16 \pi \Lambda^{\alpha 0} 
- \delta^{ij}{\bar h}^{\alpha \, \, \, \, ,0}_{\, i,j} = 0 , 
\label{Constraint}
\end{equation}
where $\Delta$ is the Laplacian in the flat space. 
We shall choose  $ {\bar h}^{ij} $ and $ {\bar h}^{ij}_{\, \, ,0}$ as 
free data and solve Eq.$(\ref{Constraint})$ 
for ${\bar h}^{\alpha 0} (\alpha =0, ..,3)$ and Eq.$(\ref{Harmonic})$ for 
${\bar h}^{\alpha 0}\,_{,0}$. 
Of course these constraints cannot be solved explicitly, 
since $\Lambda^{\alpha 0}$ contains ${\bar h}^{\alpha 0}$, 
but they can be solved iteratively as explained below.
As discussed above, we shall assume that 
the free data $ {\bar h}^{ij} $ and $ {\bar h}^{ij}_{\, \, ,0}$ 
for the field vanish. One can show that such initial data satisfy 
the O'Murchadha and York criterion for the absence of radiation far away 
from the source\cite{oy}.

In the actual calculation, it is convenient to use an expression with 
explicit dependence of $\epsilon$. The harmonic condition allows 
us to have such expression in terms of the retarded integral.
\begin{equation}
{\bar h}^{\mu\nu} ( \epsilon, \tau, x^i) 
= 4 \int_{C(\epsilon,\tau, x^i)} d^3 y
\Lambda^{\mu\nu}( \tau - \epsilon r, y^i, \epsilon) /r 
+ h^{\mu\nu}_H(\epsilon,\tau, x^i) , 
\label{Kirch}
\end{equation}
where $ r=|y^i - x^i|$ and $C(\epsilon, \tau, x^i)$ is the past flat 
light cone of event $(\tau, x^i)$ in the spacetime given by
$\epsilon$, 
truncated where it intersects with the initial hypersurface $\tau =0$. 
${\bar h}^{\mu\nu}_H$ is the unique solution of the homogeneous equation 
\begin{equation}
\eta^{\alpha\beta}{\bar h}^{\mu\nu}_{\, \, \, \, ,\alpha\beta}=0 . 
\end{equation}
We shall henceforce ignore the homogeneous solution 
because they play no important roles. 
Because of the $\epsilon$ dependence of the integral region, 
the domains of integration are finite as long as $\epsilon \neq 0$ 
and their diameter increases as $\epsilon^{-1}$ as $\epsilon \to 0$. 

Given the formal expression $(\ref{Kirch})$ in terms of 
initial data $(\ref{Initial})$, we can take the Lie derivative 
and evaluate these derivatives at $\epsilon = 0$. 
The Lie derivative is nothing but partial derivative 
with respect to $\epsilon$ in the coordinate system for the fiber bundle 
given by $(\epsilon, \tau, x^i )$. 
Accordingly one should convert all the time indices to $\tau$ indices. 
For example, $T^{\tau\tau} = \epsilon^2 T^{tt}$ 
which is of order $\epsilon^4$ since $T^{tt} \sim \rho$ 
is of order $\epsilon^2$. 
Similarly the other components of stress energy tensor 
$T^{\tau i} = \epsilon T^{ti}$ and $T^{ij}$ are of order $\epsilon^4$ 
as well. 
Thus we expect the first non-vanishing derivative 
in $(\ref{Kirch})$ will be the forth-derivatives. 
In fact we find
\begin{eqnarray}
{}_{(4)}{\bar h}^{\tau\tau}(\tau, x^i) 
&=& 4 \int_{R^3} {{{}_{(2)} \rho(\tau, y^i)}\over r} d^3y , 
\label{Tautau} \\
{}_{(4)}{\bar h}^{\tau i}(\tau, x^i) 
&=& 4 \int_{R^3} {{{}_{(2)} \rho(\tau, y^k) {}_{(1)} v^i(\tau, x^k)}\over r}
d^3y , \\
{}_{(4)}{\bar h}^{ij}(\tau, x^k) 
&=& 4 \int_{R^3} 
{{}_{(2)} \rho(\tau, y^k) {}_{(1)} v^i(\tau,y^k) {}_{(1)} v^j({\tau, y^k) 
+ {}_{(4)} t^{ij}_{LL}(\tau, y^k)}\over r} d^3y , \nonumber\\
&&
\end{eqnarray}
where we have adopted the notation
\begin{equation}
{}_{(n)} f(\tau, x^i) 
= {1\over {n!}} \lim_{\epsilon \to 0} 
{\partial^n\over {\partial \epsilon^n}} f(\epsilon, \tau, x^i) , 
\end{equation}
and 
\begin{equation}
{}_{(4)} t^{ij}_{LL} 
= {1\over {64\pi}} 
( {}_{(4)} {\bar h}^{\tau\tau,i} {}_{(4)} {\bar h}^{\tau\tau,j} 
- {1\over 2} \delta^{ij} 
{}_{(4)} {\bar h}^{\tau\tau,k}{}_{(4)} {\bar h}^{\tau\tau}_{\, \, ,k}) . 
\end{equation}

In the above calculation we have taken the point of view 
that ${\bar h}^{\mu\nu}$ is a tensor field, defined 
by giving its components in the assumed harmonic coordinate 
as the difference between the tensor 
density $\sqrt{-g} g^{\mu\nu}$ and $\eta^{\mu\nu}$.  

The conservation law $(\ref{Consv})$ also  
has its first nonvanishing derivative at this order, which are  
\begin{equation}
{}_{(2)}\rho(\tau, x^k)_{,\tau} 
+ ({}_{(2)}\rho(\tau, x^k) {}_{(1)} v^i(\tau, x^k) )_{,i} = 0 , 
\label{Newcon}
\end{equation}
\begin{equation}
({}_{(2)}\rho {}_{(1)} v^i )_{,\tau}  
+ ({}_{(2)}\rho {}_{(1)} v^i{}_{(1)} v^j )_{,j} 
+ {}_{(4)} P^{,i} -  {1\over 4} {}_{(2)} \rho {}_{(4)} 
{\bar h}^{\tau\tau,i}= 0 . 
\label{Neweq} 
\end{equation}
Equations $(\ref{Tautau})$, $(\ref{Newcon})$, and $(\ref{Neweq})$ 
consist of Newtonian theory of gravity.
Thus the lowest non-vanishing derivative with respect to $\epsilon$ 
is indeed Newtonian theory, and 
the 1PN and 2PN equations emerge from sixth and eighth derivatives, 
respectively, in the conservation law $(\ref{Consv})$. 
At the next derivative, the quadrupole radiation reaction term
comes out.

\subsection{Strong point particle limit}

If we wish to apply the post-Newtonian approximation to 
the inspiraling phase of binary neutron stars, 
the strong internal gravity must be taken into account. 
The usual post-Newtonian approximation assumes explicitly the weakness  of 
gravitational field everywhere including inside the material source.
It is argued by appealing the strong equivalence principle that the external 
gravitational field which governs the orbital motion of the binary system 
is independent of internal structure of the components up to 
tidal interaction. Thus it is expected that the results obtained under 
the assumption of weak gravity also apply for the case of neutron star binary. 
Experimental evidence for the strong equivalence principle is obtained 
only for the system with weak gravity\cite{will87,will94}, but    
no experiment is available in case with strong internal gravity at present. 

In theoretical aspect, the theory of extended object in general relativity 
\cite{dixon} is still in preliminary stage for the application 
to realistic systems. 
Matched asymptotic expansion technique has been used to treat the system with 
strong gravity in certain situations
\cite{death75a,death75b,kates80a,kates80b,damour82}.
Another way to handle with strong internal gravity is  
by the use of Dirac's delta function type source with a fixed mass\cite{eih}. 
However, this makes Einstein's equations mathematically meaningless 
because of their nonlinearity. 
Physically, there is no such source in general relativity 
because of the existence  of black holes. 
Before a body shrinks to a point, it forms a black hole 
whose size is fixed by its mass. 
For this reason, it has been claimed that no point particle exists 
in general relativity. 

This conclusion is not correct, however. We can shrink the body 
keeping the compactness $(M/R)$, i.e., the strength of 
the internal field fixed. 
Namely we should scale the mass M just like the radius R. 
This can be fitted nicely into the concept of regular asymptotic 
Newtonian sequence defined above because there the mass also scales 
along the sequence of solutions. 
In fact, if we take the masses of two stars as M, and 
the separation between two stars as L, then $\Phi \sim M/L$. 
Thus the mass $M$ scales as $\epsilon^2$ if we fix the separation. 
In the above we have assumed that the density scales as $\epsilon^2 $ to 
guarantee this scaling for the mass understanding the size of the body fixed. 
Now we shrink the size as $\epsilon^2 $ to keep the compactness 
of each component. 
Then the density should scale as $\epsilon^{-4}$. 
We shall call such a scheme as strong point particle limit 
since the limit keeps the strength of internal gravity. 
The above consideration suggests the following initial data 
to define a regular asymptotic 
Newtonian sequence which describes nearly Newtonian system with 
strong internal gravity\cite{futamase87}. 
The initial data are two uniformly rotating fluids with compact 
spatial support whose stress-energy tensor and size scales as 
$\epsilon^{-4}$ and $\epsilon^2$, respectively. 
We also assume that each of these fluid configurations would be 
a stationary equilibrium solution of Einstein's equation 
if other were absent. 
This is neccesary for the suppression of irrelevant internal motions 
of each star. 
Any remaining motions are the tidal effects caused by the other body, 
which will be of order $\epsilon^6$ smaller than the internal self-force.  
This data allows us to use the Newtonian time $\tau = \epsilon t$ as a 
natural time coordinate everywhere including inside the stars.

This choice of the data leads naturally to the introduction of 
the body zones $B_A$ and the 
body zone coordinates ${\bar x}^{k}_A$ defined by
\begin{eqnarray}
B_A &=& \{ x^k ; | x^k - \xi^k_A |  < \epsilon R  \} , \\
{\bar x}^k_A &=& \epsilon^{-2} ( x^k - \xi^k_A) , 
\label{Bzone}
\end{eqnarray}
where $R$ is some constant, and $\xi^k_A(\tau), A = I,II$ are 
the coordinates of the origin of the two stars, 
where we have used letters with bar to distinguish the body-zone
coordinates from their counterparts.
The scaling by $\epsilon^{-2}$ means that as the star shrinks 
with respect to the coordinate $x^k$, it remains of fixed size in 
the body-zone coordinate ${\bar x}^k$. 
The boundary of the body zone shrinks to a point 
with respect to $x^k$, and expands to infinity with respect to ${\bar x}^k$ 
as  $\epsilon \to 0$. This makes a clean separation of the body zone 
from the exterior geometry generated by other star. 
\begin{equation}
g_{\mu\nu} = (g_B)_{\mu\nu} + (g_{C-B})_{\mu\nu} , 
\end{equation}
where $ (g_B)_{\mu\nu} $ is the contribution from the 
body zones, $ (g_{C-B})_{\mu\nu}$ from elsewhere. 

Actual calculations are most easily performed in the harmonic coordinate. 
Take two stationary solutions of Einstein's equations for the perfect 
fluid $ \{ T^{\mu\nu}_A(x^i), g^{\mu\nu}_A(x^i) \}$ as our 
initial data in the body zone. As explained above every component of the 
stress-energy tensor $T^{\mu\nu}_A$ has the same $\epsilon^{-4}$ scaling 
in the inertia coordinates $(t, x^k)$ for a rapidly rotating star. 
In the body zone coordinates $(\tau, {\bar x}^k )$, these have the following 
scalings: 
\begin{eqnarray}
{\bar T}^{\tau\tau}_A = \epsilon^2 T^{tt}_A \sim \epsilon^{-2}, \\
{\bar T}^{\tau i}_A = \epsilon^{-1} T^{ti}_A \sim \epsilon^{-5}, \\
{\bar T}^{ij}_A = \epsilon^{-4} T^{ij}_A \sim \epsilon^{-8} . 
\end{eqnarray}
Transformed to the near zone coordinates $(\tau, x^k)$, 
$x^k = \xi^i_A + \epsilon^2 {\bar x}^k$, they take the form
\begin{eqnarray}
T^{\tau\tau}_N 
&=& {\bar T}^{\tau\tau}_A , \\
T^{\tau i}_N 
&=& \epsilon^2 {\bar T}^{\tau i}_A 
                 + v^i_A {\bar T}^{\tau\tau}_A, \\
T^{ij}_N 
&=& \epsilon^4{\bar T}^{ij}_A 
           + 2 \epsilon^2 v^{(i}_A {\bar T}^{j)\tau}_A 
           + v^i_A v^j_A {\bar T}^{\tau\tau}_A , 
\end{eqnarray}
where $v^i_A = d\xi^i_A/d\tau$ is the velocity of the origin of the body $A$.
If there were only one body, these data would produce a stationary solution 
in the body zone, which moves with uniform velocity $v^i_A$ in the 
near-zone coordinates. Now we know the ordering of the source, we can solve 
Eq.$(\ref{Kirch})$,iteratively as in the weak field case. The difference is 
that we transform the integration variables to the body zone 
coordinates in the expression of the fields whose contributions come from the 
body zone.
\begin{equation}
{\bar h}^{\mu\nu}_B ( \epsilon; \tau, x^i) 
= 4 \epsilon^6 \sum_A \int_{B_A} d^3{\bar y}_A
|x_A - \epsilon^2 {\bar y}_A |^{-1} 
\Lambda^{\mu\nu}( \tau - \epsilon | x_A - \epsilon^2 {\bar y}|, 
{\bar y}^i, \epsilon)  , 
\label{Body}
\end{equation}
where $x^i_A = x^i - \xi^i_A$. 
Expanding these expressions in terms of $\epsilon$, we obtain
\begin{eqnarray}
\bar h^{\tau\tau}_B(\epsilon; \tau, x^i) 
&=& 4 \epsilon^4 \sum_A {M_A\over {|x_A|}} + O(\epsilon^6), \\
\bar h^{\tau i}_B(\epsilon; \tau, x^i) 
&=& 4 \epsilon^3 \sum_A {J_A\over {|x_A|}} 
  +  4 \epsilon^4 \sum_A {{M_A v^i_A}\over {|x_A|}} + O(\epsilon^5), \\
{\bar h}^{ij}_B(\epsilon; \tau, x^k) 
&=& 4 \epsilon^2 \sum_A {{Z^{ij}_A}\over {|x_A|}} 
  + 8 \epsilon^3 \sum_A {{v^{(i}_A J^{j)}_A}\over {|x_A|}} 
  +  4 \epsilon^4 \sum_A {{M_A v^i_A v^j_A}\over {|x_A|}} + O(\epsilon^5),
\hspace*{1cm}
\end{eqnarray}
where 
\begin{eqnarray}
M_A&=& \lim_{\epsilon \to 0} \epsilon^2 
\int_{B_A} d^3{\bar y} \Lambda^{\tau\tau}_A(\epsilon; \tau, {\bar y}), \\
J^i_A&=& \lim_{\epsilon \to 0} \epsilon^3
\int_{B_A} d^3{\bar y} \Lambda^{\tau i}_A(\epsilon; \tau, {\bar y}), \\
Z^{ij}_A&=& \lim_{\epsilon \to 0} \epsilon^4
\int_{B_A} d^3{\bar y} \Lambda^{ij}_A(\epsilon; \tau, {\bar y}) .  
\end{eqnarray}
The $M_A$ defined above is the conserved ADM mass the body $A$ would 
have if it were isolated. Without loss of generality one can set the linear momentum $J^i_A$ of the overall internal motion of each body to zero 
by choosing appropriate origin of the coordinates. If we did not assume the internal stationarity in the initial data, then $Z^{ij}_A$ would be finite 
and no approximation would be possible. However under the condition of 
internal stationarity one can show that $Z^{ij}_A$ vanishes\cite{futamase87}.
Above expressions for the body zone contribution are used to 
calculate the pseudotensor in the near zone and then to evaluate the contribution ${\bar h}^{\mu\nu}_{C-B}$ outside the body zone. 
As the result we shall obtain the metric variables up to $O(\epsilon^6)$.
\begin{eqnarray}
\bar h^{\tau\tau}(\tau, x^i) 
&=& 4 \epsilon^4 \sum_A {M_A\over {|x_A|}} + O(\epsilon^6), \\
\bar h^{\tau i}(\tau, x^i) 
&=& 4 \epsilon^4 \sum_A {{M_A v^i_A}\over {|x_A|}} + 
   + 2 \epsilon^5 \sum_A {{x^k_A}\over |x_A|} M^{ki}_A + O(\epsilon^6), \\
\bar h^{ij}(\tau, x^k) 
&=& 4 \epsilon^4 \sum_A {{M_A v^i_A v^j_A}\over {|x_A|}} 
  - 2 \epsilon^5 I^{(3)ij}_{orb} + O(\epsilon^6) , 
\end{eqnarray}
where
\begin{eqnarray}
M^{ij}_A&=& \lim_{\epsilon \to 0} \epsilon^4 
2 \int_{B_A} d^3{\bar y} {\bar y}^{[i} \Lambda^{j]\tau}_A
(\epsilon; \tau, {\bar y}), \\
I^{ij}_{orb} &=& \sum_A \xi^i_A \xi^j_A M_A J^i_A . 
\end{eqnarray}
These are the angular momentum of the body $A$ and the quadrupole moment 
for the orbital motion, respectively. This kind of calculation is also 
performed at 2.5PN order to get the standard quadrupole 
formula except that the mass in the definition of the quadrupole moment is replaced with the ADM mass\cite{futamase87}. 
The above method has been also applied 
for the calculation of spin precession and the same form of the equation 
in the case of weak gravity is obtained. Thus we can extend the strong 
equivalence principle for bodies with strong internal gravity to 
the generation of gravitational waves and the spin precession.  
It is an open question if the strong point particle limit 
may be taken to lead a well defined equation of motion at higher 
post-Newtonian orders.

\section{Post-Newtonian Approximation in the (3+1) Formalism}
 
In the evolution of binary neutron stars, 
the orbital separation eventually becomes comparable to the radius of the 
neutron star due to radiation reaction. Then, the point particle picture 
does not apply and each star of the binary begins to behave as a hydrodynamic object because of tidal effect.
As described in the introduction  we need not only 
a hydrodynamic treatment, but also general relativistic one 
in order to study the final stage of the evolution of binary neutron stars.
Full general relativistic simulation will be a direct way to answer 
such a question, but it is one of the hardest problem in astrophysics.
Although there is much progress in this direction\cite{nakamura}, 
it will take a long time until numerical relativistic calculations become 
reliable. One of the main reasons for this would be that 
we do not know the behavior of geometric variables in the strong 
gravitational field around coalescing binary neutron stars. 
Owing to this, we do not know what sort of gauge condition is useful and 
how to give an appropriate general relativistic 
initial condition for coalescing binary neutron stars. 

The other reason is a technical one: In the case of coalescing 
binary neutron stars, the wavelength is of order of  
$\lambda \sim \pi R^{3/2}M^{-1/2}$, where $R$ and $M$ are the orbital radius 
and the total mass of binary, respectively. 
Thus, in numerical simulations, we need 
to cover a region $L >\lambda \propto R^{3/2}$ with numerical grids 
in order to attain good accuracy. 
This is in contrast with the case of Newtonian and/or PN
simulations, in which we only need to cover a region $\lambda > L > R$. 
Since the circular orbit of binary neutron stars becomes unstable at 
$R \leq 10M$ owing to the tidal effects \cite{lrs93,lrs94}  
and/or the strong general relativistic gravity \cite{kww93a}, 
we must set an initial condition of binary at $R \geq 10M$. 
For such a case the grid must cover a region $L>\lambda \sim 100M$ in 
numerical  simulations to perform an accurate simulation. 
When we assume to cover each neutron star of its radius $\sim 5M$ 
with $\sim 30$ homogeneous grids \cite{on90,on91,on92,sno92,sno93}, 
we need to take grids of at least $\sim 500^3$, 
but it seems impossible to take such a large amount of mesh points 
even for the present power of supercomputer. 
At present, we had better search other methods to prepare 
a precise initial condition for binary neutron stars. 

In the case of PN simulations, the situation is completely 
different because we do not have to treat gravitational waves 
explicitly in numerical simulations, and as a result, 
only need to cover a region at most $L \sim 20-30M$. 
In this case, it seems that $\sim 200^3$ grid numbers are enough. 
Furthermore, we can take into account general 
relativistic effects with a good accuracy: In the case of coalescing binary 
neutron stars, the error will be at most $\sim M/R \sim {\rm a~few} \times 
10\%$ for the first PN
approximation, and $\sim (M/R)^2 \sim {\rm several}~ \%$ for 2PN 
approximation. 
Hence, if we can take into account up through 2PN terms, we 
will be able to give a highly accurate initial condition (the error 
$\leq {\rm several}~\%$).

For these reasons, we present in this section the 2.5PN hydrodynamic 
equations including the 2.5PN radiation reaction potential 
in such a way that one can apply the formulation directly 
in numerical simulations. 
As for the PN hydrodynamic equation, Blanchet, Damour and Sch\"afer 
\cite{bds} have already obtained the (1+2.5) PN equations. 
In their formulation, the source terms of all Poisson equations 
take nonvanishing values only on the matter, 
like in the Newtonian hydrodynamics. 
Although their formulation is very useful for PN hydrodynamic simulations 
including the radiation reaction \cite{on90,on91,on92,sno92,sno93,rjs}, 
they did not take into account 2PN terms. 
In their formulation, they also fixed the gauge conditions to the ADM 
gauge, but in numerical relativity, it has not been known yet 
what sort of gauge condition is suitable for simulation of 
the coalescing binary neutron stars and estimation of gravitational waves 
from them. 
First, we develop the formalism for the hydrodynamics 
using the PN approximation. 
In particular, we use the (3+1) formalism of general relativity 
so that we can adopt more general class of slice conditions
\cite{futamase92,asf}. 
Next, we present methods to obtain numerically terms at the 2PN order
\cite{asf}.

\subsection{(3+1) Formalism for the post-Newtonian Approximation} 

According to the above discussion, we shall apply the (3+1) formalism 
to formulate the PN approximation. In the (3+1) formalism \cite{adm,wald,nok}, 
spacetime is foliated by a family of spacelike 3D hypersurfaces whose 
normal one-form is taken as 
\begin{equation}
\hat n_{\mu}=(-\alpha,~{\bf 0}) . 
\end{equation}
Then the line element takes the following form 
\begin{equation}
ds^2=-(\alpha^2-\beta_i \beta^i)dt^2+2\beta_i dt dx^i+
\gamma_{ij}dx^i dx^j , 
\end{equation}
where $\alpha, \beta^i$ and $\gamma_{ij}$ are the lapse function, shift 
vector  and metric on the 3D hypersurface, respectively. 
Using the (3+1) formalism, the Einstein's equation
\begin{equation}
G_{\mu\nu}=8\pi T_{\mu\nu}, 
\end{equation}
is also split into the constraint equations and 
the evolution equations. The formers are the so-called Hamiltonian and 
momentum constraints which respectively become 
\begin{equation}
{\rm tr}R-K_{ij}K^{ij}+K^2=16\pi \rho_{H}, \label{hami}
\end{equation}
\begin{equation}
D_i K^i_{~j}-D_j K=8\pi J_j, \label{cons}
\end{equation}
where  $K_{ij}$, $K$, ${\rm tr}R$ and $D_i$ are the extrinsic curvature, 
the trace part of $K_{ij}$, the scalar curvature of 3D hypersurface 
and the covariant derivative with respect of $\gamma_{ij}$. 
$\rho_{H}$ and $J_j$ are defined as
\begin{eqnarray}
\rho_{H}&=&T_{\mu\nu}\hat n^{\mu}\hat n^{\nu},\\
J_j&=&-T_{\mu\nu}\hat n^{\mu}\gamma^{\nu}_{~j}.
\end{eqnarray}
The Evolution equations for the spatial metric and the extrinsic curvature 
are respectively
\begin{eqnarray}
{\pa \over \pa t} \gamma_{ij}&=&-2\alpha K_{ij}+D_i \beta_j 
+D_j \beta_i, 
\label{dtgamij}\\
{\pa \over \pa t} K_{ij}&=&\alpha(R_{ij}+K K_{ij}-2K_{il}K^l_{~j})
 -D_iD_j \alpha 
+(D_j \beta^m) K_{mi}+(D_i \beta^m) K_{mj}
\nonumber\\
&&+\beta^m D_m K_{ij} 
-8\pi\alpha\Bigl(S_{ij}+{1 \over 2}\gamma_{ij}(\rho_{H}-S^l_{~l})\Bigr),
\label{dtKij}\\
{\pa \over \pa t} \gamma&=&2\gamma(-\alpha K+D_i \beta^i), 
\label{dtgam}\\
{\pa \over \pa t} K&=&\alpha ({\rm tr}R+K^2)-D^iD_i \alpha 
+\beta^j D_j K +4\pi\alpha (S^l_{~l}-3\rho_{H}),{\hskip 3.4cm}
\label{dtK}
\end{eqnarray}
where $R_{ij}$, $\gamma$ and $S_{ij}$ are, respectively, 
the Ricci tensor with respect of 
$\gamma_{ij}$, the determinant of $\gamma_{ij}$ and
\begin{equation}
S_{ij}=T_{kl}\gamma^k_{~i}\gamma^l_{~j}. 
\end{equation}
Hereafter we use the conformal factor 
$\psi=\gamma^{1/12}$ instead of $\gamma$ for simplicity. 

To distinguish the wave part from the non-wave part (for example, 
Newtonian potential) in the metric, we use 
$\tilde \gamma_{ij}=\psi^{-4} \gamma_{ij}$ instead of $\gamma_{ij}$. 
Then ${\rm det}(\tilde \gamma_{ij})=1$ is satisfied. 
We also define $\tilde A_{ij}$ as 
\begin{equation}
\tilde A_{ij} \equiv \psi^{-4} A_{ij}
\equiv \psi^{-4}\Bigl(K_{ij}-{1 \over 3}\gamma_{ij} K\Bigr). 
\end{equation}
We should note that in our notation, indices of 
$\tilde A_{ij}$ are raised and lowered by $\tilde \gamma_{ij}$, so that 
the relations, 
$\tilde A^i_{~j}=A^i_{~j}$ and $\tilde A^{ij}=\psi^4A^{ij}$, hold. 
Using these variables, the evolution equations (\ref{dtgamij})-(\ref{dtK}) 
can be rewritten as follows;
\begin{eqnarray}
{\pa \over \pa n} \tilde \gamma_{ij}&=&-2\alpha \tilde A_{ij} 
+\tilde \gamma_{il}{\pa \beta^l \over \pa x^j}
+\tilde \gamma_{jl}{\pa \beta^l \over \pa x^i}
-{2 \over 3}\tilde \gamma_{ij}{\pa \beta^l \over \pa x^l}, 
\hspace*{5cm}
\label{ggg} \\
{\pa \over \pa n} \tilde A_{ij}&=&{1 \over \psi^4} \Bigl[
\alpha \Bigl(R_{ij}-{1 \over 3}\gamma_{ij}{\rm tr}R \Bigr)
-\Bigl(\tilde D_i\tilde D_j \alpha-{1 \over 3}\tilde\gamma_{ij}
\tilde\Delta \alpha \Bigr) 
\nonumber\\
&&-{2 \over \psi}\Bigl(\psi_{,i}\alpha_{,j}+\psi_{,j}\alpha_{,i}
-{2 \over 3}\tilde\gamma_{ij}\tilde\gamma^{kl}\psi_{,k}\alpha_{,l} 
\Bigr)\Bigr] \nonumber\\
&&+\alpha(K \tilde A_{ij}-2\tilde A_{il}\tilde A^l_{~j})
+{\pa \beta^m \over \pa x^i} \tilde A_{mj}
+{\pa \beta^m \over \pa x^j} \tilde A_{mi}
-{2 \over 3}{\pa \beta^m \over \pa x^m} \tilde A_{ij}
\nonumber\\
&&-8\pi{\alpha \over \psi^4}
\Bigl(S_{ij}-{1 \over 3}\gamma_{ij}S^l_{~l}\Bigr), 
\label{kkk}\\
{\pa \over \pa n} \psi&=&{\psi \over 6}\Bigl(-\alpha K+
{\pa \beta^i \over \pa x^i}\Bigr), 
\label{kkg} \\
{\pa \over \pa n} K&=&\alpha \Bigl(\tilde A_{ij}\tilde A^{ij}
+{1 \over 3}K^2 \Bigr) 
-{1 \over \psi^4}\tilde\Delta\alpha-{2 \over \psi^5}\tilde\gamma^{kl} 
\psi_{,k}\alpha_{,l}+4\pi\alpha(S^i_{~i}+\rho_{H}) ,
\label{kkt}
\end{eqnarray}
where $\tilde D_i$ and $\tilde\Delta$ are 
the covariant derivative and Laplacian with respect to 
$\tilde\gamma_{ij}$ and 
\begin{equation}
{\pa \over \pa n}={\pa \over \pa t}
-\beta^i{\pa \over \pa x^i}. 
\end{equation}

The Hamiltonian constraint equation then takes the following form.
\begin{equation}
\tilde \Delta \psi={1 \over 8}{\rm tr}\tilde R\psi-2\pi \rho_{H}\psi^5
-{\psi^5 \over 8}\Bigl(\tilde A_{ij} \tilde A^{ij}-{2 \over 3}K^2 \Bigr)
,\label{hamisec}
\end{equation}
where ${\rm tr}\tilde R$ is the scalar curvature with respect to 
$\tilde\gamma_{ij}$. 
The momentum constraint is also written as 
\begin{equation}
\tilde D_j (\psi^6 \tilde A^j_{~i})-{2 \over 3}\psi^6 \tilde D_i K=
8 \pi \psi^6 J_i . 
\end{equation}

Now let us consider $R_{ij}$ in Eq.$(\ref{kkk})$, 
which is one of the main source terms of the evolution 
equation for $\tilde A_{ij}$. First we split $R_{ij}$ into two parts as 
\begin{equation}
R_{ij}=\tilde R_{ij}+R^{\psi}_{ij}, \label{ric}
\end{equation}
where $\tilde R_{ij}$ is the Ricci tensor with respect to $\tilde 
\gamma_{ij}$ and $R^{\psi}_{ij}$ is defined as 
\begin{equation}
R^{\psi}_{ij}=-{2 \over \psi}\tilde D_i \tilde D_j \psi
-{2 \over \psi} \tilde \gamma_{ij} \tilde D^k \tilde D_k \psi 
+{6 \over \psi^2}(\tilde D_i \psi)(\tilde D_j \psi)
-{2 \over \psi^2}\tilde \gamma_{ij}(\tilde D_k \psi)(\tilde D^k \psi).
\end{equation}
Using  ${\rm det}(\tilde \gamma_{ij})=1$, 
$\tilde R_{ij}$ is written as 
\begin{eqnarray}
\tilde R_{ij}&=&{1 \over 2}\Bigl[-h_{ij,kk}+h_{jl,li}+h_{il,lj}
+f^{kl}_{~~,k}(h_{lj,i}+h_{li,j}-h_{ij,l})\nonumber\\
&&~~~~+f^{kl}(h_{kj,il}+h_{ki,jl}-h_{ij,kl})\Bigr]
-\tilde \Gamma^l_{kj} \tilde \Gamma^k_{li}~, \label{ricci}
\end{eqnarray}
where $_{,i}$ denotes $\pa/\pa x^i$, $\tilde \Gamma^k_{ij}$ is 
the Christoffel symbol, 
and we split $\tilde \gamma_{ij}$ and $\tilde \gamma^{ij}$ 
as $\delta_{ij}+h_{ij}$ and $\delta^{ij}+f^{ij}$, 
by writing the flat metric as $\delta_{ij}$. 

We shall consider only the linear order in $h_{ij}$ and $f_{ij}$ 
assuming $|h_{ij}|,~|f_{ij}| \ll 1$. (As a result, $h_{ij}=-f^{ij}+O(h^2)$.) 
Such a treatment is justified because $h_{ij}$ turns out to be  2PN quantity 
in our choice of gauge condition (see below). 
Here, to clarify the wave property of $\tilde \gamma_{ij}$, 
we impose a kind of transverse gauge to $h_{ij}$ as 
\begin{equation}
h_{ij,j}=0, 
\end{equation}
which was first proposed by Nakamura in relation to numerical relativity  
\cite{nakamura}. 
Hereafter, we call this condition merely the transverse gauge. 
The equation $(\ref{ggg})$ shows that this condition is guaranteed if $\beta^i$ satisfies 
\begin{equation}
-\beta^{k}_{,~j} \tilde \gamma_{ij,k}=\Bigl(-2\alpha \tilde A_{ij} 
+\tilde \gamma_{il} \beta^l_{~,j}+\tilde \gamma_{jl} \beta^l_{~,i}
-{2 \over 3}\tilde \gamma_{ij} \beta^l_{~,l}
 \Bigr)_{,j}.\label{bbb}
\end{equation}
Using the transverse gauge, Eq.$(\ref{ricci})$ becomes 
\begin{equation}
\tilde R_{ij}=-{1 \over 2}\Delta h_{ij}+O(h^2), \label{ricci2}
\end{equation}
where $\Delta$ is the Laplacian with respect to $\delta_{ij}$. 
Note that ${\rm tr}\tilde R=O(h^2)$ is guaranteed in the transverse 
gauge because the traceless property of $h_{ij}$ holds 
in the linear order. 

We show the equations for the isentropic perfect fluid 
$(\ref{stressenergy})$. 
The conservation law of mass density, $(\ref{Nconsv})$, may be written as
\begin{equation}
{\pa \rho_{\ast} \over \pa t}+{\pa (\rho_{\ast}v^i) \over \pa x^i}=0, 
\label{continuity}
\end{equation}
where $\rho_{\ast}$ is the conserved density defined as 
\begin{equation}
\rho_{\ast}=\alpha\psi^6\rho u^0. 
\label{consdensity}
\end{equation}
The equation of motion and the energy equation are obtained from the
conservation law $(\ref{emconsv})$ which  takes the 
following forms.
\begin{equation}
{\pa S_i \over \pa t}+{\pa (S_i v^j) \over \pa x^j}
=-\alpha \psi^6 P_{,i}-\alpha \alpha_{,i} S^0 + S_j \beta^j_{~,i}
-{1 \over 2S^0}S_j S_k \gamma^{jk}_{~~,i},
\label{EOM}
\end{equation}
and 
\begin{equation}
{\pa H \over \pa t}+{\pa (H v^j) \over \pa x^j}
=-P\Bigl({\pa (\alpha \psi^6 u^0) \over \pa t}
+{\pa (\alpha \psi^6 u^0 v^j) \over \pa x^j}\Bigr),
\label{Energy}
\end{equation}
where 
\begin{eqnarray}
S_i&=&\alpha \psi^6 (\rho+\rho \Pi +P)u^0 u_i
=\rho_{\ast} \Bigl(1+\Pi+{P \over \rho} \Bigr)u_i (=\psi^6 J_i), 
\nonumber\\ 
S^0&=&\alpha \psi^6 (\rho+\rho \Pi +P)(u^0)^2 \Bigl(=
{(\rho_H +P)\psi^6 \over \alpha}\Bigr),\nonumber\\
H&=&\alpha \psi^6 \rho \Pi u^0=\rho_{\ast}\Pi ,\nonumber\\
v^i&\equiv& {u^i \over u^0}=-\beta^i+{\gamma^{ij} S_j \over S^0}. 
\label{vel} 
\end{eqnarray}
Finally, we note that in the above equations, only $\beta^i$ appears, 
and $\beta_i$ does not, so that, in the subsequent section, we only 
consider the PN expansion of $\beta^i$, not of $\beta_i$.

\subsection{Post-Newtonian approximation in the (3+1) formalism}

Next, we consider the PN approximation of 
the above set of equations. First of all, we review the PN expansion 
of the variables. 
Each metric variable may be expanded up to the relevant order as
\begin{eqnarray}
\psi&=&1+\two\psi+\four\psi+\six\psi+\seven\psi+\dots,\nonumber\\
\alpha&=&1+\two\alpha+\four\alpha+\six\alpha+\seven\alpha+\dots,\nonumber\\
&&=1-U+\Bigl({U^2 \over 2}+X\Bigr)+\six\alpha+\seven\alpha+\dots, \nonumber\\
\beta^i&=&\three\beta^i+\five\beta^i+\six\beta^i+\seven\beta^i+
\eight\beta^i+\dots,\nonumber\\
h_{ij}&=&\four h_{ij}+\five h_{ij}+\dots,\nonumber\\
\tilde A_{ij}&=&\three\tilde A_{ij}+\five\tilde A_{ij}+\six\tilde A_{ij}
+\dots,\nonumber\\
K&=&\three K+\five K+\six K+\dots, 
\end{eqnarray}
where subscripts denote the PN order($\epsilon^n$) and $U$ is 
the Newtonian potential satisfying 
\begin{equation}
\Delta U=-4\pi \rho.\label{Npot}
\end{equation}
$X$ depends on the slice condition, and in the standard PN gauge 
\cite{chandra65}, we usually use $\Phi=-X/2$, which satisfies 
\begin{equation}
\Delta \Phi=-4\pi\rho\Bigl(v^2+U+{1 \over 2}\Pi
+{3 \over 2}{P \over \rho}\Bigr).  
\end{equation}
Note that the terms relevant to 
the radiation reaction appear in $\seven\psi$, 
$\seven\alpha$, $\eight\beta^i$ and $\five h_{ij}$, and 
the quadrupole formula is derived from 
$\seven\alpha$ and $\five h_{ij}$.

The four velocity is expanded as 
\begin{eqnarray}
u^\tau&=&\epsilon \Bigl[1+\epsilon^2 \Bigl({1 \over 2}v^2+U\Bigr)+
\epsilon^4 \Bigl({3 \over 8}v^4+{5 \over 2}v^2U
+{1 \over 2}U^2+\three\beta^i v^i-X \Bigr)+O(\epsilon^6)\Bigr],\nonumber\\
u_\tau&=&-\epsilon^{-1}\Bigl[ 1+\epsilon^2 \Bigl({1 \over 2}v^2-U\Bigr)
+\epsilon^4 \Bigl({3 \over 8}v^4+{3 \over 2}v^2U
+{1 \over 2}U^2+X \Bigr)+O(\epsilon^6)\Bigr],\nonumber\\
u^i&=&v^i
\Bigl[1+\epsilon^2 \Bigl({1 \over 2}v^2+U\Bigr)
+\epsilon^4 \Bigl({3 \over 8}v^4+{5 \over 2}v^2U
+{1 \over 2}U^2+\three\beta^i v^i-X \Bigr)\Bigr]+O(\epsilon^7),\cr
u_i&=&v^i+\epsilon^3\Bigl\{\three\beta^i
+v^i\Bigl({1 \over 2}v^2+3 U\Bigr)\Bigr\} 
+\epsilon^5\Bigl[\five\beta^i+\three\beta^i\Bigl({1 \over 2}v^2+3U\Bigr)
+\four h_{ij}v^j \nonumber\\
&&~~~+v^i\Bigl({3 \over 8}v^4+{7 \over 2}v^2 U+4U^2-X+4\four\psi 
+\three\beta^j v^j\Bigr)\Bigr]+
\epsilon^6 \Bigl(\six \beta^i+\five h_{ij}v^j\Bigr) 
\nonumber\\
&&+O(\epsilon^7), 
\label{fourvel}
\end{eqnarray}
where $v^i = O(\epsilon)$ and $v^2=v^i v^i$. 
{}From $u^{\mu}u_{\mu}=-1$, we obtain the useful relation 
\begin{eqnarray}
(\alpha u^\tau)^2&=&1+\gamma^{ij}u_i u_j\cr
&=&1+\epsilon^2 v^2
+ \epsilon^4 \left( v^4+4v^2U+2\three\beta^iv^i\right)+O(\epsilon^6). 
\end{eqnarray}
Thus $\rho_{H}$, $J_i$ and $S_{ij}$ are respectively expanded as
\begin{eqnarray}
\rho_{H}&=&\epsilon^2\rho\Bigl[1+\epsilon^2\Bigl(v^2+\Pi\Bigr)
+\epsilon^4\Bigl\{v^4+v^2\Bigl(
4U+\Pi+{P \over \rho}\Bigr)+2\three\beta^i v^i\Bigr\}\Bigr]
+O(\epsilon^8), \nonumber\\
J_i&=&\epsilon^3\rho\Bigl[v^i\Bigl(1+\epsilon^2 \left(v^2+3U+\Pi
+{P \over \rho} \right) \Bigr)
+\epsilon^3\three\beta^i\Bigr]+O(\epsilon^7), \nonumber\\ 
S_{ij}&=&\epsilon^4\rho\Bigl[\Bigl(v^i v^j+{P \over \rho}\delta_{ij}\Bigr) 
+\epsilon^2 \Bigl\{\Bigl(v^2+6U+\Pi+{P \over \rho}\Bigr)v^i v^j
+v^i\three\beta^j+v^j\three\beta^i 
\nonumber\\
&&~~~~~~+2{UP \over \rho}\delta_{ij} \Bigr\} \Bigr] +O(\epsilon^8), 
\nonumber\\
S_l^{~l}&=&\epsilon^4\rho\Bigl[v^2+{3P \over \rho}
+\epsilon^2\Bigl\{2 \three \beta^i v^i+
v^2\Bigl(v^2+4U+\Pi+{P \over \rho}\Bigr)\Bigr\}\Bigr]
+O(\epsilon^8) . 
\end{eqnarray}

The conformal factor $\psi$ (and $\alpha$ in the conformal slice) is 
determined by the Hamiltonian constraint. 
In the PN approximation, the Laplacian with respect to 
${\tilde \gamma}^{ij} $ for the scalar is expanded as 
\begin{equation}
\tilde\Delta=\Delta-(\epsilon^4\four h_{ij}+\epsilon^5\five h_{ij})\pa_i\pa_j 
+O(\epsilon^6) . 
\end{equation}
At the lowest order, the Hamiltonian constraint becomes 
\begin{equation}
\Delta \two\psi=-2\pi \rho .  
\end{equation}
Thus, $\two\alpha=-2\two\psi=-U$ is satisfied in this paper. 
At the 2PN and 3PN orders, the Hamiltonian constraint equation 
becomes, respectively, 
\begin{equation}
\Delta \four\psi=-2\pi\rho\Bigl(v^2+\Pi+{5 \over 2}U \Bigr),
\label{psifo}
\end{equation}
and
\begin{eqnarray}
\Delta \six\psi&=&
-2\pi\rho\Bigl\{v^4+v^2\Bigl(\Pi+{P \over \rho}+{13 \over 2}U
\Bigr)+2 \three\beta^i v^i+{5 \over 2}\Pi U+
{5 \over 2}U^2+5\four\psi\Bigr\} \nonumber\\
&&+{1 \over 2}\four h_{ij}U_{,ij}-{1 \over 8}
\Bigl(\three\tilde A_{ij} \three\tilde A_{ij}-{2 \over 3}\three K^2\Bigr). 
\label{sixpsi}
\end{eqnarray}
The term relevant to the radiation reaction first appears in 
$\seven\psi$ whose equation becomes 
\begin{equation} 
\Delta \seven\psi={1 \over 2}\five h_{ij}U_{,ij}. \label{sevenpsi}
\end{equation} 
Hence, $\seven\alpha$ may be also relevant to the radiation reaction, 
depending on the slice condition. 

Let us now derive equations for $\beta^i$.
{}From Eq.$(\ref{bbb})$, the relation between 
$\three\tilde A_{ij}$ and $\three\beta^i$ becomes 
\begin{equation} 
-2\three\tilde A_{ij}+\three\beta^i_{,j}
+\three\beta^j{,i}-{2 \over 3}\delta_{ij}\three\beta^l_{,l}=0.\label{ppb}
\end{equation}
where we used the boundary condition that $\three\tilde A_{ij}$ and 
$\three\beta^i$ vanishes at the spatial infinity.  
$\three\tilde A_{ij}$ must also satisfy the momentum constraint. 
Since $\three\tilde A_{ij}$ does not contain the 
transverse-traceless (TT) part and only contains 
the longitudinal part, it can be written as 
\begin{equation} 
\three\tilde A_{ij} =\three W_{i,j}+\three W_{j,i}
-{2 \over 3}\delta_{ij}\three W_{k,k}~,  
\end{equation} 
where $\three W_i$ is a vector on the 3D hypersurface and satisfies 
the momentum constraint at the first PN order as follows; 
\begin{equation}
\Delta \three W_i+{1 \over 3}\three W_{j,ji}
-{2 \over 3}\three K_{,i}=8\pi \rho v^i.
\label{momo}
\end{equation}
{}From Eq.$(\ref{ppb})$, the relation, 
\begin{equation}
\three\beta^i=2\three W_i~, 
\end{equation} 
holds and at the first PN order, Eq.$(\ref{kkg})$ becomes 
\begin{equation} 
3\dot U=-\three K+\three\beta^l{,l}~,\label{dotu}
\end{equation} 
where $\dot U$ denotes the derivative of $U$ with respect to time. 
Thus Eq.$(\ref{momo})$ is rewritten as 
\begin{equation} 
\Delta \three\beta^i=16\pi \rho v^i+\Bigl(\three K_{,i}-\dot U_{,i}
\Bigr).\label{betth}
\end{equation} 
This is the equation for the vector potential 
at the first PN order. 

{}At the higher order, $\hbox{$_{(n)}$}\beta^i$ is also 
determined by the gauge 
condition, $h_{ij,j}=0$. 
The equation for $\five\beta^i$ is obtained by using the momentum 
constraint and the 2PN order of Eq.$(\ref{kkg})$ as follows, 
\begin{eqnarray} 
\Delta \five\beta^i&=&16\pi\rho\Bigl[v^i \Bigl(v^2+2U+\Pi
+{P \over \rho}\Bigr)+\three\beta^i\Bigr]-8U_{,j}\three\tilde A_{ij} 
\nonumber\\
&&+\five K_{,i}-U \three K_{,i}+{1 \over 3}U_{,i}\three K-2\four\dot\psi_{,i}
+{1 \over 2}(U \dot U)_{,i}+(\three\beta^l U_{,l})_{,i}~. 
\nonumber\\&&
\label{betfi}
\end{eqnarray} 
The equation for $\six\beta^i$ takes the purely geometrical form since 
the material contribution $J_i$ at the 1.5PN order vanishes. 
\begin{equation}
\Delta \six\beta^i=\six K_{,i}~.
\label{betsi}
\end{equation} 

Then, let us consider the wave equation for $h_{ij}$. 
{}From Eqs.$(\ref{ggg})$, $(\ref{kkk})$, $(\ref{ric})$ and $(\ref{ricci2})$, 
the wave equation for $h_{ij}$ is written as 
\begin{eqnarray} 
\Box h_{ij}&=&\Bigl(1-{\alpha^2 \over \psi^4}\Bigr)\Delta h_{ij}
+\Bigl({\pa^2 \over \pa n^2}-{\pa^2 \over \pa t^2}\Bigr)h_{ij} \nonumber\\
&&+{2\alpha \over \psi^4}
\Bigl[-{2\alpha \over \psi}\Bigl(\tilde D_i\tilde D_j
-{1 \over 3}\tilde\gamma_{ij}\tilde\Delta\Bigr)\psi
+{6\alpha \over \psi^2}\Bigl(\tilde D_i\psi \tilde D_j\psi
-{1 \over 3}\tilde\gamma_{ij}\tilde D_k\psi \tilde D^k\psi\Bigr) \nonumber\\
&&~~~~~
-\Bigl(\tilde D_i\tilde D_j-{1 \over 3}\tilde\gamma_{ij}\tilde\Delta
\Bigr)\alpha
-{2 \over \psi}\Bigl(\tilde D_i\psi\tilde D_j\alpha
+\tilde D_j\psi\tilde D_i\alpha-{2 \over 3}\tilde\gamma_{ij}
\tilde D^k \psi\tilde D_k\alpha\Bigr)\Bigr] \nonumber\\
&&+2\alpha^2 \Bigl(K \tilde A_{ij}-2\tilde A_{il} \tilde A^l_{~j}\Bigr)
+2\alpha \Bigl( \beta^m_{,~i} \tilde A_{mj}
+\beta^m_{~,j} \tilde A_{mi}-{2 \over 3} \beta^m_{~,m} \tilde A_{ij}
\Bigr)\nonumber\\
&&-16\pi{\alpha^2 \over \psi^4}
\Bigl(S_{ij}-{1 \over 3}\gamma_{ij}S^l_{~l}\Bigr)
-{\pa \over \pa n}\Bigl( \beta^m_{~,i} \tilde \gamma_{mj}
+ \beta^m_{~,j} \tilde \gamma_{mi}
-{2 \over 3}\beta^m_{~,m} \tilde \gamma_{ij} \Bigr)
+2{\pa\alpha \over \pa n}\tilde A_{ij} \nonumber\\
&\equiv&\tau_{ij}, \label{waveeq}
\end{eqnarray}
where $\Box$ is the flat spacetime wave operator defined as 
\begin{equation}
\Box=-{\pa^2 \over \pa t^2}+\Delta. 
\end{equation}
We should note that $\four\tau_{ij}$ has the TT property, i.e., 
$\four\tau_{ij,j}=0$ and $\four\tau_{ii}=0$. 
This is a natural consequence of the transverse gauge, $h_{ij,j}=0$ and 
$h_{ii}=O(h^2)$. 
Thus $\four h_{ij}$ is determined from 
\begin{equation}
\Delta \four h_{ij}=\four\tau_{ij}. 
\label{fourh}
\end{equation}
Since $O(h^2)$ turns out to be $O(\epsilon^8)$, it is enough to consider 
only the linear order of $h_{ij}$ in the case when we perform 
the PN approximation up to the 3.5PN order. 
We can obtain $\five h_{ij}$ by evaluating 
\begin{equation}
\five h_{ij}(\tau)={1 \over 4\pi}
{\pa \over \pa \tau}\int \four\tau_{ij}(\tau, {\bf y})d^3y, 
\label{quadr}
\end{equation}
and the quadrupole mode of gravitational waves in the wave zone 
is written as 
\begin{equation}
h_{ij}^{rad}(\tau, {\bf x})=-{1 \over 4\pi} \lim_{|{\bf x}| \rightarrow \infty}
\int {\four\tau_{ij}(\tau-\epsilon|{\bf x}-{\bf y}|, {\bf y}) \over 
|{\bf x}-{\bf y}|}d^3y.
\label{quadra}
\end{equation}
Later, we shall derive the quadrupole radiation-reaction 
metric in the near zone using Eq.$(\ref{quadr})$. 

Finally, we show the evolution equation for $K$. 
Since we adopt slice conditions which do not satisfy $K=0$ (i.e. the 
maximal slice condition), the evolution equation for $K$ is necessary. 
The evolution equations appear at the 1PN, 2PN and 
2.5PN orders which become respectively  
\begin{eqnarray}
{\pa \over \pa \tau}\three K&=&4 \pi\rho\Bigl(2v^2+\Pi+2U+3 {P \over \rho}
\Bigr)-\Delta X, \\
\label{Kth}
{\pa \over \pa \tau}\five K&=&4 \pi\rho\Bigl[2v^4+v^2\Bigl(6U+2\Pi
+2{P \over \rho}\Bigr)-\Bigl(\Pi +{3P \over \rho}\Bigr)U 
-4 U^2+4\four\psi+X \nonumber\\
&&~~~~~~+4\three\beta^i v^i \Bigr]
+\three \tilde A_{ij}\three \tilde A_{ij}
+{1 \over 3}\three K^2-\four h_{ij}U_{,ij}
+\three\beta^i \three K_{,i} \nonumber\\
&&-{3 \over 2}UU_{,k} U_{,k}-U_{,k}X_{,k}+2U_{,k}\four\psi_{,k}
-\Delta \six\alpha+2U\Delta X , \\ 
{\pa \over \pa \tau}\six K&=&-\Delta \seven\alpha-\five h_{ij}U_{,ij}. 
\label{dtsixK}
\end{eqnarray}

We note that for the PN equations of motion up to the 2.5PN order, 
we need $\two\alpha$, $\four\alpha$, $\six\alpha$, $\seven\alpha$, 
$\two\psi$, $\four\psi$, $\three\beta^i$, $\five\beta^i$, $\six\beta^i$, 
$\four h_{ij}$, $\five h_{ij}$, $\three K$, $\five K$ and $\six K$. 
Therefore, if we solve the above set of the equations, we can obtain 
the 2.5 PN equations of motion. 
Up to the 2.5PN order, the hydrodynamic equations become 
\begin{eqnarray}
{\pa S_i \over \pa \tau}+{\pa (S_i v^j) \over \pa x^j}
&=&-\Bigl(1+2 U+{5 \over 4}U^2+6\four\psi+X\Bigr)P_{,i} \nonumber\\
&&+\rho_{\ast}\Bigl[U_{,i} \Bigl\{1+\Pi+{P \over \rho}
+{3 \over 2}v^2-U+{5 \over 8}v^4+4 v^2U \nonumber\\
&&~~~~~~~~~~~~
+\Bigl({3 \over 2}v^2-U\Bigr)\Bigl(\Pi+{P \over \rho}\Bigr)
+3\three\beta^j v^j \Bigr\} \nonumber\\
&&~~~~~-X_{,i}\Bigl(1+\Pi+{P \over \rho}+{v^2 \over 2}\Bigr) 
+2v^2\four\psi_{,i}-\six\alpha_{,i}-\seven\alpha_{,i} \nonumber\\
&&~~~~~+v^j \Bigl\{ \three\beta^j{,i}\Bigl(1+\Pi+{P \over \rho}
+{v^2 \over 2}+3U\Bigr)+\five\beta^j{,i}+\six\beta^j_{,i} \Bigr\}
\nonumber\\
&&~~~~~+\three\beta^j\three\beta^j_{,i} 
+{1 \over 2}v^jv^k(\four h_{jk,i}+\five h_{jk,i}) \Bigr]
+O(\epsilon^8) , \\
{\pa H \over \pa \tau}+{\pa (H v^j) \over \pa x^j} 
&=&-P\Bigl[v^j_{~,j}+{\pa \over \pa t}\Bigl({1 \over 2}v^2+3 U\Bigr)
+{\pa \over \pa x^j} \Bigl\{\Bigl({1 \over 2}v^2+3 U\Bigr)v^j \Bigr\} 
+O(\epsilon^5) \Bigr], \nonumber\\
\end{eqnarray}
where we have omitted $\epsilon^n$ coefficients for the sake of
simplicity,  and  used  relations
\begin{eqnarray}
\alpha S^0&=&\rho_* \Bigl[1+\Pi+{P \over \rho}+{v^2 \over 2}
+{v^2 \over 2}\Bigl(\Pi+{P \over \rho}\Bigr)+{3 \over 8}v^4
+2v^2U+\three\beta^j v^j )\Bigr]+O(\epsilon^6),\cr
S_i&=&\rho_*\Bigl[v^i 
\Bigl(1+\Pi+{P \over \rho}+{v^2 \over 2}+3U\Bigr)+\three \beta^i \Bigr]
+O(\epsilon^5). 
\end{eqnarray}

\subsection{Strategy to obtain 2PN tensor potential}

Although the 2PN tensor potential is formally solved as 
\begin{equation}
\four h_{ij}(\tau,{\bf x})=-{1 \over 4\pi}\int{\four\tau_{ij}(\tau,{\bf y}) 
\over \vert {\bf x}-{\bf y} \vert}d^3y, \label{formalhij}
\end{equation} 
it is difficult to estimate this integral numerically  since
$\four \tau_{ij} \rightarrow O(r^{-3})$ for $r \rightarrow \infty$ and the
integral is taken all over the space.
Thus it is desirable to replace this equation by some tractable forms 
in numerical evaluation. 
We shall present two methods to do so. 
One is to change Eq.$(\ref{formalhij})$ into the form in which 
the integration is performed only over the matter distribution 
like as in the Newtonian potential. 
The other is to solve Eq.$(\ref{formalhij})$ 
as the boundary value problem\cite{asf}.

{\bf Strategy1: Direct integration method}

The explicit form of $\four \tau_{ij}$ is 
\begin{eqnarray}
\four\tau_{ij}&=&-2\hat\pa_{ij}(X+2\four\psi)+U\hat\pa_{ij} U 
-3U_{,i}U_{,j}+\delta_{ij}U_{,k}U_{,k}
-16\pi\Bigl(\rho v^iv^j-{1 \over 3}\delta_{ij}\rho v^2\Bigr) \nonumber\\
&&-\Bigl( \three\dot\beta^i_{,j}+\three\dot\beta^j_{,i}
-{2 \over 3}\delta_{ij}\three\dot\beta^k_{,k} \Bigr), 
\end{eqnarray}
where
\begin{equation}
\hat\pa_{ij} \equiv {\pa^2 \over \pa x^i \pa x^j}
-{1 \over 3}\delta_{ij}\Delta.
\end{equation}
Although $\four\tau_{ij}$ looks as if it depends on the slice condition, 
the independence is shown as follows.
Eq.($\ref{betth}$) is solved formally as
\begin{equation}
\three\beta^i=p_i-{1 \over 4\pi}\Bigl( \int {\three K \over 
\vert {\bf x}-{\bf y} \vert} d^3y \Bigr)_{,i}~, 
\label{thbet}
\end{equation}
where
\begin{equation}
p_i=-4\int{\rho v^i \over \vert {\bf x}-{\bf y} \vert} d^3y
-{1 \over 2}\Bigl(\int \dot\rho \vert {\bf x}-{\bf y} \vert d^3y 
\Bigr)_{,i}~. 
\end{equation}

{}From Eqs.$(\ref{psifo})$ and $(\ref{Kth})$, we obtain
\begin{equation}
\three\dot K=-\Delta (X+2\four\psi)+4\pi\rho\Bigl( v^2
+3{P \over \rho}-{U \over 2} \Bigr). 
\label{abceq}
\end{equation}
Combining Eq.$(\ref{thbet})$ with Eq.$(\ref{abceq})$, the equation for 
$\three\dot\beta^i$ is written as
\begin{equation}
\three\dot\beta^i=\dot  p_i-(X+2\four\psi)_{,i}
-\Bigl[\int{\Bigl( \rho v^2+3P-\rho U / 2\Bigr) 
\over \vert {\bf x}-{\bf y} \vert}d^3y \Bigr]_{,i} ~.
\end{equation}
Using this relation, 
the source term, $\four\tau_{ij}$, is split into five parts 
\begin{equation}
\four\tau_{ij}=
\four\tau_{ij}^{(S)}+\four\tau_{ij}^{(U)}+\four\tau_{ij}^{(C)}
+\four\tau_{ij}^{(\rho)}+\four\tau_{ij}^{(V)}, 
\label{fiveparts}
\end{equation}
where we introduced the following notations. 
\begin{eqnarray}
\four\tau_{ij}^{(S)}&=&-16\pi\Bigl( \rho v^iv^j-{1 \over 3}\delta_{ij}
\rho v^2 \Bigr), \nonumber\\
\four\tau_{ij}^{(U)}&=&UU_{,ij}-{1 \over 3}\delta_{ij}U\Delta U
-3U_{,i}U_{,j}+\delta_{ij}U_{,k}U_{,k}, \nonumber\\
\four\tau_{ij}^{(C)}&=&4{\pa \over \pa x^j}\int{(\rho v^i)^{\cdot}
\over \vert {\bf x}-{\bf y} \vert}d^3y
+4{\pa \over \pa x^i}\int{(\rho v^j)^{\cdot}
\over \vert {\bf x}-{\bf y} \vert}d^3y
-{8 \over 3}\delta_{ij}{\pa \over \pa x^k}\int{(\rho v^k)^{\cdot}
\over \vert {\bf x}-{\bf y} \vert}d^3y, \nonumber\\
\four\tau_{ij}^{(\rho)}&=&\hat\pa_{ij}\int\ddot\rho \vert {\bf x}-{\bf y} 
\vert d^3y, \nonumber\\
\four\tau_{ij}^{(V)}&=&2\hat\pa_{ij}\int{\Bigl( \rho v^2+3P
-\rho U / 2\Bigr) \over \vert {\bf x}-{\bf y} \vert}d^3y.  
\end{eqnarray}
Thus it becomes clear that $\four h_{ij}$ and $\five h_{ij}$ 
as well as $\four\tau_{ij}$ are expressed in terms of matter variables 
only and thus their forms do not depend on slicing conditions, 
though values of matter variables depend  on gauge conditions. 

Then, we define 
$\four h_{ij}^{(S)}=\Delta^{-1}\four\tau_{ij}^{(S)}$, 
$\four h_{ij}^{(U)}=\Delta^{-1}\four\tau_{ij}^{(U)}$, 
$ \four h_{ij}^{(C)}=\Delta^{-1}\four\tau_{ij}^{(C)}$, 
$ \four h_{ij}^{(\rho)}=\Delta^{-1}\four\tau_{ij}^{(\rho)}$ and 
$ \four h_{ij}^{(V)}=\Delta^{-1}\four\tau_{ij}^{(V)}$, 
and consider each term separately.
First, since $\four\tau_{ij}^{(S)}$ is a compact source, 
we immediately obtain  
\begin{equation}
\four h_{ij}^{(S)}=4\int{\Bigl(\rho v^i v^j-{1 \over 3}\delta_{ij}
\rho v^2 \Bigr) \over \vert {\bf x}-{\bf y} \vert}d^3y. 
\end{equation}
Second, we consider the following equation 
\begin{equation}
\Delta G({\bf x},{\bf y_1},{\bf y_2})={1 \over \vert {\bf x}-{\bf y_1} 
\vert \vert {\bf x}-{\bf y_2} \vert}. 
\label{G}
\end{equation}
It is possible to write $\four h_{ij}^{(U)}$ using integrals over the matter 
if this function, $G$, is used.
Eq.$(\ref{G})$ has solutions \cite{fock,ookh74b}, 
\begin{equation}
G({\bf x},{\bf y_1},{\bf y_2})=\ln(r_1+r_2 \pm r_{12}), 
\end{equation}
where
\begin{equation}
r_1=\vert {\bf x}-{\bf y_1} \vert, \:
r_2=\vert {\bf x}-{\bf y_2} \vert, \: 
r_{12}=\vert {\bf y_1}-{\bf y_2} \vert. 
\end{equation}
Note that $\ln(r_1+r_2 - r_{12})$ is not regular on 
the interval between ${\bf y_1}$ and ${\bf y_2}$, 
while $\ln(r_1+r_2+r_{12})$ is regular on the matter. 
Thus we use $\ln (r_1+r_2+r_{12})$ as a kernel. 
Using this kernel, $UU_{,ij}$ and $U_{,i}U_{,j}$ are rewritten as 
\begin{eqnarray}
UU_{,ij}&=&
\Bigl[ {\pa^2 \over \pa x^i \pa x^j}\Bigl(\int{\rho({\bf y_1}) \over 
|{\bf x}-{\bf y_1}|} d^3y_1 \Bigr) \Bigr] 
\Bigl(\int{\rho({\bf y_2}) \over |{\bf x}-{\bf y_2}|} d^3y_2 
\Bigr) \nonumber\\
&=&\int d^3y_1 d^3y_2 \rho({\bf y_1})\rho({\bf y_2})
{\pa ^2 \over \pa y_1^i \pa y_1^j}\Bigl({1 \over |{\bf x}-{\bf y_1}|
|{\bf x}-{\bf y_2}|} \Bigr) \nonumber\\
&=&\Delta \int d^3y_1 d^3y_2 \rho({\bf y_1}) \rho({\bf y_2}) 
{\pa^2 \over \pa y_1^i \pa y_1^j} \ln(r_1+r_2+r_{12}), \nonumber\\
U_{,i}U_{,j}&=&
\Bigl({\pa \over \pa x^i}\int{\rho({\bf y_1}) \over 
|{\bf x}-{\bf y_1}|} d^3y_1 \Bigr) 
\Bigl({\pa \over \pa x^j}\int{\rho({\bf y_2}) \over 
|{\bf x}-{\bf y_2}|} d^3y_2 \Bigr) \nonumber\\
&=&\int d^3y_1 d^3y_2 \rho({\bf y_1})\rho({\bf y_2})
{\pa ^2 \over \pa y_1^i \pa y_2^j}\Bigl({1 \over |{\bf x}-{\bf y_1}|
|{\bf x}-{\bf y_2}|} \Bigr) \nonumber\\
&=&\Delta \int d^3y_1 d^3y_2 \rho({\bf y_1}) \rho({\bf y_2}) 
{\pa^2 \over \pa y_1^i \pa y_2^j} \ln(r_1+r_2+r_{12}). 
\end{eqnarray}
Thus we can express $\four h_{ij}^{(U)}$ using the integral 
over the matter as 
\begin{eqnarray}
\four h_{ij}^{(U)}=\int d^3y_1 d^3y_2 \rho({\bf y_1}) \rho({\bf y_2}) 
&&\Bigl[ \Bigl({\pa^2 \over \pa y_1^i \pa y_1^j}
-{1 \over 3}\delta_{ij}\triangle_1 \Bigr)
-3\Bigl({\pa^2 \over \pa y_1^i \pa y_2^j}
-{1 \over 3}\delta_{ij}\triangle_{12} 
\Bigr) \Bigr] 
\nonumber\\
&&\times\ln(r_1+r_2+r_{12}), 
\end{eqnarray}
where we introduced 
\begin{equation}
\triangle_1={\pa^2 \over \pa y_1^k \pa y_1^k}, \:
\triangle_{12}={\pa^2 \over \pa y_1^k \pa y_2^k}. 
\end{equation} 
Using relations $\Delta |{\bf x}-{\bf y}|=2/ |{\bf x}-{\bf y}|$ and 
$\Delta  |{\bf x}-{\bf y}|^3=12 |{\bf x}-{\bf y}|$,
$\four h_{ij}^{(C)}$, $\four h_{ij}^{(\rho)}$ and $\four h_{ij}^{(V)}$ are 
solved as 
\begin{equation}
\four h_{ij}^{(C)}=
2{\pa \over \pa x^i}\int(\rho v^j)^{\cdot}\vert {\bf x}-{\bf y} \vert d^3y
+2{\pa \over \pa x^j}\int(\rho v^i)^{\cdot}\vert {\bf x}-{\bf y} \vert d^3y
+{4 \over 3}\delta_{ij}\int\ddot\rho \vert {\bf x}-{\bf y} \vert d^3y, 
\label{hijC}
\end{equation}
\begin{equation}
\four h_{ij}^{(\rho)}=
{1 \over 12}{\pa^2 \over \pa x^i \pa x^j}\int\ddot\rho 
\vert {\bf x}-{\bf y} \vert^3 d^3y
-{1 \over 3}\delta_{ij}\int\ddot\rho \vert {\bf x}-{\bf y} \vert d^3y, 
\end{equation}
\begin{equation}
\four h_{ij}^{(V)}=
{\pa^2 \over \pa x^i \pa x^j}\int\Bigl(\rho v^2+3P-{\rho U \over 2} \Bigr)
\vert {\bf x}-{\bf y} \vert d^3y 
-{2 \over 3}\delta_{ij}\int{\Bigl(\rho v^2+3P-\rho U / 2 \Bigr)
\over \vert {\bf x}-{\bf y} \vert} d^3y. \label{hijV}
\end{equation}
Thus we find that the 2PN tensor potentials can be expressed as the
integrals only over the matter. 

{\bf Strategy2: Partial use of boundary value approach}

Although the above expression for $\four h_{ij}$ is quite interesting 
and might play an important role in some theoretical applications, 
it will take a very long time to evaluate {\it double} 
integration numerically. 
Therefore, we  propose another strategy where Eq.$(\ref{fourh})$ 
is solved as the boundary value problem.
Here, we would like to emphasize that the boundary condition should be
imposed at $r(=\vert {\bf x} \vert) \gg \vert {\bf y_1} \vert,
\vert {\bf y_2} \vert$, but $r$ does not have to be greater than 
$\lambda$,
where $\lambda$ is a typical wave length of gravitational waves. We only
need to impose $r > R$ (a typical size of matter).
This means that we do not need a large amount of grid numbers compared with 
the case of fully general relativistic simulations, 
in which we require $r > \lambda \gg R$.

First of all, we consider the equation 
\begin{equation}
\Delta \Bigl(\four h^{(S)}_{ij}+\four h^{(U)}_{ij} \Bigr)=
\four \tau_{ij}^{(S)}+\four \tau_{ij}^{(U)}. 
\end{equation}
Since the source $\four \tau_{ij}^{(U)}$ behaves as $O(r^{-6})$ 
at $r \rightarrow 
\infty$, this equation can be accurately solved 
under the boundary condition at $r > R$ as 
\begin{eqnarray}
\four h_{ij}^{(S)}+\four h_{ij}^{(U)}&=&{2 \over r}
\Bigl(\ddot I_{ij}-{1 \over 3}\delta_{ij}\ddot I_{kk}\Bigr)\nonumber\\
&&+{2 \over 3r^2}\Bigl(n^k \ddot I_{ijk}-{1 \over 3}\delta_{ij}n^k 
\ddot I_{llk}+2n^k(\dot S_{ikj}+\dot S_{jki})-{4 \over 3}\delta_{ij}n^k 
\dot S_{lkl} \Bigr) 
\nonumber\\
&&+O(r^{-3}), 
\end{eqnarray} 
where
\begin{eqnarray}
I_{ijk}&=&\int\rho x^ix^jx^k d^3x,\nonumber\\
S_{ijk}&=&\int\rho (v^ix^j-v^jx^i)x^k d^3x.  
\end{eqnarray}

Next, we consider the equations for $\four h_{ij}^{(C)}$, 
$\four h_{ij}^{(\rho)}$ and $\four h_{ij}^{(V)}$. Using the identity, 
\begin{equation}
\ddot \rho=-(\rho v^i)_{,i}^{\cdot}
=(\rho v^iv^j)_{,ij}+\Delta P-(\rho U_{,i})_{,i}~,  
\end{equation}
we find the following relations;
\begin{eqnarray}
\int \ddot \rho |{\bf x}-{\bf y}|d^3y&=&
-\int d^3y 
{x^i-y^i \over  \vert {\bf x}-{\bf y} \vert}(\rho v^i)^{\cdot},\nonumber\\
\int \ddot \rho |{\bf x}-{\bf y}|^3d^3y&=&
3\int d^3y\biggl[ \rho v^i v^j{(x^i-y^i)(x^j-y^j) \over  |{\bf x}-{\bf y}|}
\nonumber\\
&&+\Bigl(4 P+\rho v^2
-\rho U_{,i} (x^i-y^i)\Bigr) |{\bf x}-{\bf y}|\biggr]. 
\label{rereeq}
\end{eqnarray}
Using Eqs.$(\ref{rereeq})$, 
$\four h_{ij}^{(C)}$, $\four h_{ij}^{(\rho)}$ and 
$\four h_{ij}^{(V)}$ in Eqs.$(\ref{hijC})$-$(\ref{hijV})$ can be written as 
\begin{equation}
\four h_{ij}^{(C)}=
2\int(\rho v^j)^{\cdot}{x^i-y^i \over \vert {\bf x}-{\bf y} \vert }d^3y
+2\int(\rho v^i)^{\cdot}{x^j-y^j \over \vert {\bf x}-{\bf y} \vert}d^3y 
-{4 \over 3}\delta_{ij}\int 
(\rho v^k)^{\cdot} {x^k-y^k \over  \vert {\bf x}-{\bf y} \vert} d^3y,
\end{equation}
\begin{eqnarray}
\four h_{ij}^{(\rho)}&=&
{1 \over 4}{\pa^2 \over \pa x^i \pa x^j}\int 
\rho v^k v^l {(x^k-y^k)(x^l-y^l) \over  |{\bf x}-{\bf y}| } d^3y
+{1 \over 3}\delta_{ij}\int(\rho v^k)^{\cdot} 
{x^k-y^k \over  \vert {\bf x}-{\bf y} \vert} d^3y 
\hspace*{1.7cm}\nonumber\\
&&~~~+{1 \over 2}\biggl\{
{\pa \over \pa x^i }\int P' { (x^j-y^j) \over  |{\bf x}-{\bf y}| }d^3y 
+{\pa \over \pa x^j }\int P'{ (x^i-y^i) \over  |{\bf x}-{\bf y}| }d^3y\biggr\}
\nonumber\\
&&~~~-{1 \over 8}\biggl\{
2\int \rho { U_{,j}(x^i-y^i)+U_{,i}(x^j-y^j) \over 
|{\bf x}-{\bf y}| } d^3y \nonumber\\
&&~~~~~~~~+x^k{\pa \over \pa x^i }\int \rho {U_{,k}(x^j-y^j) \over
|{\bf x}-{\bf y}|}  d^3y
+x^k{\pa \over \pa x^j }\int \rho {U_{,k}(x^i-y^i) \over
|{\bf x}-{\bf y}|} d^3y \biggr\}, 
\end{eqnarray}
and
\begin{eqnarray}
\four h_{ij}^{(V)}&=&
{1 \over 2}\Bigl[
{\pa \over \pa x^i}\int\Bigl(\rho v^2+3P-{\rho U \over 2} \Bigr)
{x^j-y^j \over \vert {\bf x}-{\bf y} \vert} d^3y \nonumber\\
&&~~~~
+{\pa \over \pa x^j}\int\Bigl(\rho v^2+3P-{\rho U \over 2} \Bigr)
{x^i-y^i \over \vert {\bf x}-{\bf y} \vert} d^3y \Bigr] \nonumber\\
&&-{2 \over 3}\delta_{ij}\int{\Bigl(\rho v^2+3P-\rho U / 2 \Bigr)
\over \vert {\bf x}-{\bf y} \vert } d^3y, 
\end{eqnarray}
where $P'=P+\rho v^2/4+\rho U_{,l}y^l/4$. 
Furthermore they are rewritten in terms of the potentials 
as follows.
\begin{eqnarray}
\four h_{ij}^{(C)}&=&2(x^i \three \dot P^j+x^j \three \dot P^i-Q_{ij})
+{4 \over 3}\delta_{ij}\Bigl({Q_{kk} \over 2}-x^k \three \dot P^k \Bigr),
\nonumber
\end{eqnarray}
\begin{eqnarray}
\four h_{ij}^{(\rho)}
&=&{1 \over 4}{\pa^2 \over \pa x^i\pa x^j}\Bigl(
V_{kl}^{(\rho v)}x^kx^l-2V_k^{(\rho v)} x^k+V^{(\rho v)} \Bigr)
+{1 \over 3}\delta_{ij} \Bigl(x^k \three \dot P_k-{Q_{kk}\over 2}\Bigr) 
\nonumber\\
&&~~~~~+{1 \over 2}\Bigl\{{\pa \over \pa x^i}\Bigl(V^{(P)}x^j-V^{(P)}_j \Bigr)
+{\pa \over \pa x^j}\Bigl(V^{(P)}x^i-V^{(P)}_i \Bigr) \Bigr\}\nonumber\\
&&~~~~~-{1 \over 8}\Bigl\{ 2\Bigl(x^iV^{(\rho U)}_j+x^iV^{(\rho U)}_i
-V^{(\rho U)}_{ij}-V^{(\rho U)}_{ji}\Bigr) \nonumber\\
&&~~~~~~~~~~+x^k{\pa \over \pa x^i}\Bigl(x^j V^{(\rho U)}_k -V^{(\rho U)}_{kj}
\Bigr)
+x^k{\pa \over \pa x^j}\Bigl(x^i V^{(\rho U)}_k -V^{(\rho U)}_{ki}\Bigr)
\Bigr\}, \nonumber
\end{eqnarray}
\begin{eqnarray}
\four h_{ij}^{(V)}&=&{1 \over 2}\Bigl(Q^{(I)}_{,j}x^i+Q^{(I)}_{,i}x^j-
Q^{(I)}_{i,j}-Q^{(I)}_{j,i}\Bigr)+{1 \over 3}Q^{(I)} \delta_{ij}, 
\end{eqnarray}
where the potentials are defined as 
\begin{eqnarray}
&&\Delta \three P_i=-4\pi \rho v^i,\nonumber\\
&&\Delta Q_{ij}=-4\pi\Bigl( x^j (\rho v^i)^{\cdot}
                                +x^i (\rho v^j)^{\cdot} \Bigr),\nonumber\\
&&\Delta Q^{(I)}=-4\pi \Bigl(\rho v^2+3P-{1 \over 2}\rho U\Bigr), 
\nonumber\\
&&\Delta Q^{(I)}_i
     =-4\pi \Bigl(\rho v^2+3P-{1 \over 2}\rho U \Bigr)x^i,\nonumber\\
&&\Delta V_{ij}^{(\rho v)}=-4\pi \rho v^i v^j,\nonumber\\
&&\Delta V_i^{(\rho v)}=-4\pi \rho v^i v^j x^j,\nonumber\\
&&\Delta V^{(\rho v)}=-4\pi \rho (v^j x^j)^2,\nonumber\\
&&\Delta V^{(P)}=-4\pi P',\nonumber\\
&&\Delta V^{(P)}_i=-4\pi P' x^i ,\nonumber\\
&&\Delta V^{(\rho U)}_i=-4\pi\rho U_{,i},\nonumber\\
&&\Delta V^{(\rho U)}_{ij}=-4\pi\rho U_{,i}x^j. \label{poiseq}
\end{eqnarray}
It should be noted that these Poisson equations have compact sources. 

In this strategy $\four h_{ij}^{(S)}$ and $\four h_{ij}^{(U)}$ are 
solved as the 
boundary value problem, while other parts are obtained by the same
method 
as in Newtonian gravity.

{\bf Strategy 3: Boundary Value approach}

Instead of the above procedure, 
we may solve the Poisson equation for $\four h_{ij}$ as a whole  
carefully imposing the boundary condition for $r \gg R$ as 
\begin{eqnarray}
\four h_{ij}&=&
{1 \over r}
\Bigl\{ {1 \over 4}I_{ij}^{(2)}+{3 \over 4}n^k\Bigl(
n^iI_{kj}^{(2)}+n^jI_{ki}^{(2)}\Bigr) \nonumber\\
&&~~~~~~~-{5 \over 8}n^in^jI_{kk}^{(2)}+{3 \over 8}n^in^jn^kn^lI_{kl}^{(2)} 
+{1 \over 8}\delta_{ij}I_{kk}^{(2)}-{5 \over 8}\delta_{ij}n^kn^lI_{kl}^{(2)} 
\Bigr\} \nonumber\\
&&+{1 \over r^2}
\Bigl[ \Bigl\{
-{5 \over 12}n^kI_{ijk}^{(2)}-{1\over 24}(n^iI_{jkk}^{(2)}+n^jI_{ikk}^{(2)})
+{5 \over 8}n^kn^l(n^iI_{jkl}^{(2)}+n^jI_{ikl}^{(2)}) \nonumber\\
&&~~~~~~~-{7 \over 8}n^in^jn^kI_{kll}^{(2)}+{5 \over 8}n^in^jn^kn^ln^m
I_{klm}^{(2)}+{11 \over 24}\delta_{ij}n^kI_{kll}^{(2)}
-{5 \over 8}\delta_{ij}n^kn^ln^mI_{klm}^{(2)} 
\Bigr\} \nonumber\\
&&~~~~~~~+\Bigl\{
{2 \over 3}n^k(\dot S_{ikj}+\dot S_{jki})-{4 \over 3}(n^i\dot S_{jkk} 
+n^j\dot S_{ikk}) \nonumber\\
&&~~~~~~~+2n^kn^l(n^i\dot S_{jkl}+n^j\dot S_{ikl})+2n^in^jn^k\dot S_{kll} 
+{2 \over 3}\delta_{ij}n^k\dot S_{kll} \Bigr\} \Bigr]+O(r^{-3}). \nonumber\\
\label{bc}
\end{eqnarray}
It is verified that $O(r^{-1})$ and $O(r^{-2})$ parts 
satisfy the traceless and divergence-free conditions respectively. 
It should be noted that $\four h_{ij}$ obtained in this way 
becomes meaningless at the far zone because Eq.$(\ref{fourh})$, from which 
$\four h_{ij}$ is derived, is valid only in the near zone.

\subsection{The Radiation Reaction due to Quadrupole Radiation} 

This topic has been already investigated by using some gauge conditions 
in previous papers \cite{ce,schafer,bds}. 
However, if we use the combination of the conformal slice and the 
transverse gauge, calculations are simplified. 

\subsubsection{conformal slice}

In combination of the conformal slice \cite{sn92} and 
the transverse gauge, Eq.$(\ref{quadr})$ becomes \cite{asf} 
\begin{eqnarray}
\five h_{ij}(\tau)&=&{1 \over 4\pi}{\pa \over \pa \tau}\int \Bigl[ -16\pi \Bigl
(\rho v^i v^j -{1 \over 3}\delta_{ij}\rho v^2 \Bigr) \nonumber\\
&&~~~~~~~~+\Bigl(U U_{,ij}-{1 \over 3}\delta_{ij}U \Delta U
-3U_{,i} U_{,j}+\delta_{ij}U_{,k} U_{,k} \Bigr) \Bigr] d^3y \nonumber\\
&&+{1 \over 4 \pi}{\pa \over \pa \tau}\int\Bigl(\three\dot\beta^i{,j}
+\three\dot\beta^j_{,i}-{2 \over 3}\delta_{ij}\three\dot\beta^k_{,k} \Bigr)
d^3y .
\label{quadrup}
\end{eqnarray}

{}From a straightforward calculation, we find \cite{asf} that the sum of 
the first and second lines becomes $-2\bI_{ij}^{(3)}$ and 
the third line becomes $6\bI_{ij}^{(3)} /5$, where 
$\bI_{ij}^{(3)}=d^3 \bI_{ij}/dt^3$ and 
\begin{equation}
\bI_{ij}=I_{ij}-{1 \over 3}\delta_{ij} I_{kk} . 
\end{equation}
Thus, $\five h_{ij}$ in the near zone becomes 
\begin{equation}
\five h_{ij}=-{4 \over 5}\bI_{ij}^{(3)} . \label{nearh}
\end{equation}
Since $h_{ij}$ has the transverse and traceless property, 
it is likely that $\five h_{ij}$ 
remains the same for other slices.
However it is not clear whether the TT property of $h_{ij}$ is satisfied 
even after the PN expansion is taken in the near zone and, as a result, 
whether $\five h_{ij}$ is independent of slicing conditions or not. 
The fact that slicing conditions never affect $\five h_{ij}$ is understood 
on the ground that $\four\tau_{ij}$ does not depend on slices, 
as already shown in Eq.($\ref{fiveparts}$). 

Then the Hamiltonian constraint at the 2.5PN order, Eq.($\ref{sevenpsi}$), 
turns out to be
\begin{equation}
\Delta \seven\psi=-{2 \over 5} \bI_{ij}^{(3)} U_{,ij}
={1 \over 5} \bI_{ij}^{(3)} \Delta \chi_{,ij} \; ,  \label{psiseven}
\end{equation}
where $\chi$ is the superpotential \cite{chandra65} defined as 
\begin{equation}
\chi=-\int\rho\vert {\bf x}-{\bf y} \vert d^3y, \label{chi}
\end{equation}
which satisfies the relation $\Delta \chi=-2U$. 
{}From this, we find $\seven\psi$ takes the following form,
\begin{eqnarray}
\seven\psi&=&-{1 \over 5} \bI_{ij}^{(3)} \int\rho_{,i} 
{(x^j-y^j)\over \vert {\bf x}-{\bf y} \vert} d^3y \nonumber\\
&=&{1 \over 5} \bI_{ij}^{(3)}\biggl(-x^j U_{,i}
+\int{\rho_{,i}y^j \over \vert {\bf x}-{\bf y} \vert} d^3y\biggr). 
\label{psisev}
\end{eqnarray}
Therefore, the lapse function at the 2.5PN order, 
$\seven\alpha=-2\seven\psi$, is derived from $U$ and $U_r$, where 
$U_r$ satisfies \cite{bds} 
\begin{equation}
\Delta U_r=-4\pi \bI_{ij}^{(3)}\rho_{,i}x^j. \label{alpsev}
\end{equation}
Since the right-hand side of Eq.$(\ref{dtsixK})$ cancels out, $\six K$ 
disappears if the $\six K$ does not exist on the initial hypersurface, 
which seems reasonable under the condition that there are 
no initial gravitational waves. 
Also, $\six\beta^i$ vanishes according to Eq.$(\ref{betsi})$. 
Hence, the quadrupole 
radiation reaction metric has the same form as that derived 
in the case of the maximal slice \cite{schafer,bds}. 

{}From Eq.$(\ref{EOM})$, the PN equation of motion becomes 
\begin{equation}
\dot v^i+v^j v^i_{,~j}=-{P_{,i} \over \rho}+U_{,i}+F_i^{1PN}
+F_i^{2PN}+F_i^{2.5PN}+O(\epsilon^8) ,
\end{equation}
where $F_i^{1PN}$ and $F_i^{2PN}$ are, respectively, the 1PN and 2PN forces. 
Since the radiation reaction potentials, $\five h_{ij}$ and $\seven\alpha$, 
are the same as those by Sch\"afer (1985) and 
Blanchet, Damour and Sch\"afer (1990) in which they use the ADM gauge, 
the radiation reaction force per unit mass, 
$F_i^{2.5PN} \equiv F_i^{r}$, is the same as their force and 
\begin{eqnarray}
F_i^{r}&=&-\Bigl( (\five h_{ij} v^{j})^{\cdot}+v^k v^j_{,k} 
\five h_{ij}+\seven\alpha_{,i} \Bigr) \nonumber\\
&=&\Bigl[ {4\over5}(\bI_{ij}^{(3)} v^j)^{\cdot}+{4\over5} \bI_{ij}^{(3)} 
v^k v^j_{,k}+{2\over5}\bI_{kl}^{(3)} {\pa \over \pa x^i}
\int\rho(t,{\bf y}) { (x^k-y^k)(x^l-y^l) \over \vert 
{\bf x}-{\bf y} \vert^3} d^3y \Bigr]. \nonumber\\
\label{rrforce}
\end{eqnarray}
Since the work done by the force ($\ref{rrforce}$) is given by 
\begin{eqnarray}
W &\equiv&\int \rho F_i^{r} v^i d^3x \nonumber\\
&=&{4\over5}{d \over d\tau}\Bigl(\bI_{ij}^{(3)} \int \rho v^i v^j 
d^3x \Bigr) -{1\over5}\bI_{ij}^{(3)}\bI^{(3) ij},  
\label{dEquadr}
\end{eqnarray}
we obtain the so-called quadrupole formula of the energy loss 
by averaging Eq.$(\ref{dEquadr})$ 
with respect to time as 
\begin{equation}
\left\langle {dE_N \over d\tau} \right\rangle=-{1\over5}
\left\langle \bI_{ij}^{(3)} \bI^{(3) ij} \right\rangle+O(\epsilon^6). 
\end{equation}

\subsubsection{Radiation reaction in other slice conditions}

In this subsection, we do not specify the slice condition. 
In this case, the reaction force takes the following form 
\begin{equation}
F_i^{r}={4\over5}(\bI_{ij}^{(3)} v^j)^{\cdot}
+{4\over5} \bI_{ij}^{(3)} v^k v^j_{,k}-\seven\alpha_{,i}-\six\dot\beta^i
+v^j\six\beta^j\,_{,i}-v^j\six\beta^i\,_{,j} . 
\label{rrforce2}
\end{equation}
Here, $\seven\alpha$ corresponds to the slice condition. 
{}From Eq.$(\ref{rrforce2})$, we obtain the work done by the reaction force 
as 
\begin{eqnarray}
W &\equiv&\int \rho F_i^{r} v^i d^3x \nonumber\\
&=&\int d^3x \rho v^i \Bigl[ {4\over5}\Bigl(\bI_{ij}^{(3)}v^j\Bigr)^{\cdot} 
+{4\over5}\bI_{ij}^{(3)}v^j\,_{,l}v^l \nonumber\\
&&~~~~~~~~~~~-\seven\alpha_{,i}-\six\dot\beta^i
+v^j\six\beta^j\,_{,i}-v^j\six\beta^i\,_{,j} \Bigr] . \label{dEquadr1}
\end{eqnarray}

Explicit calculations are done separately: 
For the first and second terms of Eq.($\ref{dEquadr1}$), we obtain  
\begin{eqnarray}
{4\over5}\int d^3x \rho\Bigl(\bI_{ij}^{(3)}v^j\Bigr)^{\cdot}v^i&=&
{4\over5}{d\over dt}\Bigl[\bI_{ij}^{(3)}\int d^3x \rho v^i v^j \Bigr] 
-{1\over5}\bI_{ij}^{(3)}\bI_{ij}^{(3)} \nonumber\\ 
&&-{2\over5}\bI_{ij}^{(3)}\int d^3x \rho(x) v^k{\pa \over \pa x^k} 
\int d^3y {\rho(y) (x^i-y^i)(x^j-y^j)\over|{\bf x}-{\bf y}|^3} \nonumber\\
&&-{2\over5}\bI_{ij}^{(3)}\int d^3x \dot\rho v^i v^j , 
\label{calc1}
\end{eqnarray}
and 
\begin{equation}
{4\over5}\int d^3x \rho\bI_{ij}^{(3)}v^j_{,l}v^lv^i 
={2\over5}\bI_{ij}^{(3)}\int d^3x \dot\rho v^iv^j . 
\label{calc2}
\end{equation}
As for the fourth term in the integral of Eq.($\ref{dEquadr1}$), 
we find the relation 
\begin{equation}
\six\dot\beta^i=-\seven\alpha_{,i}-{2\over 5}\bI_{kl}^{(3)}{\pa\over\pa x^i} 
\int d^3y \rho{(x^k-y^k)(x^l-y^l)\over|{\bf x}-{\bf y}|^3} , 
\label{calc3}
\end{equation}
which is given by Eqs.$(\ref{betsi})$, $(\ref{dtsixK})$ 
and $(\ref{nearh})$. 

Using Eqs.($\ref{dEquadr1}$), ($\ref{calc1}$), ($\ref{calc2}$) and 
($\ref{calc3}$), we obtain  
\begin{eqnarray}
W&=&{4\over5}{d \over d\tau}\Bigl(\bI_{ij}^{(3)} \int \rho v^i v^j 
d^3x \Bigr) -{1\over5}\bI_{ij}^{(3)}\bI^{(3) ij} . 
\label{dEquadr2}
\end{eqnarray}
This expression for $W$ does not depend on the slice condition. 
However, this never means that the value of $W$ is invariant 
under the change of the slice condition, since the meaning of 
the time derivative depends on the slice condition.

\subsection{Conserved quantities}

The conserved quantities are gauge-invariant so that, 
in general relativity, they play important roles to characterize 
various systems described in different gauge conditions. 
{}From the practical point of view, these are also useful for checking 
the numerical accuracy in simulations. 
Thus, we show several conserved quantities in the 2PN approximation. 

{\bf Conserved Mass And Energy}

In general relativity, we have the following conserved mass; 
\begin{equation}
M_{\ast}=\int \rho_{\ast}d^3x. 
\end{equation}
In the PN approximation, $\rho_{\ast}$ defined by
Eq.($\ref{consdensity}$) is expanded as \cite{asf} 
\begin{eqnarray}
\rho_{\ast}=\rho \Bigl[&&1+\Bigl({1 \over 2}v^2+3 U\Bigr) \nonumber\\
&&+\Bigl({3 \over 8}v^4+{7 \over 2}v^2 U
+{15 \over 4}U^2+6 \four\psi+\three\beta^i v^i\Bigr)
+\six\delta_{\ast})\Bigr]+O(\epsilon^7), 
\hspace*{2cm}
\end{eqnarray}
where $\six\delta_{\ast}$ denotes the 3PN contribution to $\rho_{\ast}$. 
This term $\six\delta_{\ast}$ will be calculated later. 

Next, we consider the ADM mass which is conserved. 
Since the asymptotic behavior of the conformal factor becomes 
\begin{equation}
\psi=1+{M_{ADM} \over 2 r}+O\Bigl({1 \over r^2}\Bigr), 
\end{equation}
the ADM mass in the PN approximation becomes 
\begin{eqnarray}
M_{ADM}&=&-{1 \over 2\pi} \int \Delta \psi d^3x \nonumber\\
&=&\int d^3x \rho \Bigl[\Bigl\{1+\Bigl(v^2+\Pi+{5 \over 2}U
\Bigr) 
+\Bigl(v^4+{13 \over 2}v^2 U+v^2\Pi+{P \over \rho}v^2
+{5 \over 2}U\Pi \nonumber\\
&&~~~~~+{5 \over 2}U^2+5\four\psi+2\three\beta^i v^i\Bigr)\Bigr\} 
+{1 \over 16\pi\rho}\Bigl(\three\tilde A_{ij} \three\tilde A_{ij}
-{2 \over 3}\three K^2 \Bigr) \nonumber\\
&&~~~~~+\six \delta_{ADM}+O(\epsilon^7) \Bigr], 
\end{eqnarray}
where $\six \delta_{ADM}$ denotes the 3PN contribution. 
This term $\six \delta_{ADM}$ will be calculated later. 

Using these two conserved quantities, we can define 
the conserved energy as follows; 
\begin{eqnarray}
E&\equiv&M_{ADM}-M_{\ast} \nonumber\\
&=&\int d^3x \rho \Bigl[\Bigl\{\Bigl({1 \over 2}v^2+\Pi
-{1 \over 2}U\Bigr) \nonumber\\
&&~~~~~~+\Bigl({5 \over 8}v^4+3v^2 U+v^2\Pi+{P \over \rho}v^2
+{5 \over 2}U\Pi-{5 \over 4}U^2-\four\psi+\three\beta^i v^i
\Bigr)\Bigr\} \nonumber\\
&&~~~~~~+{1 \over 16\pi\rho}\Bigl(\three\tilde A_{ij} 
\three\tilde A_{ij}-{2 \over 3}\three K^2\Bigr)
+\Bigl(\six\delta_{ADM}-\six\delta_{\ast}\Bigr)+O(c^{-7})\Bigr] \nonumber\\
&\equiv & E_{N}+E_{1PN}+E_{2PN}+\cdots . 
\end{eqnarray} 
We should note that the following equation holds
\begin{equation}
\int\three\tilde A_{ij} \three\tilde A_{ij} d^3x=
-8\pi\int\rho v^i \three\beta^i d^3x+\int\Bigl({2 \over 3}\three K^2
+2 \dot U \three K\Bigr)d^3x ,
\end{equation}
where we used Eqs.$(\ref{dotu})$ and $(\ref{betth})$. 
Then, we obtain the Newtonian and the first PN energies as 
\begin{equation}
E_{N}=\int\rho\Bigl({1 \over 2}v^2+\Pi-{1 \over 2}U\Bigr)d^3x ,
\end{equation}
and
\begin{eqnarray}
E_{1PN}&=&\int d^3x\Bigl[\rho\Bigl({5 \over 8}v^4+{5 \over 2}v^2 U
+v^2\Pi+{P \over \rho}v^2+2 U\Pi-{5 \over 2}U^2
+{1 \over 2}\three\beta^i v^i \Bigr) \nonumber\\ 
&&~~~~~~~~~+{1 \over 8 \pi}\dot U\three K \Bigr]. 
\end{eqnarray}
$E_{1PN}$ can be rewritten immediately in the following form 
used by Chandrasekhar \cite{chandra69a}; 
\begin{equation}
E_{1PN}=\int d^3x\rho \Bigl[{5 \over 8}v^4+{5 \over 2}v^2 U+v^2 
\Bigl(\Pi+{P \over \rho}\Bigr)+2 U\Pi-{5 \over 2}U^2
-{1 \over 2}v^i q_i \Bigr] , 
\end{equation}
where $q_i$ is the first PN shift vector in the standard PN gauge 
\cite{chandra69a}, which turns out to be $\three K=0$ 
in the (3+1) formalism and it satisfies 
\begin{equation}
\Delta q_i=-16\pi\rho v^i+\dot U_{,i} .
\end{equation}

The total energy at the 2PN order $E_{2PN}$ is calculated 
from the 3PN quantities $\six\delta_{\ast}$ and $\six\delta_{ADM}$. 
In a straightforward manner, we obtain 
\begin{equation}
\six M_{\ast}=\int \rho \six\delta_{\ast} d^3x , 
\end{equation}
where 
\begin{eqnarray}
\six\delta_{\ast}&=&
{5 \over 16}v^6+{33 \over 8}v^4 U+v^2 \Bigl( 5\four\psi
+{93 \over 8}U^2+{3 \over 2}\three\beta^i v^i-X \Bigr) 
+6\six\psi+15U\four\psi \nonumber\\
&&+{5 \over 2}U^3+7\three\beta^i v^i U
+{1 \over 2}\four h_{ij}v^iv^j
+{1 \over 2}\three\beta^i\three\beta^i+\five\beta^iv^i . 
\end{eqnarray}
Next, we consider $\six\delta_{ADM}$. 
The Hamiltonian constraint at $O(\epsilon^8)$ becomes 
\begin{eqnarray}
&&\Delta\eight\psi-\four h_{ij}\four\psi_{,ij}
-{1 \over 2}\six h_{ij}U_{,ij} \nonumber\\
&&=-{1 \over 32}(2\four h_{kl,m}\four h_{km,l}
+\four h_{kl,m}\four h_{kl,m}) 
-2\pi\six\rho_{\psi} 
\nonumber\\
&&-{1 \over 4} \Bigl( \three\tilde A_{ij}\five\tilde A_{ij}
-{2 \over 3}\three K\five K \Bigr)
-{1 \over 16}U \Bigl( \three\tilde A_{ij}\three\tilde A_{ij}
-{2 \over 3}\three K^2 \Bigr), 
\end{eqnarray}
where we define $\six\rho_{\psi}$ as 
\begin{eqnarray}
\six\rho_{\psi}&=&\rho \Bigl[ v^6+v^4 \Bigl( \Pi+{P \over \rho}
+{21 \over 2}U \Bigr) +v^2 \Bigl\{ {13 \over 2}U \Bigl( \Pi+
{P \over \rho} \Bigr) +9\four\psi-2X+20U^2 \Bigr\} \nonumber\\
&&~~~~~+\Pi \Bigl( 5\four\psi+{5 \over 2}U^2 \Bigr)+
5\six\psi+10U\four\psi+{5 \over 4}U^3 
+\four h_{ij}v^iv^j \nonumber\\
&&~~~~~+2\three\beta^iv^i \Bigl\{ 2v^2+\Pi
+{P \over \rho}+{13 \over 2}U \Bigr\} +2\five\beta^iv^i
+\three\beta^i\three\beta^i \Bigr]. 
\end{eqnarray}
Making use of the relations $\four h_{ij,j}=0$ and 
$\six h_{ij,j}=0$, we obtain 
\begin{eqnarray}
\six M_{ADM}&=&\int d^3x \six\rho_{\psi} 
+{1 \over 8\pi}\int d^3y 
\Bigl( \three\tilde A_{ij}\five\tilde A_{ij}
-{2 \over 3}\three K\five K \Bigr) \nonumber\\
&&+{1 \over 32\pi}\int d^3y 
U \Bigl( \three\tilde A_{ij}\three\tilde A_{ij}
-{2 \over 3}\three K^2 \Bigr), 
\end{eqnarray}
where we assumed $\six h_{ij} \to O(r^{-1})$ as $r \to \infty$. 
Although this assumption must be verified by performing the 3PN expansions 
which have not been done here, it seems reasonable in the asymptotically 
flat spacetime. From $\six M_{ADM}$ and $\six M_{\ast}$, 
we obtain the conserved energy at the 2PN order
\begin{eqnarray}
E_{2PN}&=&\six M_{ADM}-\six M_{\ast} \nonumber\\
&=&\int d^3x \rho \Bigl[ {11 \over 16}v^6+v^4 \Bigl( \Pi
+{P \over \rho}+{47 \over 8}U \Bigr) \nonumber\\ 
&&~~~~~~~~~~~~+v^2 \Bigl\{ 4\four\psi-X
+6U \Bigl( \Pi+{P \over \rho} \Bigr) +{41 \over 8}U^2
+{5 \over 2}\three\beta^iv^i \Bigr\} \nonumber\\
&&~~~~~~~~~~~~+\Pi \Bigl( 5\four\psi+{5 \over 4}U^2 \Bigr)
-{15 \over 2}U\four\psi-{5 \over 2}U^3  \nonumber\\
&&~~~~~~~~~~~~+{1 \over 2}\four h_{ij}v^iv^j+2\three\beta^iv^i \Bigl\{ 
\Bigl( \Pi+{P \over \rho} \Bigr) +5U \Bigr\} +\five\beta^iv^i
+{ 1\over 2}\three\beta^i\three\beta^i \Bigr] \nonumber\\
&&+{1 \over 8\pi}\int d^3y 
\Bigl( \four h_{ij}UU_{,ij}+\three\tilde A_{ij}\five\tilde A_{ij}
-{2 \over 3}\three K\five K \Bigr) . 
\end{eqnarray}
Here we used Eq.($\ref{sixpsi}$) and the relation, 
$\int d^3x \rho\six\psi=-{1 \over 4\pi}
\int d^3x U\Delta\six\psi$, in order to eliminate $\six\psi$.

{\bf Conserved linear momentum}

When we use the center of mass system as usual, 
the linear momentum of the system should vanish.
However, it may arise from numerical errors 
in numerical calculations.
Since it is useful to check the numerical accuracy, 
we mention the linear momentum derived from 
\begin{eqnarray}
P_i&=&{1 \over 8\pi}\lim_{r \to \infty}\oint
\Bigl( K_{ij}n^j-K n^i \Bigr) dS \nonumber\\
&=&{1 \over 8\pi}\lim_{r \to \infty}\oint
\Bigl( \psi^4 \tilde A_{ij} n^j-{2 \over 3}K n^i \Bigr) dS , 
\end{eqnarray}
where the surface integrals are taken over a sphere of constant $r$.
Since the asymptotic behavior of $\tilde A_{ij}$ is determined by
\begin{equation}
\three\tilde A_{ij}={1 \over 2}\bigl(\three\beta^i_{,j}+\three\beta^j_{,i}
-{2 \over 3}\delta_{ij}\three\beta^l_{,l}\Bigr)+O(r^{-3}) ,
\end{equation}
and
\begin{equation}
\five\tilde A_{ij}={1 \over 2}\Bigl(\five\beta^i_{,j}+\five\beta^j_{,i}
-{2 \over 3}\delta_{ij}\five\beta^l_{,l}\Bigr)+\five\tilde A_{ij}^{TT}
+O(r^{-3}) ,
\end{equation}
the leading term of the shift vector is necessary.
Using the asymptotic behavior derived from Eq.($\ref{betth}$) 
\begin{equation}
\three\beta^i=-{7 \over 2}{l_i \over r}
-{1 \over 2}{n^i n^j l_j \over r}+O(r^{-2}) ,
\end{equation}
we obtain 
\begin{equation}
\int\Bigl(\three\beta^i_{,j}+\three\beta^j_{,i}
-{2 \over 3}\delta_{ij}\three\beta^l_{,l}\Bigr) n^j dS
=16\pi l_i ,  
\end{equation}
where we defined $l_i=\int \rho v^i d^3x$. 
Therefore the Newtonian linear momentum is 
\begin{equation}
P_N\,^i=\int d^3x \rho v^i .
\end{equation}
Similarly the first PN linear momentum is obtained as 
\begin{equation}
P_{1PN}\,^i=\int d^3x \rho \Bigl[v^i\Bigl(v^2+\Pi+6U+{P \over \rho}
\Bigr)+\three\beta^i\Bigr].
\end{equation}
We obtain $P_{2PN}\,^i$ by the similar procedure as 
\begin{eqnarray}
P_{2PN}\,^i&=&\int d^3x \rho v^i\Bigl[ 2\three\beta^i v^i+10\four\psi
+\Bigl (6U+v^2 \Bigr) \Bigl( \Pi+{P \over \rho} \Bigr) \nonumber\\
&&~~~~~~~~~~~~~~+{67 \over 4}U^2+10Uv^2+v^4-X \Bigr] . 
\end{eqnarray}

\section {Formulation for Nonaxisymmetric Uniformly Rotating
Equilibrium Configurations}

We now consider the construction of the spacetime involving a 
close binary neutron stars as an application of PN approximation.
It is assumed that the binary stars are regarded as uniformly rotating 
equilibrium configurations. 
Here, we mention the importance of this investigation. 
To interpret the implication of the signal of gravitational waves, 
we need to understand the theoretical mechanism of merging in detail. 
When the orbital separation of binary neutron stars is $\lsim 10 GM/c^2$, 
where $M$ is the total mass of binary neutron stars, 
they move approximately in circular orbits because 
the timescale of the energy loss due to gravitational radiation 
is much longer than the orbital period. 
However, when the orbital separation becomes $6-10GM/c^2$, 
the circular orbit cannot be maintained  because of instabilities 
due to the GR gravity\cite{kww93a} or 
the tidal field \cite{lrs93,lrs94}. 
As a result of such instabilities, the circular orbit of 
binary neutron stars changes into the plunging orbit to merge. 
Gravitational waves emitted at this transition may 
present us important information about the structure of NS's 
since the location where the instability occurs will 
depend sensitively on the equation of state (EOS) of NS 
\cite{lrs93,lrs94,zcm}. 
Thus, it is very important to investigate the location of 
the innermost stable circular orbit (ISCO) of binary neutron stars. 

In order to search the ISCO, we can take the following procedure: 
First, neglecting the evolution due to gravitational radiation, 
we construct equilibrium configurations. 
Next, we take into account the radiation reaction as a correction 
to the equilibrium configurations. 
The ISCO is the orbit, where the dynamical instability 
for the equilibrium configurations occurs. 
Hence  we shall present a formalism to obtain equilibrium configurations 
of uniformly rotating fluid in 2PN order as a first step \cite{as}, 
though in reality due to the conservation of the circulation for 
gravitational radiation reaction, tidally locked configuration 
such as uniformly rotating fluid is not a good approximation. 

\subsection{Formulation} 

We shall use in this section the maximal slice condition $K_i^{~i}=0$. 
As the spatial gauge condition, we adopt 
the transverse gauge $\tilde \gamma_{ij,j}=0$ in order to remove the 
gauge modes from $\tilde \gamma_{ij}$. 
In this case, up to the 2PN approximation, each metric variable 
is expanded as \cite{asf} 
\begin{eqnarray}
\psi&=&1+{1 \over c^2}{U \over 2}+{1 \over c^4}\four\psi+O(c^{-6}),\\
\alpha&=&1-{1 \over c^2}U+{1 \over c^4}\Bigl({U^2 \over 2}+X\Bigr)
+{1 \over c^6}\six\alpha+O(c^{-7}), \\
\beta^i&=&{1 \over c^3}\three\beta^i+{1 \over c^5}\five\beta^i+
O(c^{-7}),\\
\tilde \gamma_{ij}&=&\delta_{ij}+{1 \over c^4}h_{ij}+O(c^{-5}).
\end{eqnarray}
In this section, we use $1/c$ instead of $\epsilon$ 
as the expansion parameter because of its convenience 
in numerical applications. 

For simplicity, we assume that the 
matter obeys the polytropic equation of state (EOS);
\begin{equation}
P=(\Gamma-1)\rho \Pi=K \rho^{\Gamma}, 
\end{equation}
where $\Gamma$ and $K$ are the polytropic exponent and polytropic constant, 
respectively. Up to the 2PN order, 
the four velocity is given by Eq.($\ref{fourvel}$) \cite{cn,asf,as}. 
Since we need $u^0$ up to 3PN order to obtain the 2PN equations of motion, 
we derive it here. 
Using Eq.($\ref{fourvel}$), we can calculate $(\alpha u^0)^2$ 
up to 3PN order as 
\begin{eqnarray}
(\alpha u^0)^2&=&1 + \psi^{-4}\tilde \gamma^{ij}u_i u_j \nonumber\\
&=&1+{v^2 \over c^2}+{1 \over c^4}\Bigl(2\three\beta^j v^j+4Uv^2+v^4\Bigr)
+{1 \over c^6}\Bigl\{\three\beta^j\three\beta^j+8\three\beta^j v^j U 
\nonumber\\ 
&&+h_{ij}v^i v^j +2\five\beta^iv^i
+\Bigl(4\three\beta^j v^j+4\four\psi+{15 \over 2}U^2-2X\Bigr)v^2
+8Uv^4+v^6\Bigr\} 
\nonumber\\
&&+O(c^{-7}) , 
\end{eqnarray}
where we used $\tilde \gamma^{ij}=\delta_{ij}-c^{-4}h_{ij}+O(c^{-5})$. 
Thus, we obtain $u^0$ up to the 3PN order as 
\begin{eqnarray}
u^0&=&1+{1 \over c^2}\Bigl({1 \over 2}v^2+U\Bigr)
+{1 \over c^4}\Bigl({3 \over 8}v^4+{5 \over 2}v^2U
+{1 \over 2}U^2+\three\beta^i v^i-X \Bigr) \nonumber\\
&&+{1 \over c^6}\Bigl\{-\six\alpha
+{1 \over 2}\Bigl( \three\beta^j\three\beta^j+h_{ij}v^iv^j\Bigr)
+\five \beta^j v^j + 5\three\beta^j v^j U -2UX \nonumber\\
&&~~~~~+\Bigl({3 \over 2}
\three\beta^j v^j+2\four\psi+6U^2-{3 \over 2} X\Bigr)v^2 
+{27 \over 8}Uv^4+{5 \over 16}v^6\Bigr\} 
\nonumber\\
&&+O(c^{-7}). \label{utime}
\end{eqnarray}

Substituting PN expansions of metric and matter variables into the 
Einstein equation, and using the polytropic EOS, 
we find that the metric variables obey the following
Poisson equations \cite{asf}; 
\begin{eqnarray}
&&\Delta U=-4\pi \rho, 
\label{pot1}\\
&&\Delta X=4\pi\rho\Bigl(2 v^2+2U+(3\Gamma-2)\Pi
\Bigr), \\
&&\Delta \four\psi=-2\pi\rho\Bigl(v^2+\Pi+{5 \over 2}U \Bigr), \\
&&\Delta \three\beta^i=16\pi \rho v^i-\dot U_{,i}, \label{rot3beta}\\
&&\Delta \five\beta^i =16\pi\rho\biggl[v^i \Bigl(v^2+2U+\Gamma\Pi
\Bigr)+\three\beta^i\biggr]-4U_{,j}
\Bigl(\three\beta^i_{,j}+\three\beta^j_{,i}
-{2 \over 3}\delta_{ij}\three\beta^k_{,k}\Bigr) \nonumber\\
&&{\hskip 2cm} -2\four\dot\psi_{,i}
      +{1 \over 2}(U \dot U)_{,i}+(\three\beta^l U_{,l})_{,i}, 
\label{rot5beta}\\
&&\Delta  h_{ij}=\Bigl(U U_{,ij}-{1 \over 3}\delta_{ij}U \Delta U
-3U_{,i} U_{,j}+\delta_{ij}U_{,k} U_{,k} \Bigr)
-16\pi \Bigl(\rho v^i v^j -{1 \over 3}\delta_{ij}\rho v^2 \Bigr) \nonumber\\
&&\hskip 1cm  -\Bigl(\three\dot\beta^i_{,j}+\three\dot\beta^j_{,i}
-{2 \over 3}\delta_{ij}\three\dot\beta^k_{,k} \Bigr) 
-2\Bigl( (X+2\four\psi)_{,ij}-{1 \over 3}\delta_{ij}
\Delta (X+2\four\psi) \Bigr), 
\nonumber\\
&&\\
&&\Delta \six\alpha=4\pi\rho\biggl[2v^4
+2v^2\Bigl(5U+\Gamma\Pi\Bigr)+(3\Gamma-2)\Pi U
+4\four\psi+X+4\three\beta^i v^i \biggr] \nonumber\\
&&{\hskip 3cm}-h_{ij}U_{,ij}
-{3 \over 2}U U_{,l}U_{,l}+U_{,l}(2 \four\psi -X)_{,l} \nonumber\\
&&{\hskip 3cm} +{1 \over 2}
\three\beta^i_{,j}\Bigl(\three\beta^i_{,j}+\three\beta^j_{,i}
-{2 \over 3}\delta_{ij}\three\beta^k_{,k}\Bigr) . 
\label{pot2}
\end{eqnarray}

Here, we consider {\it the uniformly rotating fluid around $z$-axis }
with the angular velocity $\Omega$, i.e., 
\begin{equation}
v^i=\epsilon_{ijk}\Omega^j x^k=(-y\Omega, x\Omega, 0), 
\end{equation}
where we choose $\Omega^j=(0,0,\Omega)$ and $\epsilon_{ijk}$ is 
the completely anti-symmetric unit tensor. 
In this case, the following relations hold;
\begin{equation}
\Bigl({\partial \over \partial t}+\Omega{\partial \over \partial \varphi}
\Bigr) Q
=\Bigl({\partial \over \partial t}+\Omega{\partial \over \partial \varphi}
\Bigr) Q_i
=\Bigl({\partial \over \partial t}+\Omega{\partial \over \partial \varphi} 
\Bigr)Q_{ij}=0,\label{killeq}
\end{equation} 
where $Q$, $Q_i$ and $Q_{ij}$ are arbitrary scalars, vectors, and tensors, 
respectively. 
Then, the conservation law ($\ref{emconsv}$) can be integrated as \cite{lppt} 
\begin{equation}
\int {dP \over \rho c^2+\rho\Pi+P}=\ln u^0+C,\label{euler}
\end{equation}
where $C$ is a constant. For the polytropic EOS, Eq.$(\ref{euler})$ becomes 
\begin{equation}
\ln \biggl[1+{\Gamma K \over c^2(\Gamma-1)} \rho^{\Gamma-1} \biggr]
= \ln u^0+C . \label{eulerf}
\end{equation}
Using Eq.$(\ref{utime})$, the 2PN approximation of Eq.$(\ref{eulerf})$ 
is written as
\begin{eqnarray}
H-{H^2 \over 2c^2}+{H^3 \over 3c^4}&=&{v^2 \over 2}+U
+{1 \over c^2}\biggl(2Uv^2+{v^4 \over 4}-X+\three\beta^i v^i \biggl) 
\nonumber\\
&&+{1 \over c^4}\biggl(-\six\alpha +{1 \over 2}\three\beta^i\three\beta^i
+4\three\beta^i v^i U-{U^3 \over 6}+\three\beta^i v^i v^2+2 \four \psi v^2 
\nonumber\\
&&
+{15\over 4} U^2 v^2+2Uv^4 +{1 \over 6} v^6-UX-v^2 X+\five\beta^i v^i
+{1 \over 2}h_{ij}v^iv^j\biggr) 
\nonumber\\
&&+C, 
\label{bern}
\end{eqnarray}
where $H=\Gamma K \rho^{\Gamma-1}/(\Gamma-1)$, $v^2=R^2\Omega^2$ and 
$R^2=x^2+y^2$. 
Note that Eq.$(\ref{bern})$ can be also obtained from the 2PN Euler 
equation like in the first PN case \cite{chandra67}. 
If we solve the coupled equations ($\ref{pot1}$)-($\ref{pot2}$) 
and $(\ref{bern})$, we can obtain equilibrium configurations 
of the non-axisymmetric uniformly rotating body. 

In the above Poisson equations for metric variables, 
the source terms in the Poisson equations for $\three \beta^i$, 
$\five \beta^i$, and $h_{ij}$ fall off slowly as $r \rightarrow \infty$ 
because these terms behave as $O(r^{-3})$ at $r \rightarrow \infty$. 
These Poisson equations do not take convenient forms when we try to solve 
them numerically as the boundary value problem.  
Hence in the following, we rewrite them into convenient forms. 

As for $h_{ij}$, first of all, we split the equation into three parts 
as \cite{asf} 
\begin{eqnarray}
\Delta h^{(U)}_{ij}&=&U\Bigl(U_{,ij}-{1 \over 3}\delta_{ij}\Delta U\Bigr)
-3U_{,i}U_{,j}+\delta_{ij}U_{,k}U_{,k} \equiv -4\pi S^{(U)}_{ij},   \\
\Delta h^{(S)}_{ij}&=&
-16\pi\Bigl(\rho v^iv^j-{1 \over 3}\delta_{ij}\rho v^2\Bigr),   \\
\Delta h^{(G)}_{ij}&=&-\Bigl(\three\dot\beta^i_{,j}+\three\dot\beta^j_{,i}
-{2 \over 3}\delta_{ij}\three\dot\beta^k_{,k} \Bigr) \nonumber\\
&&{\hskip 1.6cm}-2\Bigl( (X+2\four\psi)_{,ij}-{1 \over 3}\delta_{ij}
\Delta (X+2\four\psi) \Bigr). \label{hijG} 
\end{eqnarray}
The equation for $h^{(S)}_{ij}$ has a compact source, and also 
the source term of $h^{(U)}_{ij}$ behaves as $O(r^{-6})$ at 
$r \rightarrow \infty$, so that Poisson equations for them are 
solved easily as the boundary value problem. 
On the other hand, the source term of $h^{(G)}_{ij}$ behaves 
as $O(r^{-3})$ at $r \rightarrow \infty$, 
so that it seems troublesome to solve the equation  
as the boundary value problem. In order to solve the equation for 
$h^{(G)}_{ij}$ as the boundary value problem, we had better 
rewrite the equation into useful forms. 
As shown by Asada, Shibata and Futamase \cite{asf}, 
Eq.$(\ref{hijG})$ is integrated to give 
\begin{eqnarray}
h_{ij}^{(G)}&=&
2{\pa \over \pa x^i}\int(\rho v^j)^{\cdot}\vert {\bf x}-{\bf y} \vert d^3y
+2{\pa \over \pa x^j}\int(\rho v^i)^{\cdot}\vert {\bf x}-{\bf y} \vert d^3y
+\delta_{ij}\int\ddot\rho \vert {\bf x}-{\bf y} \vert d^3y \nonumber\\
&&+{1 \over 12}{\pa^2 \over \pa x^i \pa x^j}\int\ddot\rho 
\vert {\bf x}-{\bf y} \vert^3 d^3y
+{\pa^2 \over \pa x^i \pa x^j}\int\Bigl(\rho v^2+3P-{\rho U \over 2} \Bigr)
\vert {\bf x}-{\bf y} \vert d^3y \nonumber\\
&&-{2 \over 3}\delta_{ij}\int{\Bigl(\rho v^2+3P-\rho U / 2 \Bigr)
\over \vert {\bf x}-{\bf y} \vert} d^3y. \label{haeq}
\end{eqnarray}
Using the relations
\begin{eqnarray}
&&\ddot \rho=-(\rho v^j)_{,j}^{\cdot}+O(c^{-2}),\nonumber\\
&&\dot v^i=0,\nonumber\\
&&v^i x^i=0, 
\end{eqnarray}
Eq.$(\ref{haeq})$ is rewritten as
\begin{eqnarray}
 h_{ij}^{(G)}&=&{7 \over 4}
\Bigl(x^i \three \dot P_j+x^j \three \dot P_i-\dot Q^{(T)}_{ij}
-\dot Q^{(T)}_{ji}\Bigr)
-\delta_{ij}x^k \three \dot P_k 
\hspace*{3cm} \nonumber\\
&&-{1 \over 8} x^k 
\biggl[{\partial \over \partial x^i}\Bigl(x^j \three \dot P_k
-\dot Q^{(T)}_{kj} \Bigr)
+{\partial \over \partial x^j}\Bigl(x^i \three \dot P_k
-\dot Q^{(T)}_{ki} \Bigr)\biggr] \nonumber\\
&&+{1 \over 2}
\biggl[{\partial \over \partial x^i}\Bigl(x^j Q^{(I)}- Q^{(I)}_j \Bigr)
+{\partial \over \partial x^j}\Bigl(x^i Q^{(I)}- Q^{(I)}_i \Bigr) \biggr]
-{2 \over 3}\delta_{ij} Q^{(I)}, 
\end{eqnarray}
where 
\begin{eqnarray}
&&\Delta \three P_i=-4\pi \rho v^i, \\
&&\Delta Q^{(T)}_{ij}=-4\pi \rho v^i x^j, \\
&&\Delta Q^{(I)}=-4\pi \Bigl(\rho v^2+3P-{1 \over 2}\rho U\Bigr), \\
&&\Delta Q^{(I)}_i
     =-4\pi \Bigl(\rho v^2+3P-{1 \over 2}\rho U \Bigr)x^i. 
\end{eqnarray}
Therefore, $h_{ij}^{(G)}$ can be deduced from variables which satisfy 
the Poisson equations with compact sources. 

The source terms in the Poisson equations ($\ref{rot3beta}$) and 
($\ref{rot5beta}$) for $\three\beta^i$ and 
$\five\beta^i$ also fall off slowly. 
However, we can rewrite them as \cite{asf} 
\begin{eqnarray}
\three\beta^i&=&-4\three P_i-{1 \over 2}\Bigl(x^i \dot U-\dot q_i\Bigr), \\
\five \beta^i&=&-4\five  P_i-{1 \over 2}\Bigl(2x^i \four \dot \psi
-\dot \eta_i \Bigr), 
\end{eqnarray}
where
\begin{eqnarray}
\Delta q_i&=&-4\pi\rho x^i , 
\label{q}\\
\Delta \five P_i&=&-4\pi \rho\biggl[v^i\Bigl(v^2+2U+\Gamma\Pi \Bigr)
+\three \beta^i\biggr]+U_{,j}\Bigl(\three\beta^i_{,j}+\three\beta^j_{,i}
-{2 \over 3}\delta_{ij}\three\beta^k_{,k}\Bigr) \nonumber\\
&&-{1 \over 8}(\dot U U)_{,i}-{1 \over 4}(\three \beta^l U_{,l})_{,i}, 
\label{P}\\
\Delta \eta_i&=&-4\pi\rho\Bigl(v^2+\Pi+{5 \over 2}U\Bigr)x^i . 
\label{eta}
\end{eqnarray}
Thus $\three\beta^i$ and $\five\beta^i$ can be obtained by solving 
the Poisson equations ($\ref{q}$)-($\ref{eta}$) whose source terms 
fall off fast enough, $O(r^{-5})$, for numerical calculation. 

For later convenience, using the relation 
$\three P_i=\epsilon_{izk}q_k \Omega$ and Eq.$(\ref{killeq})$, 
$\three \beta^i$ and $\five\beta^i$ may be written 
in the form with explicit $\Omega$ dependence as 
\begin{eqnarray}
&&\three \beta^i =
\Omega\Bigl[-4\epsilon_{izk} q_k+{1 \over 2}\Bigl(x^i U_{,\varphi}
-q_{i,\varphi}\Bigr)\Bigr]\equiv \Omega\three \hat \beta^i, \\
&&\five \beta^i =
\Omega\Bigl[-4\five \hat P_i+{1 \over 2}\Bigl(2x^i \four\psi_{,\varphi}
-\eta_{i,\varphi}\Bigr)\Bigr], 
\end{eqnarray}
where
\begin{eqnarray}
\Delta \five \hat P_i &=&-4\pi\rho\biggl[ \epsilon_{izk}x^k\Bigl(v^2+2U
+\Gamma \Pi\Bigr)+\three \hat \beta^i\biggr]
+U_{,j}\Bigl(\three\hat \beta^i_{,j}+\three\hat \beta^j_{,i}
-{2 \over 3}\delta_{ij}\three\hat \beta^k_{,k}\Bigr) \nonumber\\
&&{\hskip 4cm} +{1 \over 8}(UU_{,\varphi})_{,i}-{1 \over 4}
(\three\hat \beta^k U_{,k})_{,i}. 
\end{eqnarray}


\subsection{Basic equations appropriate for numerical approach}

Although equilibrium configurations 
can be formally obtained by solving Eq.$(\ref{bern})$ as well as
metric potentials, $U$, $X$, $\four\psi$, $\six\alpha$, 
$\three \beta^i$, $\five\beta^i$ and $h_{ij}$, 
they do not take convenient forms for numerical calculation. 
Thus, we here change Eq.$(\ref{bern})$ into forms appropriate 
to obtain numerically equilibrium configurations. 

In numerical calculation, the standard method to obtain equilibrium 
configurations is as follows \cite{hachisu,on90}; 

\noindent
(1) We give a trial density configuration for $\rho$. 

\noindent
(2)
We solve the Poisson equations. 

\noindent
(3)
Using Eq.$(\ref{bern})$, we give a new density configuration. 

\noindent
These procedures are repeated until a sufficient convergence is achieved. 
Here, at (3), we need to specify unknown constants, $\Omega$ and $C$. 
In standard numerical methods \cite{hachisu,on90}, 
these are calculated during iteration by fixing 
densities at two points; i.e., if we put $\rho_1$ at $x_1$ and 
$\rho_2$ at $x_2$ into Eq.$(\ref{bern})$, we obtain 
two equations for $\Omega$ and $C$. 
Solving these two equations give us $\Omega$ and $C$. 
However, the procedure is not so simple in the PN case: $\Omega$ is 
included in the source of the Poisson equations for 
the variables such as $X$, $\four \psi$, $\six\alpha$, 
$\eta_i$, $\five \hat P_i$, $h_{ij}^{(S)}$, $Q^{(T)}_{ij}$, $Q^{(I)}$ and 
$Q^{(I)}_i$. 
Thus, if we use Eq.$(\ref{bern})$ as it is, equations for 
$\Omega$ and $C$ become implicit equations for $\Omega$. In such a situation, 
the convergence to a solution is very slow. 
Therefore, we transform those equations into other forms 
in which the potentials as well as 
Eq.$(\ref{bern})$ become explicit polynomial equations in $\Omega$.

First of all, we define $q_2$, $q_{2i}$, $q_4$, $q_u$, $q_e$ and 
$q_{ij}$ which satisfy
\begin{eqnarray}
\Delta q_2 &=&-4\pi\rho R^2, \\
\Delta q_{2i} &=&-4\pi\rho R^2 x^i, \\
\Delta q_4 &=&-4\pi\rho R^4, \\
\Delta q_u &=&-4\pi\rho U, \\
\Delta q_e &=&-4\pi\rho \Pi, \\
\Delta q_{ij} &=&-4\pi\rho x^i x^j. 
\end{eqnarray}
Then, $X$, $\four\psi$, $Q^{(I)}$, $Q_i^{(I)}$, $\eta_i$, 
$\five \hat P_i$, $Q_{ij}^{(T)}$, and $h^{(S)}_{ij}$ are written as 
\begin{eqnarray}
X&=&-2q_2\Omega^2-2 q_u-(3\Gamma-2)q_e, \\
\four\psi&=&{1 \over 2}\Bigl(q_2\Omega^2+q_e+{5 \over 2}q_u\Bigr), \\
Q^{(I)}&=&q_2 \Omega^2+3(\Gamma-1)q_e-{1 \over 2}q_u 
\equiv q_2\Omega^2+Q^{(I)}_0, \\
Q_i^{(I)}&=&q_{2i}\Omega^2+Q_{0i}^{(I)}, \\
\eta_i&=&q_{2i}\Omega^2+\eta_{0i}, \\
\five \hat P_i&=&\epsilon_{izk}q_{2k}\Omega^2+\five P_{0i}, \\
Q_{ij}^{(T)}&=&\epsilon_{izl}q_{lj} \Omega, \\
h^{(S)}_{ij}&=&4\Omega^2\Bigl(\epsilon_{izk}\epsilon_{jzl}q_{kl}
-{1 \over 3}\delta_{ij}q_2\Bigr), 
\end{eqnarray}
where $Q_{0i}^{(I)}$, $\eta_{0i}$ and $\five P_{0i}$ satisfy
\begin{eqnarray}
\Delta Q_{0i}^{(I)} &=&-4\pi\Bigl( 3P- { 1 \over 2} \rho U\Bigr)x^i
=-4\pi\rho\Bigl( 3(\Gamma-1)\Pi-{1 \over 2}U \Bigr)x^i, \\
\Delta \eta_{0i} &=&-4\pi\rho\Bigl(\Pi+{5 \over 2}U\Bigr)x^i, \\
\Delta \five P_{0i} &=&-4\pi \rho \biggl[\epsilon_{izk} x^k \Bigl(2U
+\Gamma \Pi \Bigr)+\three \hat \beta^i\biggr]
+U_{,j} \Bigl( \three \hat \beta^i_{,j}+
\three \hat \beta^j_{,i}-{2 \over 3}\delta_{ij}\three\hat \beta^k_{,k}\Bigr) 
\nonumber\\
&&{\hskip 3cm} +{1 \over 8}(UU_{,\varphi})_{,i}-{1 \over 4}
(\three \hat \beta^k U_{,k})_{,i} \equiv -4\pi S^{(P)}_i. 
\end{eqnarray}
Note that $\five \beta^i$ and $h_{ij}^{(G)}$ are the cubic and quadratic 
equations in $\Omega$, respectively, as
\begin{eqnarray}
\five \beta^i &=&\five \beta^{i\,(A)}\Omega +\five\beta^{i\,(B)}\Omega^3, \\
h_{ij}^{(G)} &=&h_{ij}^{(A)}+h_{ij}^{(B)} \Omega^2 , 
\end{eqnarray} 
where
\begin{eqnarray}
\five \beta^{i\,(A)}&=&-4\five P_{0i}+{1 \over 2} \Bigl\{ 
x^i \Bigl(q_e+{5 \over 2}q_u\Bigr)_{,\varphi}-\eta_{0i,\varphi}\Bigr\} , 
\hspace*{6cm}\\
\five \beta^{i\,(B)}&=&-4\epsilon_{izk}q_{2k}
+{1 \over 2}\Bigl(x^iq_{2,\varphi}-q_{2i,\varphi}\Bigr) , \\
h_{ij}^{(A)}&=&
{1 \over 2}\biggl[
{\pa \over \pa x^j}\Bigl(x^i Q^{(I)}_0-Q^{(I)}_{0i}\Bigr)+
{\pa \over \pa x^i}\Bigl(x^j Q^{(I)}_0-Q^{(I)}_{0j}\Bigr)
-{4 \over 3}\delta_{ij}Q^{(I)}_0\biggr] , \nonumber\\
h_{ij}^{(B)}&=&
\biggl[{1 \over 2}\biggl\{
{\pa \over \pa x^j}\Bigl(x^i q_2-q_{2i}\Bigr)+
{\pa \over \pa x^i}\Bigl(x^j q_2-q_{2j}\Bigr)
-{4 \over 3}\delta_{ij}q_2 \biggr\} \nonumber\\
&&
-{7 \over 4}\Bigl(x^i\epsilon_{jzk}q_{k,\varphi}+x^j\epsilon_{izk}q_{k,\varphi}
-\epsilon_{izk}q_{kj,\varphi}-\epsilon_{jzk}q_{ki,\varphi}\Bigr)
+\delta_{ij}x^k \epsilon_{kzl} q_l 
\nonumber\\
&&
+{1 \over 8}x^k\biggl\{ {\pa \over \pa x^i}
\Bigl(x^j\epsilon_{kzl}q_{l,\varphi}-\epsilon_{kzl}q_{lj,\varphi}\Bigr)
+{\pa \over \pa x^j}
\Bigl(x^i\epsilon_{kzl}q_{l,\varphi}-\epsilon_{kzl}q_{li,\varphi}\Bigr)
\biggr\} \biggr] . 
\end{eqnarray}

Finally, we write $\six\alpha$ as
\begin{equation}
\six\alpha=\six \alpha_0+\six \alpha_2 \Omega^2-2q_4\Omega^4, 
\end{equation}
where $\six \alpha_0$ and $\six \alpha_2$ satisfy
\begin{eqnarray}
\Delta \six \alpha_0&=&4\pi\rho\biggl[\Bigl(3\Gamma-2\Bigr) \Pi U
-\Bigl(3\Gamma-4\Bigr)q_e+3q_u\biggr] \nonumber\\
&&-\Bigl(h_{ij}^{(U)}+h_{ij}^{(A)}\Bigr)U_{,ij}-{3 \over 2}UU_{,l}U_{,l}
+U_{,l}{\pa \over \pa x^l}\Bigl({9 \over 2}q_u+(3\Gamma+1)q_e\Bigr) 
\nonumber\\
&\equiv& -4\pi S^{(\alpha_0)},  \\
\Delta \six \alpha_2&=&8\pi\rho R^2\Bigl(5U+\Gamma\Pi
+2\three\hat\beta_{\varphi} \Bigr)
-\Bigl(4\epsilon_{izk}\epsilon_{jzl}q_{kl}-{4 \over 3}\delta_{ij}q_2
+h_{ij}^{(B)}\Bigr)U_{,ij} \nonumber\\
&& +3q_{2,l}U_{,l}+{1 \over 2}\three\hat\beta^i_{,j} 
\Bigl(\three\hat\beta^i_{,j}+\three\hat\beta^j_{,i}
-{2 \over 3}\delta_{ij}\three\hat\beta^k_{,k}\Bigr) \nonumber\\
&\equiv& -4\pi S^{(\alpha_2)} . 
\end{eqnarray}
Using the above quantities, Eq.$(\ref{bern})$ is rewritten as
\begin{equation}
H-{H^2 \over 2c^2}+{H^3 \over 3c^4}=A+B\Omega^2+D\Omega^4
+{R^6 \over 6c^4}\Omega^6+C, \label{berneq}
\end{equation}
where
\begin{eqnarray}
A&=&U+{1 \over c^2}\Bigl(2q_u+(3\Gamma-2)q_e\Bigr)+{1 \over c^4}\Bigl\{
-\six\alpha_0-{U^3 \over 6}+U\Bigl(2q_u+(3\Gamma-2)q_e\Bigr)\Bigr\},\nonumber\\
B&=&{R^2 \over 2}+{1 \over c^2}\Bigl(2R^2U+2q_2+\three\hat\beta_{\varphi}\Bigr)
+{1 \over c^4}\Bigl\{
-\six\alpha_2+{1 \over 2}\three\hat\beta^i\three\hat\beta^i
+4\three\hat\beta_{\varphi}U \nonumber\\
&&+(3\Gamma-1)q_eR^2+{9 \over 2}q_uR^2+{15 \over 4}U^2R^2+2q_2U+
\five\beta_{\varphi}^{(A)}
+{1 \over 2}\Bigl(h_{\varphi\varphi}^{(U)}+h_{\varphi\varphi}^{(A)}\Bigr)
\Bigr\}, \nonumber\\
D&=&{R^4 \over 4c^2}+{1 \over c^4}\Bigl\{
2q_4+\three\hat\beta_{\varphi}R^2+{7 \over 3}q_2R^2+2UR^4+
\five\beta_{\varphi}^{(B)}+{1 \over 2}\Bigl(h_{\varphi\varphi}^{(B)}
+4R^2q_{RR}\Bigr)\Bigr\}. 
\nonumber\\&&
\end{eqnarray}
Note that in the above, we use the following relations which hold 
for arbitrary vector $Q_i$ and symmetric tensor $Q_{ij}$, 
\begin{eqnarray}
Q_{\varphi}&=&-yQ_{x}+xQ_{y}, \nonumber\\ 
Q_{\varphi\varphi}&=&y^2Q_{xx}-2xyQ_{xy}+x^2Q_{yy}, \nonumber\\
R^2Q_{RR}&=&x^2Q_{xx}+2xyQ_{xy}+y^2Q_{yy}. 
\end{eqnarray}
We also 
note that source terms of Poisson equations for variables which appear in 
$A$, $B$ and $D$ do not depend on $\Omega$ 
explicitly. Thus, Eq.$(\ref{berneq})$ takes the desired form 
for numerical calculation.

In this formalism, we need to solve 29 Poisson equations for 
$U$, $q_x$, $q_y$, $q_z$, $\five P_{0x}$, $\five P_{0y}$, 
$\eta_{0x}$, $\eta_{0y}$, $Q^{(I)}_{0x}$, $Q^{(I)}_{0y}$, $Q^{(I)}_{0z}$, 
$q_2$, $q_{2x}$, $q_{2y}$, $q_{2z}$, $q_u$, $q_e$, 
$h_{xx}^{(U)}$, $h_{xy}^{(U)}$, $h_{xz}^{(U)}$, $h_{yy}^{(U)}$, $h_{yz}^{(U)}$,
$q_{xx}$, $q_{xy}$, $q_{xz}$, $q_{yz}$, $\six\alpha_0$, 
$\six\alpha_2$ and $q_4$. In Table 1, we show the list of the Poisson equations
to be solved. In Table 2, 
we also summarize what variables are needed to calculate 
the metric variables $U$, $X$, $\four\psi$, $\six \alpha$, $\three \beta^i$, 
$\five\beta^i$, $h_{ij}^{(U)}$,  $h_{ij}^{(S)}$, $h_{ij}^{(A)}$ and 
$h_{ij}^{(B)}$. Note that we do not need $\five P_{0z}$, 
$\eta_{0z}$, and $q_{zz}$ because they do not appear in any equation. 
Also, we do not have to solve the Poisson 
equations for $h^{(U)}_{zz}$ and $q_{yy}$ because they can be calculated from 
$h^{(U)}_{zz}=-h^{(U)}_{xx}-h^{(U)}_{yy}$ and $q_{yy}=q_2-q_{xx}$. 

In order to derive $U$, $q_i$, $q_2$, $q_{2i}$, $q_4$, $q_e$ and $q_{ij}$, 
we do not need any other potential because 
only matter variables appear in the source terms of their 
Poisson equations. On the other hand, 
for $q_u$, $Q_{0i}^{(I)}$, $\eta_{0i}$ and $h_{ij}^{(U)}$, 
we need the Newtonian potential $U$, and 
for $\five P_{0i}$, $\six\alpha_0$ and $\six\alpha_2$, 
we need the Newtonian as well as PN potentials. 
Thus, $U$, $q_i$, $q_2$, $q_{2i}$, $q_4$, $q_e$ and $q_{ij}$ 
must be solved first, and then $q_u$, $Q_{0i}^{(I)}$, $\eta_{0i}$, 
$h_{ij}^{(U)}$, $\five P_{0i}$ and $\six\alpha_2$ should be solved. 
$\six\alpha_0$ must be solved after we obtain $q_u$ 
because its Poisson equation involves $q_u$ in the source term. 
In Table 1 we also list potentials which are included in the source 
terms of the Poisson equations for other potentials. 

The configuration which we are most interested in and would like to 
obtain is the equilibrium state 
for BNS's of equal mass. Hence, we show the boundary condition at 
$r \rightarrow \infty$ for this problem. 
When we consider equilibrium configurations for 
BNS's where the center of mass for each NS is on the $x$-axis, 
boundary conditions for potentials at $r \rightarrow \infty$ become 
\begin{eqnarray}
U   &=&{1 \over r}\int \rho d^3x+O(r^{-3}), \hskip 2cm~~~
q_{x}={n^x \over r^2}\int \rho  x^2 d^3x+O(r^{-4}), \nonumber\\
q_2 &=&{1 \over r}\int \rho R^2 d^3x+O(r^{-3}),\hskip 2cm
q_{y}={n^y \over r^2}\int \rho  y^2 d^3x+O(r^{-4}), \nonumber\\
q_e &=&{1 \over r}\int \rho \Pi d^3x+O(r^{-3}),\hskip 2cm~~
q_{z}={n^z \over r^2}\int \rho  z^2 d^3x+O(r^{-4}), \nonumber\\
q_u&=&{1 \over r}\int \rho U d^3x+O(r^{-3}),\hskip 2cm~
q_4={1 \over r}\int \rho R^4  d^3x+O(r^{-3}), 
\end{eqnarray}
\begin{eqnarray}
\five P_{0x}&=&{n^x \over r^2}\int S^{(P)}_x x d^3x
+{n^y \over r^2}\int S^{(P)}_y y d^3x +O(r^{-3}), \nonumber\\
\five P_{0y}&=&{n^x \over r^2}\int S^{(P)}_y x d^3x
+{n^y \over r^2}\int S^{(P)}_y y d^3x +O(r^{-3}), 
\end{eqnarray}
\begin{eqnarray}
\eta_{0x}&=&{n^x \over r^2}\int \rho x^2\Bigl(\Pi+{5 \over 2}U\Bigr)
d^3x+O(r^{-4}), \nonumber\\
\eta_{0y}&=&{n^y \over r^2}\int \rho y^2\Bigl(\Pi+{5 \over 2}U\Bigr)
d^3x+O(r^{-4}), 
\end{eqnarray}
\begin{eqnarray}
Q^{(I)}_{0x}&=&{n^x \over r^2}\int \rho x^2\Bigl(3(\Gamma-1)\Pi
-{1 \over 2}U\Bigr) d^3x+O(r^{-4}), \hspace{0.3cm}
q_{2x}={n^x \over r^2}\int \rho R^2 x^2 d^3x+O(r^{-4}), \nonumber\\
Q^{(I)}_{0y}&=&{n^y \over r^2}\int \rho y^2\Bigl(3(\Gamma-1)\Pi
-{1 \over 2}U\Bigr) d^3x+O(r^{-4}), \hspace{0.3cm}
q_{2y}={n^y \over r^2}\int \rho R^2 y^2 d^3x+O(r^{-4}), \nonumber\\
Q^{(I)}_{0z}&=&{n^y \over r^2}\int \rho z^2\Bigl(3(\Gamma-1)\Pi
-{1 \over 2}U\Bigr) d^3x+O(r^{-4}), \hspace{0.3cm}
q_{2z}={n^z \over r^2}\int \rho R^2 z^2 d^3x+O(r^{-4}), \nonumber\\
&&
\end{eqnarray}
\begin{eqnarray}
h^{(U)}_{xx}&=&{1 \over r}\int S^{(U)}_{xx}  d^3x+O(r^{-3}), \hspace{2.5cm}
h^{(U)}_{xy}={3n^x n^y \over r^3}\int S^{(U)}_{xy}xy  d^3x+O(r^{-5}), 
\nonumber\\
h^{(U)}_{yy}&=&{1 \over r}\int S^{(U)}_{yy}  d^3x+O(r^{-3}), \hspace{2.5cm}
h^{(U)}_{xz}={3n^x n^z \over r^3}\int S^{(U)}_{xz}xz  d^3x+O(r^{-5}), 
\nonumber\\
h^{(U)}_{yz}&=&{3n^y n^z \over r^3}\int S^{(U)}_{yz}yz  d^3x+O(r^{-5}), 
\end{eqnarray}
\begin{eqnarray}
q_{xx}&=&{1 \over r}\int \rho x^2  d^3x+O(r^{-3}), \hspace{1.6cm}
q_{xy}={3n^x n^y \over r^3}\int \rho x^2y^2  d^3x+O(r^{-5}), 
\hspace*{1.5cm}\nonumber\\
q_{xz}&=&{3n^x n^z \over r^3}\int \rho x^2 z^2  d^3x+O(r^{-5}), 
\hspace{0.5cm}
q_{yz}={3n^y n^z \over r^3}\int \rho y^2z^2  d^3x+O(r^{-5}),
\end{eqnarray}
\begin{eqnarray}
\six\alpha_0&=&{1 \over r}\int S^{(\alpha_0)}  d^3x+O(r^{-3}), \hspace{0.3cm}
\six\alpha_2 ={1 \over r}\int S^{(\alpha_2)}  d^3x+O(r^{-3}),  
\end{eqnarray}
where  $n^i=x^i / r$. 
We note that $S_i^{(P)} \rightarrow O(r^{-5})$, 
$S_{ij}^{(U)} \rightarrow O(r^{-6})$, 
$S^{(\alpha_0)} \rightarrow O(r^{-4})$ and 
$S^{(\alpha_2)} \rightarrow O(r^{-4})$ as $r \rightarrow \infty$, 
so that all the above integrals are well defined. 

We would like to emphasize that from the 2PN order, the 
tensor part of the 3-metric, $\tilde\gamma_{ij}$, cannot be neglected  
even if we ignore gravitational waves. 
Recently, Wilson and Mathews \cite{wm}, Wilson, Mathews and Marronetti 
\cite{wmm} presented numerical equilibrium configurations of 
binary neutron stars using a semi-relativistic approximation, in which 
they assume the spatially conformal flat metric as the spatial 3-metric, 
i.e., $\tilde \gamma_{ij}=\delta_{ij}$. 
Thus, in their method, a 2PN term, $h_{ij}$, was completely neglected. 
However, it should be noted that this tensor potential plays 
an important role at the 2PN order: 
This is because they appear in the equations to determine 
equilibrium configurations as shown in previous sections and they also 
contribute to the total energy and angular momentum of systems. 
This means that if we performed the stability analysis ignoring the tensor 
potentials, we might reach an incorrect conclusion. 
If we hope to obtain a general relativistic equilibrium configuration 
of binary neutron stars with a better accuracy (say less than $1\%$), 
we should take into account the tensor part of the 3-metric. 

In our formalism, we extract terms depending on the angular velocity $\Omega$ 
from the integrated 
Euler equation and Poisson equations for potentials, and rewrite 
the integrated Euler equation as an explicit equation in $\Omega$. 
This reduction will improve the convergence in numerical 
iteration procedure. 
As a result, the number of Poisson equations we need to solve in each step 
of iteration reaches 29. 
However, source terms of the Poisson equations 
decrease rapidly enough, at worst $O(r^{-4})$, 
in the region far from the source, so that we can solve accurately 
these equations as the boundary value problem like in the case of 
the first PN calculations. 

The formalism presented here will be useful to obtain equilibrium 
configurations for synchronized BNS's or the Jacobi ellipsoid. 
In this context, Bonazzola, Frieben and Gourgoulhon (1996) obtained 
an approximate nonaxisymmetric neutron star by perturbing a stationary 
axisymmetric configuration. 
However, they do not solve the exact 3D Einstein's equation. 
Thus, it is interesting to examine their conclusion on the transition 
between equilibrium configurations, which are approximately ellipsoids, 
by our PN methods.


\section{Gravitational Waves from post-Newtonian sources}

In the post-Minkowskian approximation, the background geometry is 
the Minkowski spacetime where linearized gravitational waves propagate 
\cite{thorne80,bd86,bd88}. 
The corrections to propagation of gravitational waves 
can be taken into account if one performs the post-Minkowskian 
approximation up to higher orders. 
In fact, Blanchet and Damour obtained the tail term 
of gravitational waves as the integral over the past history 
of the source \cite{bd88,bd92}. 
They introduced an unphysical complex parameter $B$, and used 
the analytic continuation as a mathematical device 
in order to evaluate the so-called $\log$ term in the tail contribution. 
The method is powerful, but it is not easy to see the origin of 
the tail term. 
Will and Wiseman (1996) have also obtained the tail term 
from the mass quadrupole moment, by improving the Epstein-Wagoner 
formalism \cite{ew}. Nakamura\cite{nakamura95} and
Sch\"afer\cite{schafer90} have obtained the quadrupole energy 
loss formula including the contribution from the tail term by studying 
the wave propagation in the Coulomb type potential. 
This indicates clearly that 
the origin of the tail term is due to the Coulomb type potential 
generated by the mass of the source. 
In the following treatment we calculate the waveform from the slow motion 
source and show explicitly 
how the tail term originates from the difference 
between the flat light cone and the true one which is due to 
the mass of the source $GM/c^2$ as the lowest order correction. 

\subsection{Gravitational waves in Coulomb type potential}

We wish to clarify that the main part of 
the tail term is due to the propagation of gravitational waves 
on the light cone which deviates slightly from the flat light cone 
owing to the mass of the source. 
For this purpose, we shall work in the harmonic coordinate since 
the deviation may be easily seen in the reduced Einstein's equation 
in this coordinates. 
\begin{equation}
 ( \eta^{\alpha\beta} - {\bar h}^{\alpha\beta} )
{\bar h}^{\mu\nu}_{\;\;\; ,\alpha\beta} 
= - 16 \pi \Theta^{\mu\nu} 
+ {\bar h}^{\mu\alpha}_{\;\;\; ,\beta}
{\bar h}^{\nu\beta}_{\;\;\; ,\alpha} . 
\label{waveeq0}
\end{equation}
Since we look for  $1/r$ part of the solution, the spatial derivative 
is not relevant in the differential operator so that it is more convenient 
for our purpose to transform Eq.($\ref{waveeq0}$) into the following form 
\begin{equation}
( \Box - {\bar h}^{00}\partial_0 \partial_0 )
{\bar h}^{\mu\nu}= - 16 \pi {\cal S}^{\mu\nu} , 
\end{equation}
where we defined ${\cal S}^{\mu\nu}$ as 
\begin{equation}
{\cal S}^{\mu\nu} = \Theta^{\mu\nu} 
-{1\over {16\pi}} ( {\bar h}^{\mu\alpha}_{\;\;\; ,\beta} 
{\bar h}^{\nu\beta}_{\;\;\; ,\alpha}
- 2{\bar h}^{0i}{\bar h}^{\mu\nu}_{\;\;\; ,0i} 
- {\bar h}^{ij}{\bar h}^{\mu\nu}_{\;\;\; ,ij} ) . 
\label{waveeq1}
\end{equation}
Substituting the lowest order expression for ${\bar h}^{00}$ 
derived in section 2, we finally obtain our basic equation to be solved.   
\begin{equation}
\Bigl[ -\Bigl(1+{4M \over r}\Bigr){\pa^2 \over \pa t^2}+\Delta \Bigr] 
{\bar h}^{\mu\nu} = \tilde\tau^{\mu\nu} ,  
\label{tildewaveeq}
\end{equation}
where the effective source $\tilde \tau^{\mu\nu}$ is defined as 
\begin{equation}
{\tilde \tau^{\mu\nu}} 
=  {\cal S}^{\mu\nu} 
- {1\over {16\pi}} \left({\bar h}^{00} 
- {{4M}\over r} \right){\bar h}^{\mu\nu}_{\;\;\; ,00} . 
\end{equation}
At the lowest order, we obtain 
$\four\tilde\tau_{\mu\nu}=\four\Theta_{\mu\nu}$. 

The solution for $(\ref{tildewaveeq})$ may be written down by using the 
retarded Green's function in the following form.
\begin{equation}
{\bar h}^{\mu\nu}(x) = \int d x' G_M^{(+)}(x, x') 
{\tilde  \tau}^{\mu\nu}(x') . 
\label{formalsol}
\end{equation}
The retarded Green's function is defined as satisfying the equation; 
\begin{equation}
\Box_M G_M^{(+)}(x,y)= \delta^4(x-y) , \label{curvedG}
\end{equation}
with an appropriate boundary condition. 
We also defined the following symbol for the differential operator 
appearing in our basic equation.
\begin{equation}
\Box_M=\Bigl[ -\Bigl( 1+{4M \over r} \Bigr) {\partial^2 \over {\partial t^2}}
+\Delta \Bigr], 
\label{box2}
\end{equation}
where $\Delta$ is the Laplacian in the flat space. 

The Green's function $G_0$ for $M=0$ i.e. in the Minkowski spacetime is 
well known. We shall present the detailed  derivation of the Green 
function in appendix B, and present here the result 
\begin{eqnarray}
G^{(+)}_M(x,x')&=&\sum_{l m} e^{i \sigma_l} 
\int d\omega sgn(\omega) \Bigl( 
\Psi^{+ \,\omega\,l\,m}(x) \Psi^{S\,\omega\,l\,m\,\ast}(x') 
\theta(r-r') \nonumber\\
&&~~~~~~~~~~~~~~~~~~~~~~~~~~
+\Psi^{S\,\omega\,l\,m}(x) \Psi^{+ \,\omega\,l\,m\,\ast}(x') 
\theta(r'-r) \Bigr) , 
\label{green}
\end{eqnarray}
where $*$ denotes the complex conjugate, $sgn(\omega)$ denotes 
a sign of $\omega$  and we defined $\Psi^{+\,\omega\,l\,m}(x)$ 
and $\Psi^{S\,\omega\,l\,m}(x)$ as 
\begin{eqnarray}
\Psi^{+\omega\,l\,m}(x)&=&\sqrt{{|\omega| \over 2\pi}}
e^{-i\omega t} \rho^{-1} u^{(+)}_{l}(\rho;\gamma)
Y_{l\,m} , \nonumber\\
\Psi^{S\,\omega\,l\,m}(x)&=&\sqrt{{|\omega| \over 2\pi}} 
e^{-i\omega t} \rho^{-1} F_{l}(\rho;\gamma) 
Y_{l\,m} , 
\end{eqnarray}
where $ \rho = \omega r $ and $u^{(+)}_l$ is 
a spherical Coulomb function satisfying the outgoing wave condition at 
null infinity, and $F_\ell$ is also a spherical 
Coulomb function which is regular at the origin.  

Since we study an asymptotic form of the waves, we use 
Eqs.($\ref{coulomb1}$) and ($\ref{coulomb2}$) in appendix B 
to obtain the asymptotic form of the Green function as
\begin{eqnarray}
G^{(+)}_M(x,x')&=&{1 \over r}\sum_{lm}{(-i)^l \over 2\pi} 
\int d\omega c_l e^{i\sigma_l} e^{-i \omega(t-r-2M\ln{2M\omega}-t')} 
(\omega r')^l Y_{lm}(\Omega) Y^{*}_{lm}(\Omega') \nonumber\\
&&+O(r^{-2}) \quad\mbox{for}\quad r \to \infty . 
\label{green2}
\end{eqnarray}
In addition, for slow motion sources, we evaluate the asymptotic form
of the Green function up to $O(M\omega)$ as 
\begin{eqnarray}
G^{(+)}_M(x,x')&=&{1 \over r}\sum_{lm}{(-i)^l \over 2\pi (2l+1)!!} 
\int d\omega \Bigl\{ 1+\pi M\omega+2iM\omega 
\Bigl( \ln{2M\omega}-\sum^l_{s=1} {1 \over s} +C \Bigr) 
\nonumber\\
&&+O(M^2\omega^2) \Bigr\}  e^{-i \omega(t-r-t')} (\omega r')^l 
Y_{lm}(\Omega) Y^{*}_{lm}(\Omega') +O(r^{-2}) \nonumber\\
&=&{1 \over r} \sum_{lm}{(-i)^l \over 2\pi (2l+1)!!}
\int d\omega \Bigl[ 1+2M\omega \Bigl\{ i \Bigl( -\sum^l_{s=1}{1 \over s} 
+\ln{2M} \Bigr)+{\pi \over 2} sgn(\omega)
\nonumber\\ 
&&+i (\ln{|\omega|}+C) \Bigr\} +O(M^2\omega^2) \Bigr] 
e^{-i \omega(t-r-t')} (\omega r')^l 
Y_{lm}(\Omega) Y^{*}_{lm}(\Omega') 
\nonumber\\ 
&&+O(r^{-2}) , 
\label{green3}
\end{eqnarray}
where $C$ is Euler's number. In deriving the above expression we have used 
the following expansion for $c_l$ and $\sigma_l$ in $M\omega$.
\begin{eqnarray}
c_l&=&{1+\pi M\omega+O(M^2\omega^2) \over (2l+1)!!} , \nonumber\\
\sigma_l&=&2M\omega \Bigl( C-\sum^{l}_{s=1}{1 \over s} \Bigr) 
+O(M^2\omega^2) . 
\end{eqnarray}

Now we apply the formula \cite{gr,bs} 
\begin{equation}
\omega\int_0^1 dv e^{i\omega v} \ln{v}+i\int_1^{\infty}
{dv \over v} e^{i\omega v}
=-{\pi \over 2}sgn(\omega)-i(\ln{|\omega|}+C), 
\label{formula} 
\end{equation}
to Eq.($\ref{green3}$), then we obtain
\begin{eqnarray}
G^{(+)}_M(x,x')&=&{1 \over r} \sum_{lm}{(-i)^l \over 2\pi (2l+1)!!}
\nonumber\\
&&\times\int d\omega 
\Bigl\{ 1-2M \Bigl( -\sum^l_{s=1}{1 \over s} +\ln{2M} \Bigr) {d \over dt} 
+2M \Bigl(\int^1_0 dv e^{i\omega v} \ln{v} \Bigr) {d^2 \over dt^2} 
\nonumber\\
&&\hspace*{1.5cm}+2M \Bigl(\int^{\infty}_1 {dv \over v} e^{i\omega v} 
\Bigr) {d \over dt} 
+O(M^2\omega^2) \Bigr\} \nonumber\\
&&\times e^{-i \omega(t-r-t')} (\omega r')^l 
Y_{lm}(\Omega) Y^{*}_{lm}(\Omega') +O(r^{-2}) . 
\label{green4}
\end{eqnarray}
It is worthwhile to point out that $\ln{2M}$ in Eq.($\ref{green4}$) 
can be removed by using the freedom to time translation. 

Here, we assume the no incoming radiation condition 
on the initial hypersurface so that we may take  
\begin{equation}
\lim_{v \to \infty} e^{-i \omega (t-r-v-t')} \ln{v} \to 0 . 
\end{equation}
Thus we can make the following replacement  
\begin{equation}
\int^{\infty}_1 {dv \over v} e^{i \omega v} \to 
-i \omega \int^{\infty}_1 dv e^{i \omega v} \ln{v}  
=\Bigl( \int^{\infty}_1 dv e^{i \omega v} \ln{v} \Bigr) {d\over dt} . 
\label{replace}
\end{equation}
Inserting Eq.($\ref{replace}$) into Eq.($\ref{green4}$), we finally 
obtain the desired expression  for the retarded Green's function
\begin{eqnarray}
G^{(+)}_M(x,x')&=&{1 \over r} \sum_{lm}{(-i)^l \over 2\pi (2l+1)!!} 
\int d\omega 
\Bigl[ 1+2M \Bigl( \sum^l_{s=1}{1 \over s} -\ln{2M} \Bigr) {d \over dt}
\nonumber\\
&&\hspace*{2cm}+2M \Bigl(\int^{\infty}_0 dv e^{i\omega v} \ln{v}
\Bigr) {d^2 \over dt^2} +O(M^2 \omega^2) \Bigr] 
\nonumber\\
&&\times e^{-i \omega(t-r-t')} (\omega r')^l 
Y_{lm}(\Omega) Y^{*}_{lm}(\Omega') +O(r^{-2}) 
\nonumber\\ 
&=&{1 \over r} \;\mbox{part of} \nonumber\\ 
&&\Bigl[ G_{0}(x,x')+2M {d^2 \over dt^2} \sum_{lm} 
\int dv \Bigl\{ \ln \Bigl( {v \over 2M} \Bigr) 
+\sum^l_{s=1} {1 \over s} \Bigr\} 
G_{0}^{l\,m}(t-v,{\bf x},x') 
\nonumber\\
&&+O(M^2) \Bigr] +O(r^{-2}) , 
\label{green5}
\end{eqnarray}
where we defined the spherical harmonic expansion coefficient of the 
flat Green's function as follows. 
\begin{equation}
G_{0}^{l\,m}(x,x')=Y_{l\,m}(\Omega) \int d\Omega' G_{0}(x,x') 
Y_{l\,m}(\Omega') . 
\end{equation}
As a result, we obtain the waveform generated by the linear part 
of the effective source which corresponds to (C1) by Blanchet \cite{blan95}. 
\begin{eqnarray}
h_{ij}^{TT}&=&{4 \over r} P_{ijpq} 
\sum^{\infty}_{l=2} {1 \over l!} 
\Bigl[ n_{L-2} \Bigl\{ \tilde M_{pqL-2}(t-r)
+2M {d^2 \over dt^2} 
\int dv \Bigl\{ \ln \Bigl({v \over 2M} \Bigr) 
+\sum^{l-2}_{s=1} {1 \over s} \Bigr\} 
\nonumber\\
&&\hspace*{2.5cm}\times \tilde M_{pqL-2}(t-r-v) \Bigr\} 
-{2l \over l+1} n_{aL-2} 
\Bigl\{ \epsilon_{ab(p} \tilde S_{q)bL-2}(t-r)
\nonumber\\
&&\hspace*{2.5cm}+2M {d^2 \over dt^2} \int dv 
\Bigl\{ \ln \Bigl({v \over 2M}\Bigr)
+\sum^{l-1}_{s=1} {1 \over s} \Bigr\}  
\epsilon_{ab(p} \tilde S_{q)bL-2}(t-r-v) \Bigr\} 
\nonumber\\
&&\hspace*{2.5cm}+O(M^2) \Bigr] 
+O(r^{-2}) , 
\label{waveform}
\end{eqnarray}
where $P_{ijpq}$ is the transverse and traceless projection tensor 
and parentheses denote symmetrization. $\tilde M_{pq L-2}$ and 
$\tilde S_{pq L-2}$ are the mass and current multipole moments
generated by the full nonlinear effective source $\tilde\tau_{\mu\nu}$. 
They take the same forms as in Thorne (1980).

\subsection{Comparison with the previous work} 

By using the post-Minkowskian approximation, Blanchet obtained 
the radiative mass moment and radiative current moment as \cite{blan95}
\begin{eqnarray}
U^{(l)}_L(u)&=&M^{(l)}_L(u)+2GM\int^{\infty}_0 dv M^{(l+2)}_L(u-v)
\Big\{ \ln \Bigl({v \over P}\Bigr)+\kappa_l \Bigr\}  
+O(G^2M^2) , 
\nonumber\\
&&
\label{blanmoment}
\end{eqnarray}
and 
\begin{eqnarray}
V^{(l)}_L(u)&=&S^{(l)}_L(u)+2GM\int^{\infty}_0 dv S^{(l+2)}_L(u-v)
\Big\{ \ln \Bigl({v \over P}\Bigr)+\kappa_l^{\prime} \Bigr\} 
+O(G^2M^2) , 
\nonumber\\
&&
\label{blanmoment2}
\end{eqnarray}
where their moments $M_L$ and $S_L$ do not contain the nonlinear 
contribution outside the matter, $P$ is a constant with 
temporal dimension, and $\kappa_l$ and 
$\kappa_l^{\prime}$ are defined as 
\begin{equation}
\kappa_l=\sum^{l-2}_{s=1}{1 \over s}+{2l^2+5l+4 \over l(l+1)(l+2)}, 
\label{kappa1}
\end{equation}
and 
\begin{equation}
\kappa_l^{\prime}=\sum^{l-1}_{s=1}{1 \over s}+{l-1 \over l(l+1)}. 
\label{kappa2}
\end{equation}
In black hole perturbation, the same tail corrections with 
multipole moments induced by linear perturbations have been obtained 
\cite{ps}. 
Compared with Eqs.($\ref{blanmoment}$)-($\ref{kappa2}$), 
Eq.($\ref{waveform}$) shows that $log$ term and 
$\sum 1/s$ in the radiative moments ($\ref{blanmoment}$) and 
($\ref{blanmoment2}$) originate from propagation 
on the slightly curved light cone determined by Eq.($\ref{box2}$). 
That is to say, $log$ term is produced by propagation of
gravitational waves in the Coulomb type potential in the external
region. 
It is worthwhile to point out the following fact: 
Only the $log$ term has a hereditary property expressed as the integral 
over the past history of the source, since the constants $\kappa_l$ 
and $\kappa_l^{\prime}$ represent merely instantaneous parts 
after performing the integral under the assumption that 
the source approaches static as the past infinity. 

However, Eq.($\ref{waveform}$) does not apparently agree 
with Eqs.($\ref{blanmoment}$) and ($\ref{blanmoment2}$), 
because the definition of the moments 
$\{\tilde M_L, \tilde S_L\}$ and $\{M_L, S_L\}$ are different. 
We shall show the equivalence between our expression and 
that of Blanchet (1995), by calculating the contribution 
from nonlinear terms like $M\times M_L$ or $M\times S_L$. 
It is noteworthy that the luminosity of gravitational waves 
obtained by the present approach agrees at the tail term i.e. $O(c^{-3})$ 
with that by the post-Minkowskian approximation. 
\begin{eqnarray}
{\cal L}&=&\sum_{l=2}^{\infty} {(l+1)(l+2) \over (l-1)l} 
{1 \over l! (2l+1)!!} 
< \mathop{U_L}^{(l+1)}\mathop{U_L}^{(l+1)} > \nonumber\\ 
&&+\sum_{l=2}^{\infty} {4l(l+2) \over (l-1)} 
{1 \over (l+1)! (2l+1)!!} 
< \mathop{V_L}^{(l+1)}\mathop{V_L}^{(l+1)} > . 
\end{eqnarray}

\subsection{Contributions from the nonlinear sources} 

In order to evaluate the waveform produced by the nonlinear sources in 
$\tilde\tau_{\mu\nu}$, it is enough to use the flat Green's function 
\begin{eqnarray}
G_{0}(x,x')&=&-i \sum_{lm} \int d\omega {\omega \over \pi}
\Bigl(e^{-i\omega t}h_l(\omega r)Y_{lm}(\Omega) 
e^{i\omega t'}j_l(\omega r')Y_{lm}^{*}(\Omega ')\theta(r-r') \nonumber\\
&&\hspace{3cm}+e^{-i\omega t}j_l(\omega r)Y_{lm}(\Omega) 
e^{i\omega t'}h_l(\omega r')Y_{lm}^{*}(\Omega ')\theta(r'-r) \Bigr) . 
\nonumber\\
&&
\end{eqnarray}

Since the nonlinear sources  have the form of 
either $M\times M_L$ or $M\times S_L$, we have to 
treat the following type of retarded integral 
\begin{eqnarray}
\Box^{-1}\Bigl[ \partial_i({1 \over r}) \hat\partial_Q 
\Bigl({F(t-r) \over r} \Bigr)\Bigr]&=&(-)^{q+1} 
\sum^q_{j=0}{ (q+j)! \over 2^j j! (q-j)! } \nonumber\\ 
&&\times\Box^{-1}\Bigl[ \Bigl(\hat n_{iQ}+{q \over 2q+1}\delta_{i<a_q} 
\hat n_{Q-1>} \Bigr) {1 \over r^{j+3}} \mathop{F}^{(q-j)}\!\!(t-r) \Bigr] .
\nonumber\\
&&
\label{c201}
\end{eqnarray}
The evaluation can be made by using the formula which will be proved 
in appendix C. 
\begin{eqnarray}
\Box^{-1} \Bigl[ {\hat n_L \over r^k}F(t-r) \Bigr] 
&=&-2^{l+1} \lim_{\lambda \to 0} \Bigl( \sum^{\infty}_{n=0} 
{(l+n+1)!\Gamma(-k+l+3+2n-\lambda) \over n! (2l+2n+2)!} \Bigr) 
\nonumber\\
&&\times{\hat n_L \over r}\mathop{F}^{(k-3)}\!\!(t-r) +O(r^{-2}) .
\label{formula3}
\end{eqnarray}
Although some terms of nonlinear sources may produce disastrous
divergence in Eq.($\ref{formula3}$), 
we expect these divergent parts cancel out in total, 
which is explicitly demonstrated in the appendix. 

Using the above formula we find 
\begin{eqnarray}
&&\Box^{-1} \Bigl[ \partial_i ({1 \over r}) \hat\partial_Q 
\Bigl( {F(t-r) \over r} \Bigr) \Bigr] \nonumber\\
&&={(-)^q \over 2(q+1)} \Bigl[ \hat n_{iQ}-{q+1 \over 2q+1}
\delta_{i<a_1} \hat n_{Q-1>} \Bigr] {1 \over r} 
\mathop{F}^{(q)}\!(t-r)+O(r^{-2}) , 
\label{c2}
\end{eqnarray}
which is same with the expression (C2) obtained by Blanchet \cite{blan95}. 

Applying Eqs.($\ref{c2}$) to the nonlinear source $\tilde\tau_{\mu\nu}$, 
we can evaluate its contribution to the waveform. 
Together with Eq.($\ref{waveform}$), we obtain the total waveform as 
\begin{eqnarray}
h_{ij}^{TT}&=&{4 \over r} P_{ijpq} 
\sum^{\infty}_{l=2} {1 \over l!} 
\Bigl[ n_{L-2} \Bigl\{ M_{pqL-2}(t-r)
+2M {d^2 \over dt^2} 
\int dv \Bigl\{ \ln \Bigl({v \over 2M} \Bigr) 
+\kappa_l \Bigr\} 
\nonumber\\
&&\hspace*{2.5cm}\times M_{pqL-2}(t-r-v) \Bigr\} 
-{2l \over l+1} n_{aL-2} 
\Bigl\{ \epsilon_{ab(p}S_{q)bL-2}(t-r)
\nonumber\\
&&\hspace*{2.5cm}+2M {d^2 \over dt^2} \int dv 
\Bigl\{ \ln \Bigl({v \over 2M}\Bigr)
+\kappa_l^{\prime} \Bigr\}  
\epsilon_{ab(p}S_{q)bL-2}(t-r-v) \Bigr\} 
\nonumber\\
&&\hspace*{2.5cm}+O(M^2) \Bigr] 
+O(r^{-2}) . 
\label{tailwave}
\end{eqnarray}

In this section, we have derived the formula ($\ref{tailwave}$) 
for gravitational waves including tail by using the formula 
($\ref{waveform}$) and ($\ref{c2}$). 
We would like to emphasize two points on the derivation: 
First, in deriving Eq.($\ref{waveform}$), spherical Coulomb functions 
are used, since we use the wave operator ($\ref{box2}$) 
which take account of the Coulomb-type potential $M/r$. 
As a consequence, $\ln (v/2M)$ appears naturally in Eq.($\ref{tailwave}$). 
This is in contrast with Blanchet and Damour's method, 
where an arbitrary constant with temporal dimension $P$ appears 
in the form of $\ln (v/P)$. 
Our derivation shows that the main part of the tail, 
which needs the past history of the source only through $\ln{v}$, 
is produced by propagation in the Coulomb-type potential. 
Our method might make it easy to clarify conditions 
for wave formula including tail. 
This remains as a future work. 

The second point relates with physical application: 
At the starting point of our derivation, the Fourier representation 
in frequency space has been used. 
Such a representation seems to simplify the calculation of gravitational 
waveforms from compact binaries in the quasi-circular orbit, 
since such a system can be described by a characteristic frequency. 
Applications to physical systems will be also done 
in the future.

\section{Conclusion}

We have discussed various aspects on  the 
post-Newtonian approximation mainly based on our own work.  
After presenting the basic structure of the PN approximation 
in the framework of Newtonian limit along a regular asymptotic 
Newtonian sequence, we reformulated it in the appropriate form for 
numerical approach. For this purpose we have adopted the (3+1) formalism 
in general relativity.  
Although we restricted ourselves within the transverse gauge 
in this paper, we can use any gauge condition and investigate 
its property relatively easily in the (3+1) formalism, 
compared with in the standard PN approximation performed so far 
\cite{chandra65,chandra67,chandra69a,cn,ce}. 
Using the developed formalism, we have written down 
the hydrodynamic equation up to 2.5PN order. 
For the sake of an actual numerical simulation, we consider 
carefully methods to solve the various metric quantities, 
especially, the 2PN tensor potential $\four h_{ij}$. 
We found it possible to solve them by using 
the numerical methods familiar in Newtonian gravity. 
Thus, the formalism discussed in this paper will be useful  
in numerical applications.  As an example of the application, 
we have presented the formalism for constructing the equilibrium 
configuration of nonaxisymmetric uniformly rotating fluid. 
We have also discussed the propagation of gravitational waves 
explicitly taking into account of the deviation of the light cone from 
that of the flat spacetime and obtained essentially the same result 
with that of Blanchet and Damour for the waveform 
from post-Newtonian systems.

\bigskip

{\bf Acknowledgements}
We would like to thank Y. Kojima, T. Nakamura, K. Nakao, K. Oohaka, 
M. Sasaki, B. F. Schutz, M. Shibata and T. Tanaka 
for fruitful discussion. 
Part of this work is based on the collaboration with M. Shibata. 
This work is supported in 
part by the Japanese Grant-in-Aid for Science Research fund of 
Ministry of Education, Science and Culture No.09640332 (T.F.) 
and in part by Soryushi Shogakukai (H.A.).


\appendix

\section{Conserved quantities for 2PN approximation of uniformly rotating 
fluid}

The conserved quantities in the 2PN 
approximation will be useful to investigate the stability 
property of equilibrium solutions obtained in numerical calculations. 
Since the definitions of the conserved quantities are given 
in section 3, we shall present only the results in this appendix. 

\noindent
(1)Conserved mass \cite{asf}; 
\begin{equation}
M_{\ast} \equiv \int \rho_{\ast}d^3x, 
\end{equation}
where 
\begin{equation}
\rho_{\ast}=\rho \biggl[1+{1 \over c^2}\Bigl({1 \over 2}v^2+3 U\Bigr)
+{1 \over c^4}\Bigl({3 \over 8}v^4+{13 \over 2}v^2 U
+{45 \over 4}U^2+3U\Pi+\three\beta^i v^i\Bigr)+O(c^{-6})\biggr].
\end{equation}
\noindent
                                
(2)ADM mass \cite{adm,wald,asf}; 
\begin{equation}
M_{ADM}=-{1 \over 2\pi}\int \Delta \psi d^3 x \equiv 
\int \rho_{ADM}d^3x, 
\end{equation}
where 
\begin{eqnarray}
\rho_{ADM}&=&\rho \biggl[1+{1 \over c^2}
\Bigl(v^2+\Pi+{5 \over 2}U\Bigr) 
\hspace*{8.5cm}
\nonumber\\ 
&&~~~
+{1 \over c^4}\Bigl(v^4+9 v^2 U+\Gamma\Pi v^2+5 U\Pi
+{35 \over 4}U^2 +{3 \over 2}\three\beta^i v^i \Bigr)  
+O(c^{-6}) \biggr]. 
\end{eqnarray}
\noindent
(3)Total energy, which is calculated from $M_{ADM}-M_*$ in the third PN 
order \cite{asf}; 
\begin{equation}
E \equiv \int \rho_E d^3x, 
\end{equation}
where 
\begin{eqnarray}
\rho_E &=&\rho\biggl[
\biggl( {1 \over 2}v^2+\Pi-{1 \over 2}U \biggr)
+{1 \over c^2}\biggl({5 \over 8}v^4+{5 \over 2}v^2 U
+\Gamma v^2\Pi+2 U\Pi-{5 \over 2}U^2
+{1 \over 2}\three\beta^i v^i \biggr)  \nonumber\\
&&
+{1 \over c^4}\biggl\{ {11 \over 16}v^6+v^4\Bigl(\Gamma \Pi
+{47 \over 8} U\Bigr) +v^2 \Bigl( 4\four\psi+6\Gamma \Pi U
+{41 \over 8}U^2+{5 \over 2}\three\beta^i v^i-X \Bigr)  
\hspace{5mm}
\nonumber\\
&&
-{5 \over 2}U^3+2\Gamma \three\beta^i v^i \Pi+5 \Pi\four\psi
+5U \three\beta^i v^i-{15 \over 2}U\four\psi+{5 \over 4}U^2\Pi  
\nonumber\\
&&
+{1 \over 2}h_{ij}v^iv^j
+{1 \over 2}\three\beta^i\three\beta^i  \nonumber\\
&&
+{U \over 16\pi \rho}\biggl(2h_{ij}U_{,ij}+
\three\beta^i_{,j} \Bigl(\three\beta^i_{,j}+\three\beta^j_{,i}
-{2 \over 3}\delta_{ij}\three\beta^k_{,k}\Bigr)
\biggr)\biggr\}+O(c^{-6}) \biggr]. 
\end{eqnarray}
It is noteworthy that terms including $\five\beta^i$ cancel out in total. 

\noindent
(4)Total linear and angular momenta: In the case $K_i^{~i}=0$, 
these are calculated from (Wald 1984) 
\begin{eqnarray}
P_i&=&{1 \over 8\pi}\lim_{r \to \infty}\oint 
K_{ij} n^j  dS  \nonumber\\
&=&{1 \over 8\pi} \int (\psi^6 K_i^{~j})_{,j} d^3x  \nonumber\\
&=& \int \Bigl( J_i +{1 \over 16\pi} \psi^4 \tilde \gamma_{jk,i}
K^{jk} \Bigr) \psi^6 d^3x, 
\label{jjjeq}
\end{eqnarray}
where $J_i=(\rho c^2+\rho \Pi+P) \alpha u^0u_i$. Up to the 
2PN order, the second term in the last line of Eq.$(\ref{jjjeq})$ becomes 
\begin{eqnarray}
&&{1 \over 16\pi} \int  h_{jk,i} \three \beta^j_{,k} d^3x \nonumber\\ 
&=&{1 \over 16\pi} \int \biggl[ \Bigl(h_{jk,i} \three \beta^{j} \Bigr)_{,k}-
h_{jk,ik} \three \beta^{j} \biggr] d^3x \nonumber\\
&=&{1 \over 16\pi} \lim_{r \to \infty}\oint 
h_{jk,i} \three \beta^{j}  n^k dS=0, 
\end{eqnarray}
where we use $h_{jk} \rightarrow O(r^{-1})$ and  
$\three \beta^{j} \rightarrow O(r^{-2})$ at $r \rightarrow \infty$, and 
the gauge condition $h_{jk,k}=0$. 
Thus, in the 2PN approximation, $P_i$ becomes 
\begin{equation}
P_i \equiv \int p_{i} d^3x ,
\end{equation}
where 
\begin{eqnarray}
p_i=\rho \biggl[ && v^i+{1 \over c^2}\biggl\{
v^i\Bigl(v^2+\Gamma\Pi+6U\Bigr)+\three\beta^i \biggr\} 
+{1 \over c^4}\biggl\{h_{ij}v^j+\five \beta^i 
\nonumber\\
&&+\three\beta^i\Bigl(v^2+6U+\Gamma \Pi\Bigr) 
+v^i\Bigl(2\three\beta^i v^i+10\four\psi+6\Gamma \Pi U 
\nonumber\\
&&+{67 \over 4}U^2+\Gamma \Pi v^2+10Uv^2+v^4-X\Bigr)\biggr\}
+O(c^{-5}) \biggr]. 
\end{eqnarray}
The total angular momentum $J$ becomes
\begin{equation}
J= \int p_{\varphi} d^3x , 
\end{equation}
where $p_{\varphi}=-yp_x+xp_y$.

\section{Construction of Green's function}
In this appendix we shall present a detailed derivation of 
Green's function for $\Box_M$. 
The following procedure is similar to that by Thorne (1980) 
for the Minkowski background spacetime \cite{thorne80}. 

The Green's function satisfying Eq.($\ref{curvedG}$) can be constructed 
by using the homogeneous solutions for the equation 
\begin{equation}
\Box_M\Psi=0 . 
\label{homoeq}
\end{equation}
The homogeneous solution for Eq.($\ref{homoeq}$) takes a form of 
\begin{equation}
e^{-i\omega t}f_l(\rho)Y_{l m}(\theta,\phi) , 
\label{homsol}
\end{equation}
where we defined 
\begin{equation}
\rho=\omega r . 
\end{equation}
Then the radial function $\tilde f_l(\rho)\equiv \rho f_l(\rho)$ 
satisfies  
\begin{equation}
\Bigl( {d^2 \over d\rho^2}+1+{4M\omega \over \rho}-{l(l+1) \over \rho^2} 
\Bigr) \tilde f_l(\rho)=0 , 
\end{equation}
so that Eq.($\ref{homsol}$) is a solution for Eq.($\ref{homoeq}$). 
Thus we can obtain homogeneous solutions for Eq.($\ref{homoeq}$) 
by choosing $\tilde f_l(\rho)$ as one of spherical Coulomb functions; 
$u^{(\pm)}_l(\rho;\gamma)$, $F_l(\rho;\gamma)$ and 
$G_l(\rho;\gamma)$ with $\gamma=-2M\omega$. 
Here, we adopted the following definition of the spherical Coulomb 
function \cite{messiah} as 
\begin{eqnarray}
F_l(\rho;\gamma)&=&c_le^{i\rho}\rho^{l+1}F(l+1+i\gamma|2l+2|-2i\rho) , 
\nonumber\\ 
u^{(\pm)}_l(\rho;\gamma)&=&\pm 2ie^{\mp i\sigma_l} c_l 
e^{\pm i\rho}\rho^{l+1} 
W_1(l+1\pm i\gamma|2l+2|\mp 2i\rho) , \nonumber\\
G_l(\rho;\gamma)&=&{1 \over 2}(u^{(+)}_l e^{i\sigma_l}
+u^{(-)}_l e^{-i\sigma_l}) , 
\end{eqnarray}
where $c_l$ and $\sigma_l$ are defined as 
\begin{eqnarray}
c_l&=&2^l e^{-\pi\gamma /2}{|\Gamma(l+1+i\gamma)| \over (2l+1)!} , 
\nonumber\\
\sigma_l&=&\mbox{arg}\Gamma(l+1+i\gamma) . 
\label{cl}
\end{eqnarray}
Here, $F$ and $W_1$ are the confluent hypergeometric function and 
the Whittaker's function respectively. 
These spherical Coulomb functions have asymptotic behavior as 
\begin{eqnarray}
F_l&\sim&\sin\Bigl(\rho-\gamma\ln{2\rho}-{1 \over 2}l\pi+\sigma_l \Bigr) , 
\nonumber\\
G_l&\sim&\cos\Bigl(\rho-\gamma\ln{2\rho}-{1 \over 2}l\pi+\sigma_l \Bigr) , 
\nonumber\\
u^{(\pm)}&\sim&\exp\Bigl[ \pm i\Bigl( \rho-\gamma\ln{2\rho}
-{1 \over 2}l\pi \Bigr) \Bigr] \quad\mbox{for}\quad r \to \infty , 
\label{coulomb1}
\end{eqnarray}
and 
\begin{eqnarray}
F_l&\sim&c_l \rho^{l+1} , \nonumber\\
G_l&\sim&{1 \over (2l+1)c_l} \rho^{-l} \quad\mbox{for}\quad r \to 0 . 
\label{coulomb2}
\end{eqnarray}

In terms of spherical Coulomb functions introduced above, we can 
construct the Green's function in the standard way.
\begin{eqnarray}
G^{(\epsilon)}_M(x,x')&=&\sum_{l m} e^{i \sigma_l} 
\int d\omega sgn(\omega) \Bigl( 
\Psi^{\epsilon \,\omega\,l\,m}(x) \Psi^{S\,\omega\,l\,m\,\ast}(x') 
\theta(r-r') \nonumber\\
&&~~~~~~~~~~~~~~~~~~~~~~~~~~
+\Psi^{S\,\omega\,l\,m}(x) \Psi^{\epsilon \,\omega\,l\,m\,\ast}(x') 
\theta(r'-r) \Bigr) , 
\label{appgreen}
\end{eqnarray}
where $*$ denotes the complex conjugate, $sgn(\omega)$ denotes 
a sign of $\omega$  and we defined $\Psi^{\epsilon\,\omega\,l\,m}(x)$ 
and $\Psi^{S\,\omega\,l\,m}(x)$ as 
\begin{eqnarray}
\Psi^{\epsilon\omega\,l\,m}(x)&=&\sqrt{{|\omega| \over 2\pi}}
e^{-i\omega t} \rho^{-1} u^{(+)}_{l}(\rho;\gamma)
Y_{l\,m} , \nonumber\\
\Psi^{S\,\omega\,l\,m}(x)&=&\sqrt{{|\omega| \over 2\pi}} 
e^{-i\omega t} \rho^{-1} F_{l}(\rho;\gamma) 
Y_{l\,m} . 
\end{eqnarray}
The retarded Green's function is obtained for $\epsilon=+$.

\section{Derivation of the formula $(5.28)$}

We evaluate the asymptotic form of the retarded integral as 
\begin{eqnarray}
\Box^{-1}\Bigl[ {\hat n_L \over r^k} F(t-r) \Bigr]&&=
\int d^4x' G_0(x,x') \Bigl[ {\hat n'_L \over r'^k} F(t'-r') \Bigr]
\nonumber\\
&&\to -i \int d\omega \omega e^{-i\omega t} h_l(\omega r) F_{\omega} 
\hat n_L \int^r_0 r'^2 dr' j_l(\omega r') {e^{i \omega r'}\over r'^k} 
\nonumber\\
&&\hspace*{6cm}\mbox{for large r} , 
\label{formula0}
\end{eqnarray}
where we defined $F_{\omega}$ as
\begin{equation}
F_{\omega}={1 \over 2\pi} \int dt \, e^{i\omega t} F(t) . 
\label{fomega}
\end{equation}
Since the Bessel function is defined as 
\begin{equation}
J_{\nu}(z)=\sum^{\infty}_{n=0} {(-)^n (z/2)^{\nu+2n} \over 
n! \Gamma(\nu+n+1)} , 
\end{equation}
we obtain formally 
\begin{eqnarray}
\int^{\infty}_0 dy J_{\nu}(by) e^{-ay} y^{\mu-1}
&=&\sum^{\infty}_{n=0} {(-)^n (b/2)^{\nu+2n} \over n! \Gamma(\nu+n+1)}
\int^{\infty}_0 dy e^{-ay} y^{\mu+\nu+2n-1} \nonumber\\
&=&\sum^{\infty}_{n=0} {(-)^n (b/2)^{\nu+2n} \Gamma(\mu+\nu+2n) \over
n! a^{\mu+\nu+2n} \Gamma(\nu+n+1)} .
\label{Besselint3}
\end{eqnarray}
Putting $a=-i$, $b=1$, $\mu=-k+5/2-\lambda$ and $\nu=l+1/2$ 
in Eq.($\ref{Besselint3}$), we obtain 
where we introduced small $\lambda \in C$ in order to 
avoid the pole of $\Gamma(-k+l+3+2n)$ for $k \ge l+3$. 
Thus we obtain 
\begin{eqnarray}
\int^r_0 r'^2 dr' j_l(\omega r') {e^{i\omega r'}\over {r'}^{k+\lambda}}
&=&\omega^{k-3} \Bigl( \sqrt{\pi \over 2}{i^{-k+l+3} \over 2^{l+{1 \over 2}}}
\sum^{\infty}_{n=0} {\Gamma(-k+l+3+2n-\lambda) \over 2^{2n} n! 
\Gamma(l+n+{3 \over 2})} +O(r^{-1}) \Bigr) . \nonumber\\
&&
\label{Besselint5}
\end{eqnarray}
{}From Eqs.($\ref{formula0}$), ($\ref{fomega}$) and
($\ref{Besselint5}$), we obtain 
\begin{eqnarray}
\Box^{-1} \Bigl[ {\hat n_L \over r^k}F(t-r) \Bigr] 
&=&-\lim_{\lambda \to 0} {\sqrt{\pi} \over 2^{l+1}} 
\Bigl( \sum^{\infty}_{n=0} {\Gamma(-k+l+3+2n-\lambda) \over 2^{2n} n! 
\Gamma(l+n+{3 \over 2})} \Bigr) 
{\hat n_L \over r^k} \mathop{F}^{(k-3)}(t-r) \nonumber\\
&&+O(r^{-2}) . 
\label{formula2}
\end{eqnarray}
This equation is valid for $k \ge 3$. 
For $3 \le k \le l+2$, we obtain 
\begin{eqnarray}
\Box^{-1}\Bigl[ {\hat n_L \over r^k} F(t-r) \Bigr]
&=&- 2^{k-3} {(k-3)! (l-k+2)! \over (l+k-2)!} 
{\hat n_L \over r} {\mathop{F}^{(k-3)}}\!\!(t-r)
+O(r^{-2}) \nonumber\\
&&\hspace*{3cm}\mbox{for} \quad 3 \le k \le l+2 , 
\label{formula1}
\end{eqnarray} 
since 
\begin{equation}
{\sqrt{\pi} \over 2^{l+1}} \Bigl( \sum^{\infty}_{n=0} 
{\Gamma(-k+l+3+2n) \over 2^{2n} n! \Gamma(l+n+{3 \over 2})} \Bigr) 
=2^{k-3}{ (k-3)! (l-k+2)! \over (l+k-2)! } 
\quad \mbox{for} \quad 3 \le k \le l+2 . 
\end{equation}
This formula for $3 \le k \le l+2$ has been derived by 
Blanchet and Damour (1988).

\section{Derivation of $(5.29)$}

{}From Eq.($\ref{c201}$), we shall derive Eq.($\ref{c2}$) in this appendix.
First, for the first term in the right hand side of Eq.($\ref{c201}$), 
we use Eq.(\ref{formula1}) to obtain
\begin{eqnarray}
&&(-)^{q+1} \sum^q_{j=0}{ (q+j)! \over 2^j j! (q-j)! } \Box^{-1} 
\Bigl[ \hat n_{iQ} {1 \over r^{j+3}} \mathop{F}^{(q-j)}\!\!(t-r) \Bigr] 
\nonumber\\
&&={(-)^q \over 2(q+1)} {\hat n_{iQ} \over r} \mathop{F}^{(q)}\!(t-r)
+O(r^{-2}) . 
\label{c211}
\end{eqnarray}
Next, the second term in the right hand side of Eq.($\ref{c201}$) 
is calculated as 
\begin{eqnarray}
&&(-)^{q+1} {q \over 2q+1} \sum^{q-2}_{j=0}{ (q+j)! \over 2^j j! (q-j)!}
\Box^{-1} \Bigl[ \delta_{i<a_q} \hat n_{Q-1>} 
{1 \over r^{j+3}} \mathop{F}^{(q-j)}\!\!(t-r) \Bigr] \nonumber\\
&&=(-)^q{q-1 \over 2q+1} {\delta_{i<a_q}\hat n_{Q-1>} \over r} 
\mathop{F}^{(q)}\!(t-r)+O(r^{-2}) , 
\label{c221}
\end{eqnarray}
and 
\begin{eqnarray}
&&(-)^{q+1} {q \over 2q+1} \sum^{q}_{j=q-1}{ (q+j)! \over 2^j j! (q-j)!}
\Box^{-1} \Bigl[ \delta_{i<a_q} \hat n_{Q-1>} 
{1 \over r^{j+3}} \mathop{F}^{(q-j)}\!\!(t-r) \Bigr] \nonumber\\
&=&(-)^q {q (2q)! \over (2q+1) q!} {\delta_{i<a_q}\hat n_{Q-1>} \over r} 
\mathop{F}^{(q)}\!(t-r) \nonumber\\
&&\times\lim_{\lambda \to 0} \Bigl[ 
\sum^{\infty}_{n=1} \Bigl\{ {(q+n)! \over n! (2q+2n)!} 
\Bigl( \Gamma(2n-\lambda)+\Gamma(2n-1-\lambda) \Bigr) \nonumber\\
&&\hspace*{2cm}+{q! \over (2q)!} \Bigl( \Gamma(-\lambda) 
+\Gamma(-1-\lambda) \Bigr) \Bigr\} \Bigr] +O(r^{-2}) \nonumber\\
&&=(-)^{q+1} {2q-1 \over 2(2q+1)} {\delta_{i<a_q}\hat n_{Q-1>} \over r} 
\mathop{F}^{(q)}\!(t-r) +O(r^{-2}). 
\label{c222}
\end{eqnarray}
Here we used Eq.($\ref{formula3}$), 
\begin{equation}
\sum^{\infty}_{n=1} {(q+n)! \over n! (2q+2n)!} 
\Bigl( \Gamma(2n)+\Gamma(2n-1) \Bigr)={q! \over 2q (2q)!} , 
\end{equation}
and 
\begin{equation}
\lim_{\lambda \to 0} \Bigl( \Gamma(-\lambda)+\Gamma(-1-\lambda) \Bigr) 
=\lim_{\lambda \to 0} (-\lambda) \Gamma(-1-\lambda) = -1 , 
\label{gamma1}
\end{equation}
which, in fact, means {\it the cancellation of the poles}. 

{}From Eqs.($\ref{c201}$) and ($\ref{c211}$)-($\ref{c222}$), we obtain 
\begin{eqnarray}
&&\Box^{-1} \Bigl[ \partial_i ({1 \over r}) \hat\partial_Q 
\Bigl( {F(t-r) \over r} \Bigr) \Bigr] \nonumber\\
&&={(-)^q \over 2(q+1)} \Bigl[ \hat n_{iQ}-{q+1 \over 2q+1}
\delta_{i<a_1} \hat n_{Q-1>} \Bigr] {1 \over r} 
\mathop{F}^{(q)}\!(t-r)+O(r^{-2}) , 
\label{appc2}
\end{eqnarray}
which equals to (C2) obtained by Blanchet \cite{blan95}. 
As for Eq.($\ref{appc2}$), we used 
unphysical but small parameter $\lambda$ in order to 
avoid the pole of the integrals. 
However, in the limit $\lambda \to 0$, we showed explicitly 
the cancellation of poles and obtained the finite values in total.

\newpage

\begin{center}
{\bf Table 1}
\end{center}

\noindent
List of potentials to be solved (column 1), 
Poisson equations for them (column 2), 
and other potential variables which appear in the source term of the 
Poisson equation (column 3). Note that 
$i$ and $j$ run $x,y,z$. Also, 
note that we do not have to solve $\eta_{0z}$, $\five P_{0z}$, 
$q_{yy}$, $q_{zz}$ and $h_{zz}^{(U)}$. 

$$\vbox{\offinterlineskip
\tabskip=0pt
\def\tablerule{\noalign{\hrule}}
\def\tablespace{\omit&height2pt&\omit&&\omit&&\omit&&
\omit&&\omit&&\omit&&\omit&\cr}
\halign{
\strut#&\vrule#&
\quad\hfil#\hfil\quad &\vrule#& \quad\hfil#\hfil\quad &\vrule#&
\quad\hfil#\hfil\quad &\vrule#& \hfil#\hfil~ &\vrule#&
\quad\hfil#\hfil\quad &\vrule#& \quad\hfil#\hfil\quad &\vrule#&
\quad\hfil#\hfil\quad &\vrule#& \quad\hfil#\hfil\quad &\vrule#&
\quad\hfil#\hfil\quad &\vrule#& \quad\hfil#\hfil\quad &\vrule# \cr
\tablerule\tablespace
&& Pot. && Eq. && Needed pots. && && Pot. && Eq. && Needed pots. 
& \cr\tablespace\tablerule\tablespace
&& $U$ && (2.11) && None && && $q_{ij}$ && (4.6) && None  
& \cr\tablespace\tablerule\tablespace
&& $q_i$ && (3.14) && None && && $Q_{0i}^{(I)}$ && (4.15) && $U$
&\cr\tablespace\tablerule\tablespace
&& $q_2$ && (4.1)&& None && && $\eta_{0i}$ && (4.16) && $U$
&\cr\tablespace\tablerule\tablespace
&& $q_{2i}$ && (4.2) && None && && $\five P_{0i}$ && (4.17) && $U,~q_i$
&\cr\tablespace\tablerule\tablespace
&& $q_4$ && (4.3) && None && && $\six\alpha_0$ && (4.21) 
&& $U,~q_e,~q_u,~h_{ij}^{(U)},~Q_{0i}^{(I)}$
&\cr\tablespace\tablerule\tablespace
&& $q_u$ && (4.4) && $U$  && && $\six\alpha_2$ && (4.22) 
&& $U,~q_2,~q_i,~q_{2i},~q_{ij}$
&\cr\tablespace\tablerule\tablespace
&& $q_e$ && (4.5) && None && && $h_{ij}^{(U)}$ && (3.1) && $U$
&\cr\tablespace\tablerule
}}$$



\begin{center}
{\bf Table 2}
\end{center}

\noindent
Variables to be solved in order to obtain the original metric variables. 

$$\vbox{\offinterlineskip
\tabskip=0pt
\def\tablerule{\noalign{\hrule}}
\def\tablespace{\omit&height2pt&\omit&&\omit&&\omit&\cr}
\halign{
\strut#&\vrule#&
\quad\hfil#\hfil\quad &\vrule#& \quad\hfil#\hfil\quad &\vrule#&
\quad\hfil#\hfil\quad &\vrule#& \quad\hfil#\hfil\quad &\vrule#&
\quad\hfil#\hfil\quad &\vrule#& \quad\hfil#\hfil\quad &\vrule#&
\quad\hfil#\hfil\quad &\vrule#& \quad\hfil#\hfil\quad &\vrule# \cr
\tablerule\tablespace
&& Metric && Variables to be solved && see Eq.
& \cr\tablespace\tablerule\tablespace
&& $U$ && $U$ && (2.11)
& \cr\tablespace\tablerule\tablespace
&& $\three\beta^i$ && $q_i$,~$U$ && (3.17)
& \cr\tablespace\tablerule\tablespace
&& $X$ && $q_2$,~ $q_u$,~ $q_e$ && (4.7)
& \cr\tablespace\tablerule\tablespace
&& $\four\psi$ && $q_2$,~ $q_u$,~ $q_e$ && (4.8)
& \cr\tablespace\tablerule\tablespace
&& $\five \beta^{i\,(A)}$ && $\five P_{0i}$,~$\eta_{0i}$,~$q_u$,~$q_e$ &&(4.18)
& \cr\tablespace\tablerule\tablespace
&& $\five \beta^{i\,(B)}$ && $q_{2i}$,~$q_2$ && (4.18) 
& \cr\tablespace\tablerule\tablespace
&& $\six\alpha$ &&  $\six\alpha_0$,~ $\six\alpha_2$,~ $q_4$ &&(4.20)
& \cr\tablespace\tablerule\tablespace
&& $h_{ij}^{(U)}$ && $h_{ij}^{(U)}$ && (3.1)
& \cr\tablespace\tablerule\tablespace
&& $h_{ij}^{(S)}$ && $q_{ij},q_2$ &&(4.14)
& \cr\tablespace\tablerule\tablespace
&& $h_{ij}^{(A)}$ && $Q_{0i}^{(I)}$,~ $q_u$,~ $q_e$ && (4.19)
& \cr\tablespace\tablerule\tablespace
&& $h_{ij}^{(B)}$ && $q_{ij}$,~ $q_2$,~ $q_{2i}$,~ $q_i$ && (4.19)
& \cr\tablespace\tablerule
}}$$


\end{document}